\documentclass[12pt, preprint]{imsart}	
\RequirePackage[colorlinks,citecolor=blue,urlcolor=blue]{hyperref}
\usepackage{subfigure}
\usepackage{amsmath,bm}
\usepackage{amsthm}
\usepackage{amsopn}
\usepackage{graphicx}
\usepackage{amssymb}
\usepackage{amsfonts}		
\usepackage{multirow}
\usepackage[usenames,dvipsnames]{color}
\usepackage[]{natbib} 
\usepackage[total={150mm,240mm}]{geometry}
\usepackage[ruled,vlined]{algorithm2e}
\usepackage[usenames]{color}
\usepackage[titletoc,title]{appendix}

\newtheorem{prop}{Proposition}	
\usepackage{soul}
\setstcolor{red}
\makeatletter 		
\newcommand{\figcaption}{\def\@captype{figure}\caption}
\newcommand{\tabcaption}{\def\@captype{table}\caption}

\makeatother
\newcommand{\ignore}[1]{}

\begin{document}
	
\begin{frontmatter}
	\title{Regularized Principal Component Analysis for Spatial Data}
	\runtitle{Regularized Principal Component Analysis for Spatial Data}	
	\author{
	    Wen-Ting Wang \\
	    Institute of Statistics \\
	    National Chiao Tung University \\
	    \emph{egpivo@gmail.com} \\
	    \and\\
	    Hsin-Cheng Huang \\
	    Institute of Statistical Science \\
	    Academia Sinica \\
	    \emph{hchuang@stat.sinica.edu.tw} \\
	}
\doublespacing
\begin{abstract}
In many atmospheric and earth sciences,
it is of interest to identify dominant spatial patterns of variation
based on data observed at $p$ locations and $n$ time points with the possibility that $p>n$.
While principal component analysis (PCA) is commonly applied to find the dominant patterns,
the eigenimages produced from PCA may exhibit patterns that are too noisy to be physically meaningful when $p$ is large relative to $n$.
To obtain more precise estimates of eigenimages, we propose a regularization approach incorporating smoothness and sparseness of eigenimages,
while accounting for their orthogonality. Our method allows data taken at irregularly spaced or sparse locations.
In addition, the resulting optimization problem can be solved using the alternating direction method of multipliers,
which is easy to implement, and applicable to a large spatial dataset.
Furthermore, the estimated eigenfunctions provide a natural basis
for representing the underlying spatial process in a spatial random-effects model,
from which spatial covariance function estimation and spatial prediction can be efficiently performed using
a regularized fixed-rank kriging method. Finally, the effectiveness of the proposed method is demonstrated by several numerical examples.\\


\end{abstract}
\end{frontmatter}
		
\doublespacing
\section{Introduction}
	
In many atmospheric and earth sciences, it is of interest to identify dominant spatial patterns of variation
based on data observed at $p$ locations with $n$ repeated measurements, where $p$ may be larger than $n$.
The dominant patterns are the eigenimages of the underlying (nonstationary) spatial covariance function
with large eigenvalues. A commonly used approach for estimating the eigenimages is the principal component analysis (PCA),
also known as the empirical orthogonal function analysis in atmospheric science.
However, when $p$ is large relative to $n$, the leading eigenimages produced from PCA may be noisy with high estimation variability,
or exhibit some bizarre patterns that are not physically meaningful.
To enhance the interpretability, a few approaches, such as rotation of components according to some criteria
(see \citet{rotate}, \citet{jolliffe1987rotation}, \citet{richman1987rotation}),
have been proposed to form more desirable patterns. However, how to obtain a desired rotation in practice is not completely clear.
Some discussion can be found in \citet{eofreview}.

Another approach to aid interpretation is to seek sparse or spatially localized patterns, which can be done by imposing an $L_1$ constraint
or adding an $L_1$ penalty to an original PCA optimization formulation (\citet{lassoeof}, \citet{spca},  \citet{Shen}, \citet{spca2}, and \citet{spca3}). However, this approach may produce a pattern with isolated zero and nonzero components, and
except \citet{lassoeof} and \citet{spca3}, the PC estimates produced from these approaches may not have orthogonal PC loadings.

For continuous spatial domains, the problem becomes even more challenging.
Instead of looking for eigenimages on a lattice, we need to find eigenfunctions by essentially solving an infinite dimensional problem
based on data observed at possibly sparse and irregularly spaced locations.
Although some approaches have been developed using functional principal component analysis
(see e.g., \citet{bfda}, \citet{fpca_long} and \citet{fda}), they typically focus on one-dimensional processes, or require data observed at dense locations.
In particular, these methods generally do not work well when data are observed at fixed but sparse locations.
Reviews of PCA on spatial data can be found in \citet{eofreview} and \citet{spatial_pca}.
	
In this research, we propose a regularization approach for estimating the dominant patterns,
taking into account smoothness and localized features that are expected in real-world spatial processes.
The proposed estimates are directly obtained by solving a minimization problem.
We call our method \emph{SpatPCA}, which not only gives effective estimates of dominant patterns,
but also provides an ideal set of basis functions for estimating the underlying (nonstationary) spatial covariance function,
even when data are irregularly or sparsely located in space.
In addition, we develop a fast algorithm to solve the resulting optimization problem
using the alternating direction method of multipliers (ADMM) (see \citet{admm}).
An R package called SpatPCA is developed and available on the Comprehensive R Archive Network (CRAN).

The rest of this paper is organized as follows. In Section~\ref{sec: proposal}, we introduce the proposed SpatPCA method,
including dominant patterns estimation and spatial covariance function estimation.
Our ADMM algorithm for computing the SpatPCA estimates is provided in Section~\ref{sec:algorithm}.
Some simulation experiments that illustrate the superiority of SpatPCA and an application of SpatPCA to a global sea surface
temperature dataset are presented in Section~\ref{sec:simulation}.

\section{The Proposed Method}
\label{sec: proposal}
	
Consider a sequence of zero-mean $L^2$-continuous spatial processes, $\{\eta_i(\bm s); \bm{s} \in D\}$; $i=1,\dots,n$,
defined on a spatial domain $D \subset \mathbb{R}^d$,
which are mutually uncorrelated, and have a common spatial covariance function,
$C_\eta(\bm{s},\bm{s}^*)=\mathrm{cov}(\eta_i(\bm{s}),\eta_i(\bm{s}^*))$.
We consider a rank-$K$ spatial random-effects model for $\eta_i(\cdot)$:
\begin{equation*}
	\eta_i(\bm{s})=(\varphi_1(\bm{s}),\dots,\varphi_{K}(\bm{s}))\bm{\xi}_i=
	\sum_{k=1}^{K}\xi_{ik}\varphi_k(\bm{s});\quad\bm{s}\in D,\quad i=1,\dots,n,
\end{equation*}
	
\noindent where $\{\varphi_k(.)\}$ are unknown orthonormal basis functions,
$\bm{\xi}_i=(\xi_{i1},\dots,\xi_{iK})'\sim(0,\bm{\Lambda})$; $i=1,\dots,n$, are uncorrelated random variables,
and $\bm{\Lambda}$ is an unknown symmetric nonnegative-definite matrix, denoted by $\bm{\Lambda}\succeq\bm{0}$.
A similar model based on given $\{\varphi_k(\cdot)\}$ was introduced by \citet{fixedrank}
and in a Bayesian framework by \citet{bayes}.

Let $\lambda_{kk'}$ be the $(k,k')$-th entry of $\bm{\Lambda}$. Then the spatial covariance function
of $\eta_i(\cdot)$ is:
\begin{equation}
	C_\eta(\bm{s},\bm{s}^*)=\mathrm{cov}(\eta_i(\bm{s}),\eta_i(\bm{s}^*))=
	\sum_{k=1}^{K}\sum_{k'=1}^{K}	\lambda_{kk'}\varphi_k(\bm{s})\varphi_{k' }(\bm{s}^*).
	\label{eq:covariance}
\end{equation}

\noindent Note that $\bm{\Lambda}$ is not restricted to be a diagonal matrix.

Let $\bm{\Lambda}=\bm{V}\bm{\Lambda^*}\bm{V}'$ be the eigen-decomposition of $\bm{\Lambda}$,
where $\bm{V}$ consists of $K$ orthonormal eigenvectors, and  $\bm{\Lambda}^*=\mathrm{diag}(\lambda^*_1,\dots,\lambda^*_{K})$
consists of eigenvalues with $\lambda^*_1\geq\cdots\geq\lambda^*_{K}$.
Let $\bm{\xi}^*_i=\bm{V}'\bm{\xi}_i$ and
\[
(\varphi^*_1(\bm{s}),\dots,\varphi^*_{K}(\bm{s}))
=(\varphi_1(\bm{s}),\dots,\varphi_{K}(\bm{s}))\bm{V};\quad\bm{s}\in D.
\]
Then $\varphi^*_k(\cdot)$'s are also orthonormal,
and $\xi^*_{ik}\sim(0,\lambda^*_k)$; $i=1,\dots,n,\,k=1,\dots,K$, are mutually uncorrelated.
Therefore, we can rewrite $\eta_i(\cdot)$ in terms of $\varphi^*_k(\cdot)$'s:
\begin{equation}
	\eta_i(\bm{s})=(\varphi^*_1(\bm{s}),\dots,\varphi^*_{K}(\bm{s}))\bm{\xi}^*_i=
	\sum_{k=1}^{K}\xi^*_{ik}\varphi^*_k(\bm{s});\quad\bm{s}\in D.
	\label{eq:eta}
\end{equation}
	
\noindent The above expansion is known as the Karhunen-Lo\'{e}ve expansion
of $\eta_i(\cdot)$ (\citet{karhunen}; \citet{loeve}) with $K$ nonzero eigenvalues,
where $\varphi^*_k(\cdot)$ is the $k$-th eigenfunction of $C_\eta(\cdot,\cdot)$ with $\lambda^*_k$ the corresponding eigenvalue.
	
Suppose that we observe data $\bm{Y}_i=(Y_i(\bm{s}_1),\dots,Y_i(\bm{s}_p))'$ with
added white noise $\bm{\epsilon}_i\sim(\bm{0},\sigma^2\bm{I})$ at $p$ spatial locations, $\bm{s}_1,\dots,\bm{s}_p \in D$,  according to
\begin{equation}
\bm{Y}_i=\bm{\eta}_i+\bm{\epsilon}_i=\bm{\Phi}\bm{\xi}_i+\bm{\epsilon}_i;\quad i=1,\dots,n,
\label{eq:measurement}
\end{equation}
	
\noindent where $\bm{\eta}_i=(\eta_{i}(\bm{s}_1),\dots,\eta_{i}(\bm{s}_p))'$,
$\bm{\Phi}=(\bm{\phi}_1,\dots,\bm{\phi}_K)$ is a $p\times K$ matrix with the $(j,k)$-th entry $\varphi_k(\bm{s}_j)$, and
$\bm{\epsilon}_i$'s and $\bm{\xi}_{i}$'s are uncorrelated. Our goal is to identify
the first $L\leq K$ dominant patterns, $\varphi_1(\cdot),\dots,\varphi_{L}(\cdot)$, with relatively large
$\lambda^*_1,\dots,\lambda^*_{L}$.
Additionally,  we are interested in estimating $C_\eta(\cdot,\cdot)$, which is essential for spatial prediction.

Let $\bm{Y}=(\bm{Y}_1,\dots,\bm{Y}_n)'$ be the $n\times p$ data matrix. Throughout the paper, we assume that the mean of $\bm{Y}$ is known as zero.
So the sample covariance matrix of $\bm{Y}$ is $\bm{S}=\bm{Y}'\bm{Y}/n$.
A popular approach for estimating $\{\varphi^*_k(\cdot)\}$ is PCA, which
estimates $(\varphi^*_k(\bm{s}_1),\dots,\varphi^*_k(\bm{s}_p))'$ by $\tilde{\bm{\phi}}_k$,
the $k$-th eigenvector of $\bm{S}$, for $k=1,\dots,K$.
Let $\tilde{\bm{\Phi}}=\big(\tilde{\bm{\phi}}_1,\dots,\tilde{\bm{\phi}}_K\big)$ be a $p\times K$ matrix
formed by the first $K$ principal component loadings.
Then $\tilde{\bm{\Phi}}$
solves the following constrained optimization problem:
\[
\min_{\bm{\Phi}}\|\bm{Y}-\bm{Y}\bm{\Phi}\bm{\Phi}'\|^2_F\quad\text{subject to }	
\bm{\Phi}'\bm{\Phi}=\bm{I}_K,
\] 
where $\bm{\Phi}=(\bm{\phi}_1,\dots,\bm{\phi}_K)$ and $\|\bm{M}\|_F=
\Big(\displaystyle\sum_{i,j}m^2_{ij}\Big)^{1/2}$ is the Frobenius norm of a matrix $\bm{M}$.
Unfortunately, $\tilde{\bm{\Phi}}$ tends to have high estimation variability when $p$ is large (leading to excessive number of parameters),
$n$ is small, or $\sigma^2$ is large.
Consequently, the patterns of $\tilde{\bm{\Phi}}$ may be too noisy to be physically interpretable.
In addition, for a continuous spatial domain $D$,
we also need to estimate $\varphi_k^*(\bm{s})$'s for locations with no data observed (i.e., $\bm{s}\notin\{\bm{s}_1,\dots,\bm{s}_p\}$);
see some discussion in Section 12.4 and 13.6 of \citet{jolliffe2002principal}.

\subsection{Regularized Spatial PCA}
	
To prevent high estimation variability of PCA, we adopt a regularization approach by minimizing the following objective function:
\begin{equation}\label{eq:objective}
\|\bm{Y}-\bm{Y}\bm{\Phi}\bm{\Phi}'\|^2_F +\tau_1\sum_{k=1}^K
J({\varphi}_k)+\tau_2\sum_{k=1}^K \sum_{j=1}^{p}\big|\varphi_k(\bm{s}_j)\big|,
\end{equation}			
	
\noindent over $\varphi_1(\cdot),\dots,\varphi_K(\cdot)$, subject to $\bm{\Phi}'\bm{\Phi}=\bm{I}_K$
and $\bm{\phi}'_1\bm{S}\bm{\phi}_1\geq\bm{\phi}'_2\bm{S}\bm{\phi}_2\geq\cdots\geq\bm{\phi}'_K
\bm{S}\bm{\phi}_K$, where
\[
J(\varphi)=\sum_{z_1+\cdots+z_d=2}\int_{\mathcal{R}^d}\left(
\frac{\partial^2 \varphi(\bm{s})}{\partial x_1^{z_1}\dots\partial x_d^{z_d}}\right)^2 d\bm{s},
\]			
is a roughness penalty, $\bm{s}=(x_1,\dots,x_d)'$, $\tau_1\geq 0$ is a smoothness parameter, and $\tau_2\geq 0$ is a sparseness parameter.
The objective function \eqref{eq:objective} consists of two penalty terms.
The first one is designed to enhance smoothness of $\varphi_k(\cdot)$ through the smoothing spline penalty $J(\varphi_k)$,
while the second one is the $L_1$ Lasso penalty (\cite{lasso}), used to promote sparse patterns by shrinking some PC loadings to zero.
While the $L_1$ penalty alone may lead to isolated zero and nonzero components with no global feature,
when it is paired with the smoothness penalty, local sparsity translates into global sparsity, resulting in connected zero and nonzero patterns.
Hence the two penalty terms together lead to desired patterns that are not only smooth but also localized.
Specifically, when $\tau_1$ is larger, $\hat{\varphi}_k(\cdot)$'s tend to be smoother and \textit{vice versa}.
When $\tau_2$ is larger, $\hat{\varphi}_k(\cdot)$'s are forced to be zero at some $\bm{s}\in D$.
On the other hand, when both $\tau_1$ and $\tau_2$ are close to zero,
the estimates are close to those obtained from PCA.
By suitably choosing $\tau_1$ and $\tau_2$, we can obtain a good compromise among
goodness of fit, smoothness of the eigenfunctions, and sparseness of the eigenfunctions, leading to more interpretable results.
Note that due to computational difficulty, the orthogonal constraint,
is not considered by many PCA regularization methods (e.g., \citet{spca}, \cite{Shen}, \cite{fused}, \cite{hong}).

Although $J(\varphi)$ involves integration, it is well known from the theory of smoothing splines (\citet{nonparametric}) that for each $k=1,\dots,K$, $\hat{\varphi}_k(\cdot)$ has to be a natural cubic spline when $d=1$,
and a thin-plate spline when
$d\in\{2,3\}$ with nodes at $\{\bm{s}_1,\dots,\bm{s}_p\}$. Specifically,
\begin{equation}\label{eq:basis}
	\hat{\varphi}_k(\bm{s})=\sum_{i=1}^p {a}_i g(\|\bm{s}-\bm{s}_i\|)+b_0+\sum_{j=1}^{d} {b}_j x_j\:, 
\end{equation}
	
\noindent where $\bm{s}=(x_1,\dots,x_d)'$,		
\[
g(r) = \left\{
\begin{array}{ll}
\displaystyle\frac{1}{16\pi}r^{2}\log{r};  & \mbox{if $d=2$,}\smallskip\\
\displaystyle\frac{\Gamma(d/2-2)}{16\pi^{d/2}}r^{4-d}; & \mbox{if }d=1,3,\\
\end{array}\right.	 
\]
and the coefficients ${\bm{a}}=\left({a}_1,\dots,{a}_p\right)'$
and ${\bm{b}}=\left({b}_0,b_1,\dots,{b}_{d}\right)'$ satisfy
\[
	\begin{bmatrix}
	\bm{G} & \bm{E} \\
	\bm{E}^T & \bm{0} \\
	\end{bmatrix}
	\begin{bmatrix}
	{\bm{a}}\\
	{\bm{b}}
	\end{bmatrix}		
	=\begin{bmatrix}
	\hat{\bm{\phi}}_k\\
	\bm{0}
	\end{bmatrix}.
\]
Here $\bm{G}$ is a $p\times p$ matrix with the $(i,j)$-th element $g(\|\bm{s}_i-\bm{s}_j\|)$,
and $\bm{E}$ is a $p\times (d+1)$ matrix with the $i$-th row $(1,\bm{s}'_{i})$.
Consequently, $\hat{\varphi}_k(\cdot)$ in (\ref{eq:basis}) can be expressed in terms of $\hat{\bm{\phi}}_k$.
Additionally, the roughness penalty can also be written as
\begin{equation}
\label{eq:smoothness}
J(\varphi_k) =\bm{\phi}'_k \bm\Omega \bm{\phi}_k,
\end{equation}

\noindent with $\bm\Omega$ a known $p\times p$ matrix determined only by $\bm{s}_1,\dots,\bm{s}_p$.
The readers are referred to \citet{nonparametric} for more details regarding smoothing splines. 
	
From (\ref{eq:objective}) and (\ref{eq:smoothness}), the proposed SpatPCA estimate of $\bm{\Phi}$ can be written as:
\begin{equation}
	\hat{\bm{\Phi}}_{\tau_1,\tau_2}=\mathop{\arg\min}_{\bm{\Phi}:\bm{\Phi}'\bm{\Phi} = \bm{I}_K}
	\|\bm{Y} - \bm{Y}\bm{\Phi}\bm{\Phi} '\|^2_F +
	\tau_1 \sum_{k=1}^K \bm{\phi}'_k\bm\Omega\bm{\phi}_k+\tau_2 \sum_{k=1}^K\sum_{j=1}^p\left|{\phi}_{jk}\right|,
	\label{eq:objective2}
\end{equation}			
	
\noindent subject to $\bm{\phi}'_1\bm{S}\bm{\phi}_1\geq\bm{\phi}'_2\bm{S}\bm{\phi}_2\geq\cdots\geq\bm{\phi}'_K
\bm{S}\bm{\phi}_K$.
The resulting estimates of $\varphi_1(\cdot),\dots,\varphi_K(\cdot)$ can be directly computed from (\ref{eq:basis}).
When no confusion may arise, we shall simply write $\hat{\bm{\Phi}}_{\tau_1,\tau_2}$ as $\hat{\bm{\Phi}}$.
Note that the SpatPCA estimate of (\ref{eq:objective2}) reduces to a sparse PCA estimate of \citet{spca}
if the orthogonal constraint is dropped and $\bm{\Omega}=\bm{I}$ (i.e., no spatial structure is considered).

The tuning parameters $\tau_1$ and $\tau_2$ are selected using $M$-fold cross-validation (CV).
First, we partition $\{1,\dots,n\}$ into $M$ parts with as close to the same size as possible.
Let $\bm{Y}^{(m)}$ be the sub-matrix of $\bm{Y}$ corresponding to the $m$-th part, for $m=1,\dots,M$.
For each part, we treat $\bm{Y}^{(m)}$ as the validation data,
and obtain the estimate $\hat{\bm{\Phi}}^{(-m)}_{\tau_1,\tau_2}$ of $\bm{\Phi}$ for $(\tau_1,\tau_2)\in\mathcal{A}$
based on the remaining data $\bm{Y}^{(-m)}$ using the proposed method,
where $\mathcal{A}\subset[0,\infty)^2$ is a candidate index set. The proposed CV criterion is given in terms of an average
residual sum of squares:
\begin{equation}
\label{eq:cv}
\mathrm{CV}_1(\tau_1,\tau_2)=\frac{1}{M}\sum_{m=1}^M\big\|\bm{Y}^{(m)}-\bm{Y}^{(m)}\hat{\bm{\Phi}}_{\tau_1, \tau_2}^{(-m)}
(\hat{\bm{\Phi}}^{(-m)}_{\tau_1, \tau_2})'  \big\|^2_F\:,
\end{equation}

\noindent where $\bm{Y}^{(m)}\hat{\bm{\Phi}}_{\tau_1, \tau_2}^{(-m)}
(\hat{\bm{\Phi}}^{(-m)}_{\tau_1, \tau_2})' $ is the projection of $\bm{Y}^{(m)}$
onto the column space of $\hat{\bm{\Phi}}^{(-m)}_{\tau_1,\tau_2}$. The final $\tau_1$ and $\tau_2$ values are
$(\hat{\tau}_1, \hat{\tau}_2)=\displaystyle\mathop{\arg\min}_{(\tau_1,\tau_2)\in\mathcal{A}}\mathrm{CV}_1(\tau_1,\tau_2)$.

\subsection{Estimation of Spatial Covariance Function}
\label{subsec: cov_fn}

To estimate $C_\eta(\cdot,\cdot)$ in (\ref{eq:covariance}), we also need to estimate the spatial covariance parameters,
$\sigma^2$ and $\bm{\Lambda}$.
We apply the regularized least squares method of \citet{regularized_covariance}:
\begin{equation}
\big(\hat{\sigma}^2,\hat{\bm{\Lambda}}\big)=\mathop{\arg\min}_{(\sigma^2,\bm{\Lambda}):  \sigma^2\geq 0,\,\bm{\Lambda}\succeq\bm{0}}
\bigg\{\frac{1}{2}\big\|\bm{S}-\hat{\bm{\Phi}}\bm{\Lambda}\hat{\bm{\Phi}}'-\sigma^2\bm{I}\big\|^2_F+\gamma\|\hat{\bm{\Phi}}\bm{\Lambda}
\hat{\bm{\Phi}}'\|_{*}\bigg\},
\label{eq:covariance.estimate}
\end{equation}

\noindent where $\gamma \geq 0$ is a tuning parameter, and  $\|\bm{M}\|_{*}=\mathrm{tr}((\bm{M}'\bm{M})^{1/2})$
is the nuclear norm of $\bm{M}$. 
The first term of (\ref{eq:covariance.estimate})
corresponds to goodness of fit by noting that
$\mathrm{var}(\bm{Y}_i)=\bm{\Phi\Lambda\Phi}'+\sigma^2\bm{I}$. The second term of (\ref{eq:covariance.estimate}) is
a convex penalty, shrinking the eigenvalues of $\hat{\bm{\Phi}}\bm{\Lambda}\hat{\bm{\Phi}}'$
to promote a low-rank structure and to avoid the eigenvalues being overestimated.
By suitably choosing a tuning parameter $\gamma$, we can control the bias, while reducing the estimation variability. This is particularly
effective when $K$ is large.

\citet{regularized_covariance} provides a closed-form solution for $\hat{\bm{\Lambda}}$, but requires an iterative procedure
for solving $\hat{\sigma}^2$.
We found that closed-form expressions for both $\hat{\sigma}^2$ and $\hat{\bm{\Lambda}}$ are available, and are shown in the following proposition
with its proof given in the Appendix.
\begin{prop}
The solutions of (\ref{eq:covariance.estimate}) are given by
\begin{align}
	\hat{\bm{\Lambda}}
=&~ \hat{\bm{V}}\mathrm{diag}\big(\hat{\lambda}_1^*,\dots,\hat{\lambda}_K^*\big)\hat{\bm{V}}',
\label{eq:Lambda.hat}\\
	\hat{\sigma}^2
=&~	\left\{
 	\begin{array}{ll}
 	\displaystyle\frac{1}{p-\hat{L}} \bigg(\mathrm{tr}(\bm{S})-\sum_{k=1}^{\hat{L}} \big(\hat{d}_k-\gamma\big)\bigg);
	& \mbox{if $\hat{d}_1 > \gamma$,}\\
 	\displaystyle \frac{1}{p} \left(\mathrm{tr}(\bm{S})\right);
	& \mbox{if $\hat{d}_1 \leq \gamma$ ,}\\
 	\end{array}\right.
	\label{eq:sigma.hat}
\end{align}

\noindent  where $\hat{\bm{V}}\mathrm{diag}(\hat{d}_1,\dots,\hat{d}_K)\hat{\bm{V}}'$ is the
eigen-decomposition of $\hat{\bm{\Phi}}'\bm{S}\hat{\bm{\Phi}}$ with $\hat{d}_1\geq\cdots\geq\hat{d}_K$,
 \begin{equation}
 \hat{L}=\max \bigg\{ L:  \hat{d}_L-\gamma >\frac{1}{p-L} \bigg(\mathrm{tr}(\bm{S})-\sum_{k=1}^L (\hat{d}_k-\gamma)\bigg), L=1,\dots,K \bigg\},
 \label{eq:rank_est}
 \end{equation}
 and $\hat{\lambda}^*_k=\max(\hat{d}_k-\hat{\sigma}^2-\gamma,0)$; $k=1,\dots,K$.
\label{prop:sigma_Lambda_hat}
\end{prop}


With $\hat{\bm{\Lambda}}=\big(\hat{\lambda}_{kk'}\big)_{K\times K}$ given by (\ref{eq:covariance.estimate})
and $\hat{\varphi}_k(\bm{s})$ given by \eqref{eq:basis}, the proposed estimate of $C_\eta(\bm{s},\bm{s}^*)$ is
\begin{equation}
\hat{C}_\eta(\bm{s},\bm{s}^*)
=\sum_{k=1}^K\sum_{k'=1}^K \hat{\lambda}_{kk'}\,\hat{\varphi}_k(\bm{s})\hat{\varphi}_{k'}(\bm{s}^*).
\label{eq:covariance.hat}
\end{equation} 

\noindent Then the proposed estimate of $(\varphi^*_1(\bm{s}),\dots,\varphi^*_K(\bm{s}))$ is
\[
(\hat{\varphi}^*_1(\bm{s}),\dots,\hat{\varphi}^*_K(\bm{s}))=
(\hat{\varphi}_1(\bm{s}),\dots,\hat{\varphi}_K(\bm{s}))\hat{\bm{V}};\quad \bm{s}\in D.
\]

We consider $M$-fold CV to select $\gamma$. As in the previous section, we partition the data
into $M$ parts, $\bm{Y}^{(1)},\dots,\bm{Y}^{(M)}$. For $m=1,\dots,M$, we estimate
$\mathrm{var}\big(\bm{Y}^{(-m)}\big)$ by $\hat{\bm{\Sigma}}^{(-m)}=\hat{\bm{\Phi}}^{(-m)}\hat{\bm{\Lambda}}^{(-m)}_{\gamma}
\big(\hat{\bm{\Phi}}^{(-m)}\big)'+\big(\hat{\sigma}^2_{\gamma}\big)^{(-m)}\bm{I}$ based on the remaining data $\bm{Y}^{(-m)}$
by removing $\bm{Y}^{(m)}$ from $\bm{Y}$,
where $\hat{\bm{\Lambda}}^{(-m)}_{\gamma}$, $\big(\hat{\sigma}^2_{\gamma}\big)^{(-m)}$ and $\hat{\bm{\Phi}}^{(-m)}$
are the estimates of $\bm{\Lambda}$, $\sigma^2$ and $\bm{\Phi}$ based on $\bm{Y}^{(-m)}$,
and for notational simplicity, their dependences on the selected $(\tau_1,\tau_2)$ and $K$ are suppressed. The proposed CV criterion is given by
\begin{equation}
\mathrm{CV}_2(K,\gamma)= \frac{1}{M}\sum_{m=1}^{M}\big\|\bm{S}^{(m)} -
\hat{\bm{\Phi}}^{(-m)}\hat{\bm{\Lambda}}^{(-m)}_{\gamma}(\hat{\bm{\Phi}}^{(-m)})'-(\hat{\sigma}_{\gamma}^2)^{(-m)}\bm{I} \big\|^2_F\:,
\label{eq:cv.gamma}
\end{equation}

\noindent where $\bm{S}^{(m)}=(\bm{Y}^{(m)})'\bm{Y}^{(m)}/n$.
Then the $\gamma$ selected by CV$_2$ based on $K$ is $\hat{\gamma}_K=\displaystyle\mathop{\arg\min}_{\gamma\geq 0}\mathrm{CV}_2(K,\gamma)$.

The dimension of eigen-space $K$, corresponding to the maximum rank of $\bm{\Phi}\bm{\Lambda}\bm{\Phi}'$,
could be selected by traditional approaches based on a given proportion of total variation explained
or the scree plot of the sample eigenvalues.
However, these approaches tend to be more subjective and may not be effective for the covariance estimation purpose.
We propose to select $K$ using CV$_2$ of (\ref{eq:cv.gamma}) by
subsequently increase the value of $K$ from $K=1,2,\dots$, until no further reduction of the CV$_2$ value.
Specifically, we select
\begin{equation}
\hat{K}=\min\{K:\mathrm{CV}_2(K,\hat{\gamma}_{K})\leq\mathrm{CV}_2(K+1,\hat{\gamma}_{K+1}),\,K=1,2,\dots\}.
\label{eq:khat}
\end{equation}

\section{Computation Algorithm}
\label{sec:algorithm}

Solving \eqref{eq:objective2} is a challenging problem especially when both the orthogonal constraint and
the $L_1$ penalty are involved simultaneously.
Consequently, many regularized PCA approaches, such as sparse PCA \citep{spca}, do not cope with the orthogonal constraint.
We adopt the ADMM algorithm by decomposing the original constrained optimization problem
into small subproblems that can be efficiently handled through an iterative procedure.
This type of algorithm was developed early in \citet{amdd_2}, and was systematically studied by \citet{admm} more recently.

First, the optimization problem of \eqref{eq:objective2} is transferred into the following equivalent problem by adding an $p\times K$
parameter matrix $\bm{Q}$: 
\begin{align}\label{eq:ADMM1}
& \min_{\bm{\Phi},\bm{Q}\in\mathbb{R}^{p\times K}} \|\bm{Y} - \bm{Y}\bm{\Phi}\bm{\Phi} '\|^2_F +\tau_1 \sum_{k=1}^K \bm{\phi}_{k}'\bm\Omega\bm{\phi}_k+
\tau_2 \sum_{k=1}^K\sum_{j=1}^{p}\left|\phi_{jk}\right|,
\end{align}			

\noindent subject to $\bm{Q}'\bm{Q}=\bm{I}_K$, $\bm{\phi}'_1\bm{S}\bm{\phi}_1\geq\bm{\phi}'_2\bm{S}\bm{\phi}_2
\geq\cdots\geq\bm{\phi}'_K\bm{S}\bm{\phi}_K$, and a new constrain, $\bm{\Phi}=\bm{Q}$. Then the resulting
constrained optimization problem of \eqref{eq:ADMM1} is solved using the augmented Lagrangian method with its Lagrangian given by
\begin{align*}
	L(\bm{\Phi}, \bm{Q},\bm{\Gamma})
=&~ \|\bm{Y}-\bm{Y}\bm{\Phi}\bm{\Phi}'\|^2_F+\tau_1\sum_{k=1}^K \bm{\phi}_k'\bm\Omega\bm{\phi}_k
	+\tau_2 \sum_{k=1}^K\sum_{j=1}^{p}|\phi_{jk}|\\
&~ + \mathrm{tr}(\bm{\Gamma}'(\bm{\Phi}-\bm{Q}))+\frac{\rho}{2}\|\bm{\Phi}-\bm{Q}\|^2_F\:,
\end{align*}

\noindent subject to $\bm{Q}'\bm{Q}=\bm{I}_K$ and
$\bm{\phi}'_1\bm{S}\bm{\phi}_1\geq\bm{\phi}'_2\bm{S}\bm{\phi}_2\geq\cdots\geq\bm{\phi}'_K\bm{S}\bm{\phi}_K$,
where $\bm{\Gamma}$ is a $p\times K$ matrix of the Lagrange multipliers, and
$\rho>0$ is a penalty parameter to facilitate convergence. Note that the value of $\rho$ does not affect the original optimization problem.
The ADMM algorithm iteratively updates one group of parameters at a time in both the primal and the dual spaces until convergence. Given the initial estimates, $\bm{Q}^{(0)}$ and $\bm{\Gamma}^{(0)}$ of $\bm{Q}$ and $\bm{\Gamma}$,
our ADMM algorithm consists of the following steps at the $\ell$-th iteration:
\begin{align}
	\bm{\Phi}^{(\ell+1)}
=&~ \mathop{\arg\min}_{\bm{\Phi}}L\big(\bm{\Phi}, \bm{Q}^{(\ell)},\bm{\Gamma}^{(\ell)}\big)
	\notag\\
=&~ \mathop{\arg\min}_{\bm{\Phi}} \sum_{k=1}^K \bigg\{\|\bm{z}_k^{(\ell)} - \bm{X}\bm{\phi}_k \|^2 + \sum_{j=1}^p \tau_2|\phi_{jk}|\bigg\}
	\label{eq:phi_1},\\
	\bm{Q}^{(\ell+1)}
=&~ \mathop{\arg\min}_{\bm{Q}: \bm{Q}'\bm{Q}=\bm{I}_K}L\big(\bm{\Phi}^{(\ell+1)},\bm{Q},
	\bm{\Gamma}^{(\ell)}\big)\,=\,\bm{U}^{(\ell)}\big(\bm{V}^{(\ell)}\big)',
	\label{eq:Q_1}\\
\bm{\Gamma}^{(\ell+1)}
=&~ \bm{\Gamma}^{(\ell) }+ \rho\left(\bm{\Phi}^{(\ell+1)}-\bm{Q}^{(\ell+1)}\right), \label{eq:Gamma}
\end{align}

\noindent where $\bm{X} =(\tau_1\bm{\Omega}-\bm{Y}'\bm{Y}+\rho\bm{I}_p/2)^{1/2}$, $\bm{z}^{(\ell)}_k$ is the $k$-th column of $\bm{X}^{-1}(\rho\bm{Q}^{(\ell)}-\bm{\Gamma}^{(\ell)})/2$, $\bm{U}^{(\ell)}\bm{D}^{(\ell)}\big(\bm{V}^{(\ell)}\big)'$ is the singular value decomposition of
$\bm{\Phi}^{(\ell+1)}+\rho^{-1}\bm{\Gamma}^{(\ell)}$, and
$\rho$ must be chosen large enough (e.g., twice the maximum eigenvalue of $\bm{Y}'\bm{Y}$) to ensure that $\bm{X}$ is positive-definite.
Note that \eqref{eq:phi_1} is simply a
Lasso problem (\cite{lasso}), which can be solved effectively using the coordinate descent algorithm \citep{coord}.

Except (\ref{eq:phi_1}), the ADMM steps given by (\ref{eq:phi_1})-(\ref{eq:Gamma}) have closed-form expressions.
In fact, we can make the algorithm involve only closed-form updates by further decomposing (\ref{eq:phi_1}) into another ADMM step.
Specifically, we can introduce another parameters $r_{jk}$'s to replace the last term of \eqref{eq:ADMM1} and
add the constraint, $\phi_{jk}=r_{jk}$ for $j=1,\dots,p$ and $k=1,\dots,K$, to form an equivalent problem:
\begin{equation*}
\label{eq:ADMM}
\min_{\bm{\Phi},\bm{Q},\bm{R}} \|\bm{Y} - \bm{Y}\bm{\Phi}\bm{\Phi} '\|^2_F +\tau_1 \sum_{k=1}^K \bm{\phi}_i'\bm\Omega\bm{\phi}_k+
\tau_2 \sum_{k=1}^K\sum_{j=1}^{p}\left|r_{jk}\right|,
\end{equation*}			

\noindent subject to $\bm{Q}'\bm{Q}=\bm{I}_K$, $\bm{\Phi}=\bm{Q}=\bm{R}$, and $\bm{\phi}'_1\bm{S}\bm{\phi}_1\geq\bm{\phi}'_2\bm{S}\bm{\phi}_2\geq\cdots\geq\bm{\phi}'_K\bm{S}\bm{\phi}_K$,  where
$r_{jk}$ is the $(j,k)$-th element of $\bm{R}$. Then the corresponding augmented Lagrangian is
\begin{align*}
	L(\bm{\Phi}, \bm{Q},\bm{R},\bm{\Gamma}_1,\bm{\Gamma}_2)
=&~ \|\bm{Y}-\bm{Y}\bm{\Phi}\bm{\Phi}'\|^2_F+\tau_1\sum_{k=1}^K \bm{\phi}_i'\bm\Omega\bm{\phi}_k+
	\tau_2 \sum_{k=1}^K\sum_{j=1}^{p}|r_{jk}|\nonumber\\
&~ + \mathrm{tr}(\bm{\Gamma}_1'(\bm{\Phi}-\bm{Q}))+ \mathrm{tr}(\bm{\Gamma}_2'(\bm{\Phi}-\bm{R}))\nonumber\\
&~ +\frac{\rho}{2}(\|\bm{\Phi}-\bm{Q}\|^2_F+\|\bm{\Phi}-\bm{R}\|^2_F),
\end{align*}

\noindent subject to $\bm{Q}'\bm{Q}=\bm{I}_K$ and
$\bm{\phi}'_1\bm{S}\bm{\phi}_1\geq\bm{\phi}'_2\bm{S}\bm{\phi}_2\geq\cdots\geq\bm{\phi}'_K\bm{S}\bm{\phi}_K$,
where $\bm{\Gamma}_1$ and $\bm{\Gamma}_2$ are $p\times K$ matrices of the Lagrange multipliers.
Then the ADMM steps at the $\ell$-th iteration are given by
\begin{align}
	\bm{\Phi}^{(\ell+1)}
=&~ \mathop{\arg\min}_{\bm{\Phi}}L\big(\bm{\Phi}, \bm{Q}^{(\ell)},\bm{R}^{(\ell)},\bm{\Gamma}^{(\ell)}_1,\bm{\Gamma}^{(\ell)}_2\big)
	\notag\\
=&~ \frac{1}{2}(\tau_1\bm\Omega+\rho\bm{I}_p-\bm{Y}'\bm{Y})^{-1}\big\{\rho\big(\bm{Q}^{(\ell)}+\bm{R}^{(\ell)}\big)-
	\bm{\Gamma}_1-\bm{\Gamma}_2\big\},\label{eq:phi}\\
	\bm{Q}^{(\ell+1)}
=&~ \mathop{\arg\min}_{\bm{Q}: \bm{Q}'\bm{Q}=\bm{I}_K}L\big(\bm{\Phi}^{(\ell+1)},\bm{Q},\bm{R}^{(\ell)},
	\bm{\Gamma}^{(\ell)}_1,\bm{\Gamma}^{(\ell)}_2\big)=\bm{U}^{(\ell)}\big(\bm{V}^{(\ell)}\big)',\label{eq:Q}\\
	\bm{R}^{(\ell+1)}
=&~ \mathop{\arg\min}_{\bm{R}} L\big(\bm{\Phi}^{(\ell+1)},\bm{Q}^{(\ell+1)},\bm{R},\bm{\Gamma}^{(\ell)}_1,\bm{\Gamma}^{(\ell)}_2\big)
	\notag\\
=&~ \frac{1}{\rho}\mathcal{S}_{\tau_2}\big(\rho\bm{\Phi}^{(\ell+1)}+\bm{\Gamma}_{2}^{(\ell)}\big),
	\label{eq:R}\\
	\bm{\Gamma}^{(\ell+1)}_1
=&~ \bm{\Gamma}^{(\ell) }_1+ \rho\left(\bm{\Phi}^{(\ell+1)}-\bm{Q}^{(\ell+1)}\right), \label{eq:Gamma1}\\
	\bm{\Gamma}^{(\ell+1)}_2
=&~ \bm{\Gamma}^{(\ell) }_2+ \rho\left(\bm{\Phi}^{(\ell+1)}-\bm{R}^{(\ell+1)}\right),\label{eq:Gamma2}
\end{align}

\noindent where $\bm{R}^{(0)}$, $\bm{\Gamma}_1^{(0)}$ and $\bm{\Gamma}_2^{(0)}$ are initial
estimates of $\bm{R}$, $\bm{\Gamma}_1$ and $\bm{\Gamma}_2$, respectively,
$\bm{U}^{(\ell)}\bm{D}^{(\ell)}\big(\bm{V}^{(\ell)}\big)'$ is the singular value decomposition of
$\bm{\Phi}^{(\ell+1)}+\rho^{-1}\bm{\Gamma}_{1}^{(\ell)}$, and
$\mathcal{S}_{\tau_2}(\cdot)$ is the element-wise soft-thresholding operator with a threshold $\tau_2$
(i.e., the $(j,k)$-th element of $\mathcal{S}_{\tau_2}(\bm{M})$ is
$\mathrm{sign}(m_{jk})\max(|m_{jk}|-\tau_2,0)$ with $m_{jk}$ the $(j,k)$-th element of $\bm{M}$).
Similarly to (\ref{eq:phi_1}), $\rho$ must be chosen large enough to ensure that $\tau_1\bm\Omega+\rho\bm{I}_p-\bm{Y}'\bm{Y}$ in \eqref{eq:phi}
is positive definite.


\section{Numerical Examples} 
\label{sec:simulation}

We conducted some simulation experiments in one-dimensional and two-dimensional spatial domains,
and applied SpatPCA to a real-world dataset. We compared the proposed SpatPCA with three methods: (1) PCA ($\tau_1=\tau_2=0$);
(2) SpatPCA with the smoothness penalty only ($\tau_2=0$);
(3) SpatPCA with the sparseness penalty only ($\tau_1=0$),
based on the two loss functions.
The first one measures the prediction ability in terms of an average squared prediction error:
\begin{equation}
\label{eq:loss_sim} 
\mathrm{Loss}(\hat{\bm{\Phi}}) = \frac{1}{n}\sum_{i=1}^n
\big\|\hat{\bm{\Phi}}\hat{\bm{\xi}}_i - \bm{\Phi}\bm{\xi}_i\big\|^2,
\end{equation}

\noindent where $\bm{\Phi}$ is the true eigenvector matrix formed by the first $K$ eigenvectors
and
\[
\hat{\bm{\xi}}_i=\hat{\bm{V}}\mathrm{diag}\bigg(
\frac{\hat{\lambda}^*_1}{\hat{\lambda}^*_1+\hat{\sigma}^2},\dots,
\frac{\hat{\lambda}^*_K}{\hat{\lambda}^*_K+\hat{\sigma}^2}\bigg)\hat{\bm{V}}'\hat{\bm{\Phi}}'\bm{Y}_i,
\]
is the empirical best linear unbiased predictor of $\bm{\xi}_i$ with the estimated parameters plugged in. The second one concerns
the goodness of covariance function estimation in terms of an average squared estimation error:
\begin{equation}
\label{eq:loss2_sim} 
\mathrm{Loss}(\hat{C}_\eta) =\frac{1}{p^2} \sum_{i=1}^{p}\sum_{j=1}^{p}\big(\hat{C}_\eta(\bm{s}_i,\bm{s}_j)-C_\eta(\bm{s}_i,\bm{s}_j)\big)^2\:.
\end{equation}


We applied the ADMM algorithm given by (\ref{eq:phi})-(\ref{eq:Gamma2}) to compute the SpatPCA estimates
with $\rho$ being ten times the maximum eigenvalue of $\bm{Y}'\bm{Y}$.
The stopping criterion for the ADMM algorithm is
\[
\frac{1}{\sqrt{p}} \max\left( \|\bm{\Phi}^{(\ell+1)}-\bm{\Phi}^{(\ell)}\|_F,\|\bm{\Phi}^{(\ell+1)}-\bm{R}^{(\ell+1)}\|_F,
\|\bm{\Phi}^{(\ell+1)}-\bm{Q}^{(\ell+1)}\|_F \right)\leq 10^{-4}\:.
\]

\subsection{One-Dimensional Experiment}
\label{sec:ex1}

In the first experiment,  we generated data according to (\ref{eq:measurement}) with $K=2$,
$\bm{\xi}_i\sim N(\bm{0}, \mathrm{diag}(\lambda_1,\lambda_2))$, $\bm{\epsilon}_{i}\sim N(\bm{0}, \bm{I})$,
$n=100$, $p=50$, $\bm{s}_1,\dots,\bm{s}_{50}$ equally spaced in $D=[-5,5]$,
and
\begin{align}
	\phi_1(\bm{s})
=&~ \frac{1}{c_1}\exp(-(x_1^2+\cdots+x_d^2)),
\label{eq:phi1_sim}\\
	\phi_2(\bm{s})
=&~ \frac{1}{c_2}x_1\cdots x_d\exp(-(x_1^2+\cdots+x_d^2)),
\label{eq:phi2_sim}
\end{align}

\noindent where $\bm{s}=(x_1,\dots,x_d)'$, $c_1$ and $c_2$ are normalization constants such that
$\|\bm{\phi}_1\|_2=\|\bm{\phi}_2\|_2=1$, and $d=1$.
We considered three pairs of $(\lambda_1,\lambda_2)\in\{(9,0),(1,0),(9,4)\}$,
and applied the proposed
SpatPCA with $K\in\{1,2,5\}$ and $\hat{K}$ selected from (\ref{eq:khat}), resulting in 12 different combinations.
For each combination, we considered $11$ values of $\tau_1$ (including $0$, and the other 10 values from $1$ to $10^3$ equally spaced on the log scale)
and $31$ values of $\tau_2$ (including $0$, and the other 30 values from $1$ to $10^3$ equally spaced on the log scale).
But instead of performing a two-dimensional optimization by selecting among all possible pairs of $(\tau_1,\tau_2)$,
we applied a more efficient two-step procedure involving only one-dimensional optimization.
First, we selected among 11 values of $\tau_1$ by fixing $\tau_2 = 0$ using 5-fold CV of \eqref{eq:cv}
with the initial estimate of $\hat{\bm{\Phi}}_{\tau_1,0}^{(0)}$ given by the first $K$ eigenvectors of $\bm{Y}'\bm{Y}-\tau_1\bm\Omega$ as its columns.
Note that this initial estimate is actually the true estimate $\hat{\bm{\Phi}}_{\tau_1,0}$ when $\bm{Y}'\bm{Y}-\tau_1\bm\Omega\succeq\bm{0}$.
Then we selected among $31$ values of $\tau_2$ with the selected $\tau_1$ using 5-fold CV of \eqref{eq:cv}.

For covariance function estimation, we selected the tuning parameter $\gamma$ 
among $11$ values of $\gamma$ using 5-fold CV of \eqref{eq:cv.gamma}, including $\gamma=0$ and the other 10 values
from $1$ to $\hat{d}_1$ equally spaced on the log scale, where $\hat{d}_1$ is the largest eigenvalues of $\hat{\bm{\Phi}}'\bm{S}\hat{\bm{\Phi}}$.

\begin{figure}\centering
$\hat{\phi}_1(\cdot)$ based on $(\lambda_1,\lambda_2)=(9,0)$ and $K=1$
\includegraphics[scale=0.39]{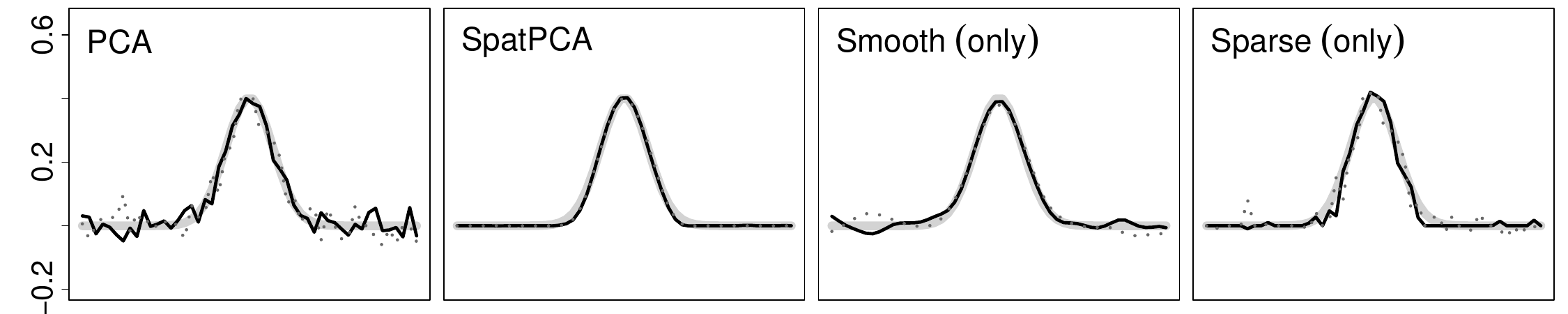}
$\hat{\phi}_1(\cdot)$ based on $(\lambda_1,\lambda_2)=(1,0)$ and $K=1$
\includegraphics[scale=0.39]{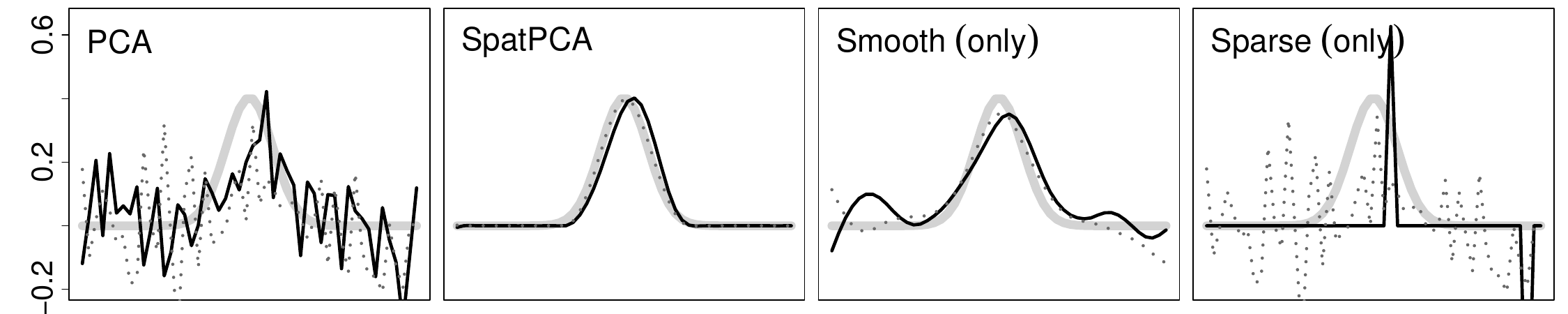}
$\hat{\phi}_1(\cdot)$ based on $(\lambda_1,\lambda_2)=(9,4)$ and $K=2$
\includegraphics[scale=0.39]{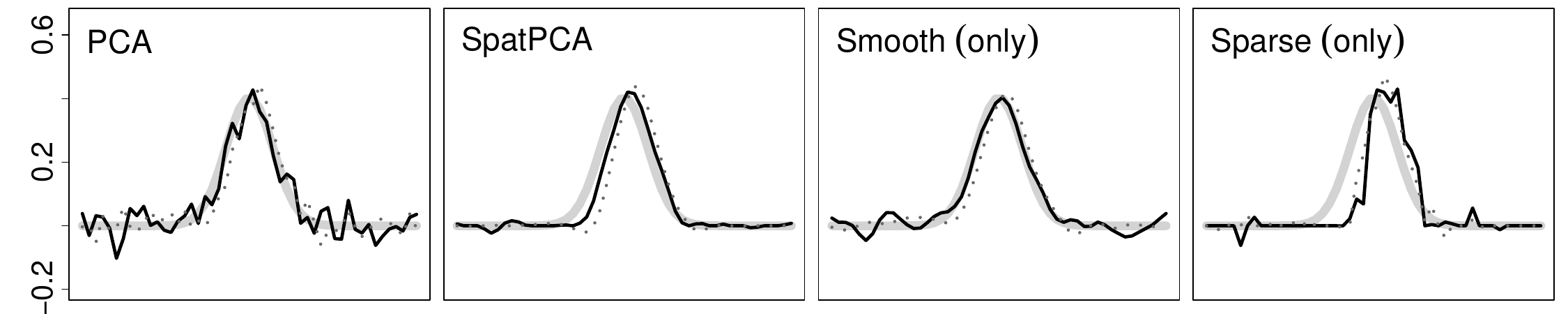}
$\hat{\phi}_2(\cdot)$ based on $(\lambda_1,\lambda_2)=(9,4)$ and $K=2$
\includegraphics[scale=0.39]{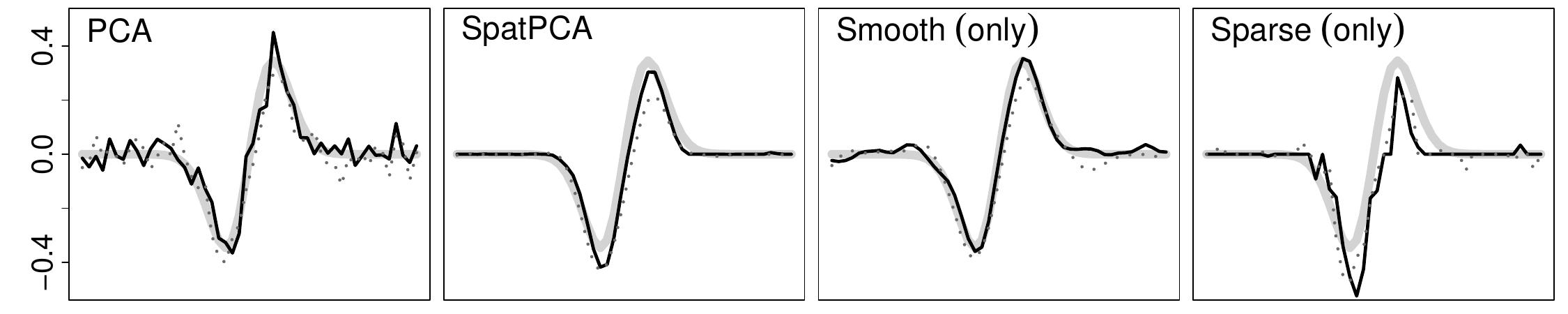}
\caption{Estimates of $\phi_1(\cdot)$ and $\phi_2(\cdot)$ obtained from various methods based on three different combinations of eigenvalues.
Each panel consists of two estimates (in two different line types) corresponding to two randomly generated datasets, where the solid grey lines are the true eigenfunctions.}
\label{fig:est_d1}
\end{figure}

Figure~\ref{fig:est_d1} shows the estimates of $\phi_1(\cdot)$ and $\phi_2(\cdot)$ for the four methods
based on three different combinations of eigenvalues. Each case contains two estimated functions based on two randomly generated datasets.
As expected, the PCA estimates, which consider no spatial structure, are very noisy, particularly when the signal-to-noise ratio is small.
Adding only the smoothness penalty (i.e., $\tau_2=0$) makes the estimates considerably less noisy.
But the resulting estimates show some obvious bias.
On the other hand, adding only the sparseness penalty (i.e, $\tau_1=0$) forces the eigenfunction estimates to be zeros at some locations.
But the estimated patterns are still very noisy.
Overall, our SpatPCA estimates reproduce the targets with little noise for all cases
even when the signal-to-noise ratio is small, indicating the effectiveness of regularization.

\begin{figure}\centering
$(\lambda_1,\lambda_2)=(9,0)$ and $K=1$
\includegraphics[scale=0.42]{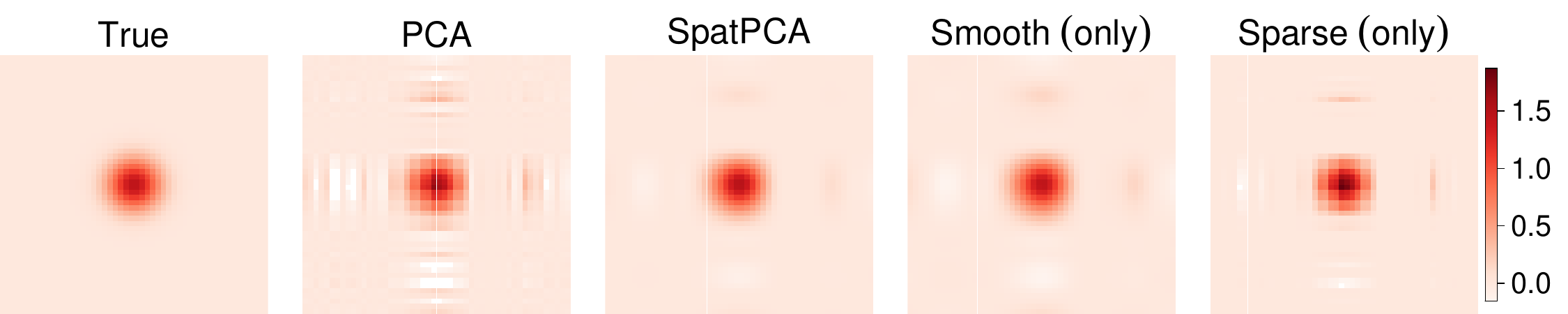}
$(\lambda_1,\lambda_2)=(1,0)$ and $K=1$
\includegraphics[scale=0.42]{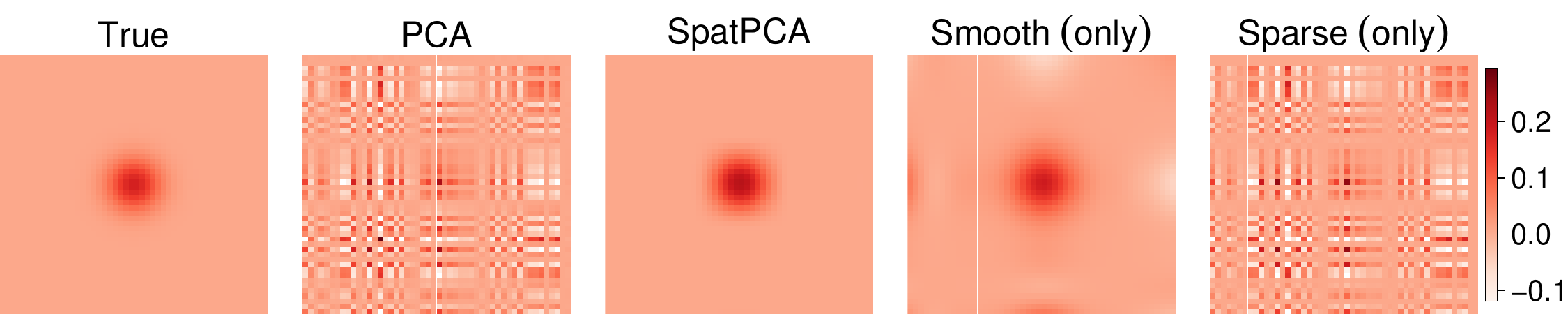}
$(\lambda_1,\lambda_2)=(9,4)$ and $K=2$
\includegraphics[scale=0.42]{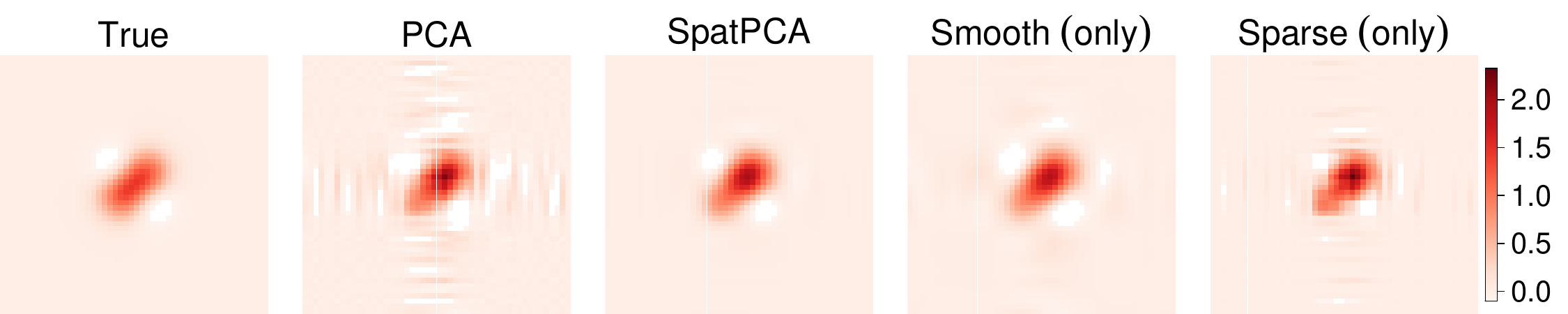}
\caption{{True covariance functions and their estimates obtained from various methods based on three different combinations of eigenvalues.}}
\label{fig:cov_d1}
\end{figure}

Figure~\ref{fig:cov_d1} shows the covariance function estimates for the four methods based on a randomly generated dataset.
The proposed SpatPCA can be seen to perform considerably better than the other methods for all cases by being able to
reconstruct the underlying nonstationary spatial covariance functions without having noticeable visual artifacts. 

\begin{figure}
{
	\begin{tabular}{ccc}
{$(\lambda_1,\lambda_2)=(9,0)$, $K=1$}&{$(\lambda_1,\lambda_2)=(1,0)$, $K=1$}&{$(\lambda_1,\lambda_2)=(9,4)$, $K=1$}\\
\includegraphics[scale=0.12]{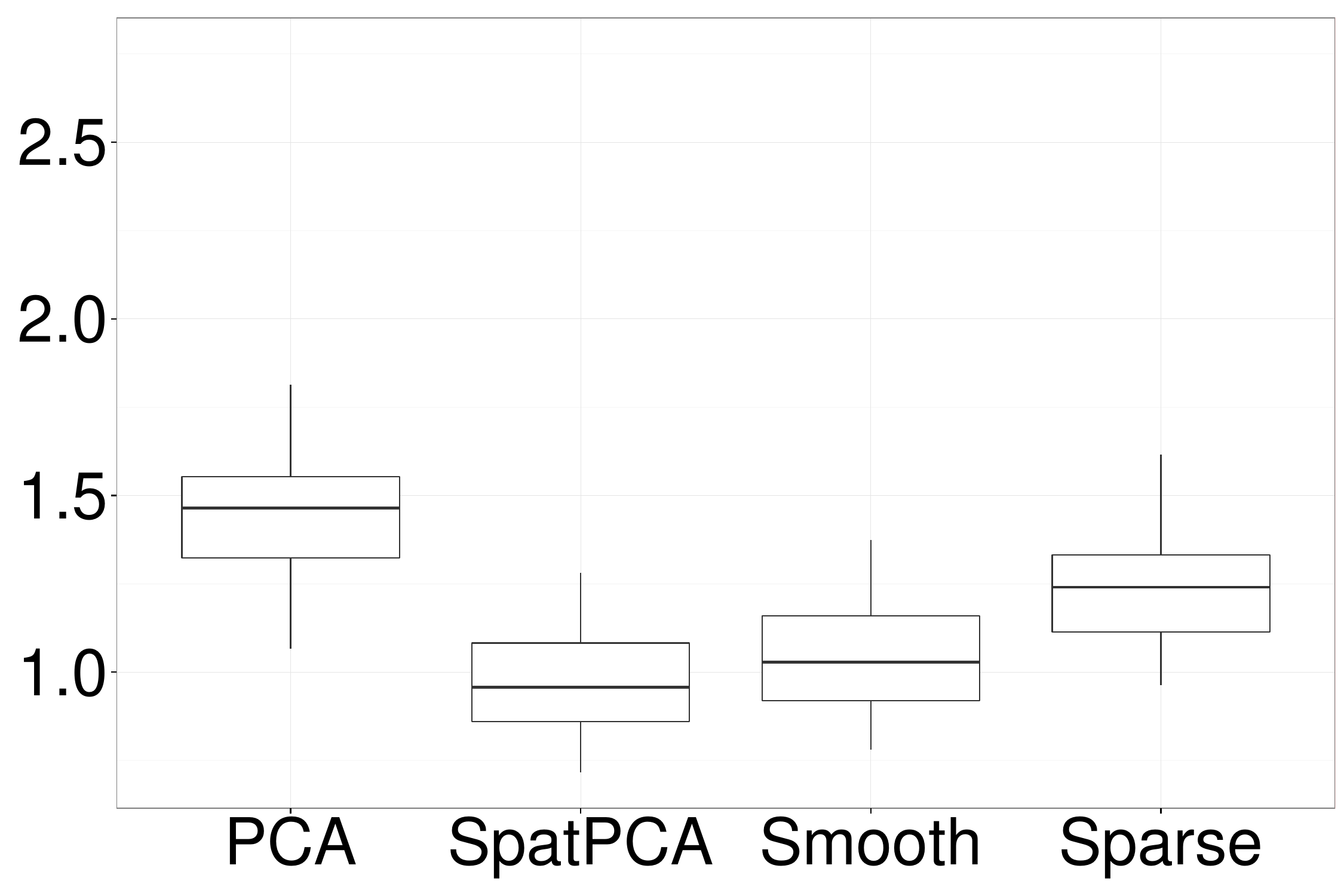}&
\includegraphics[scale=0.12]{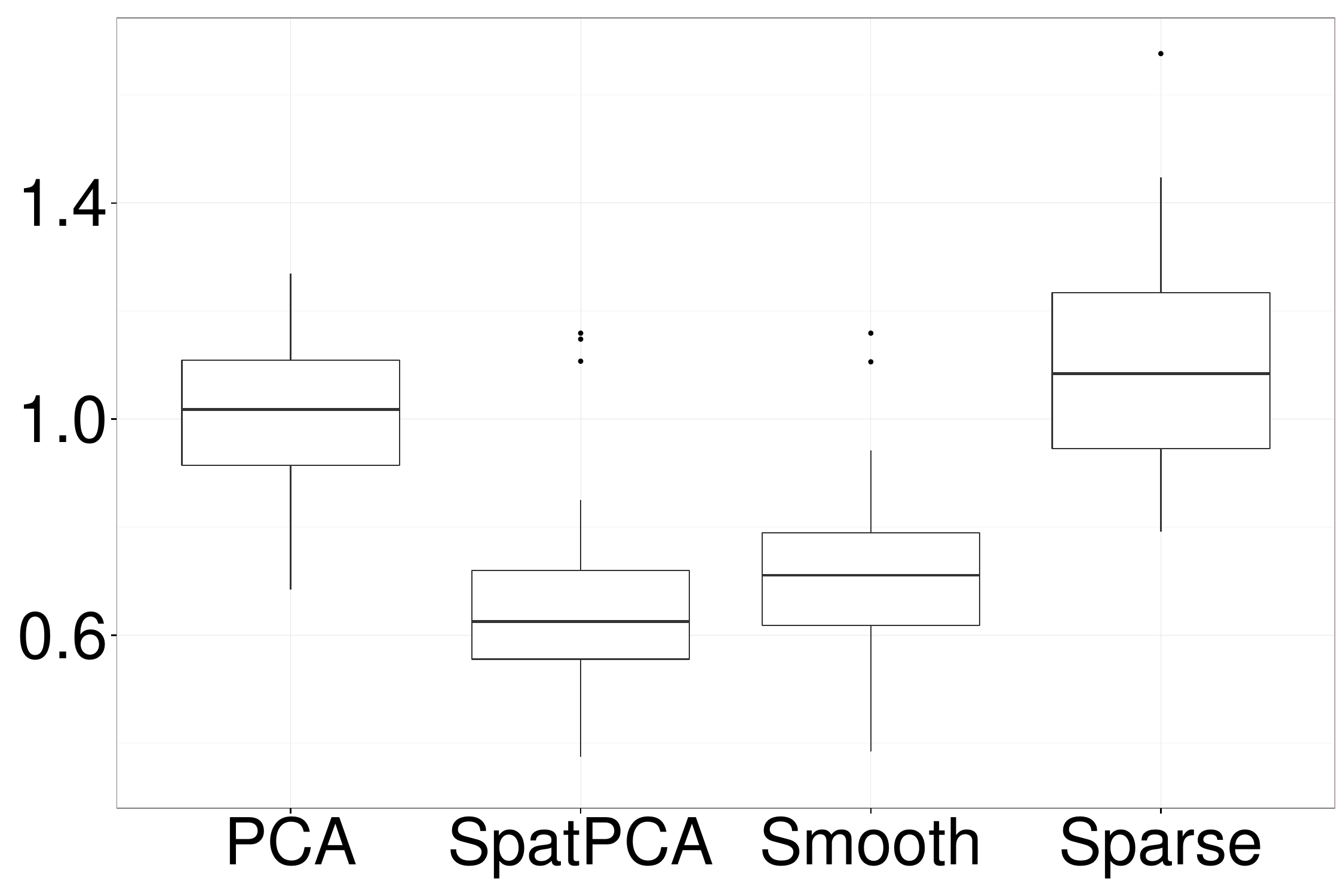}&
\includegraphics[scale=0.12]{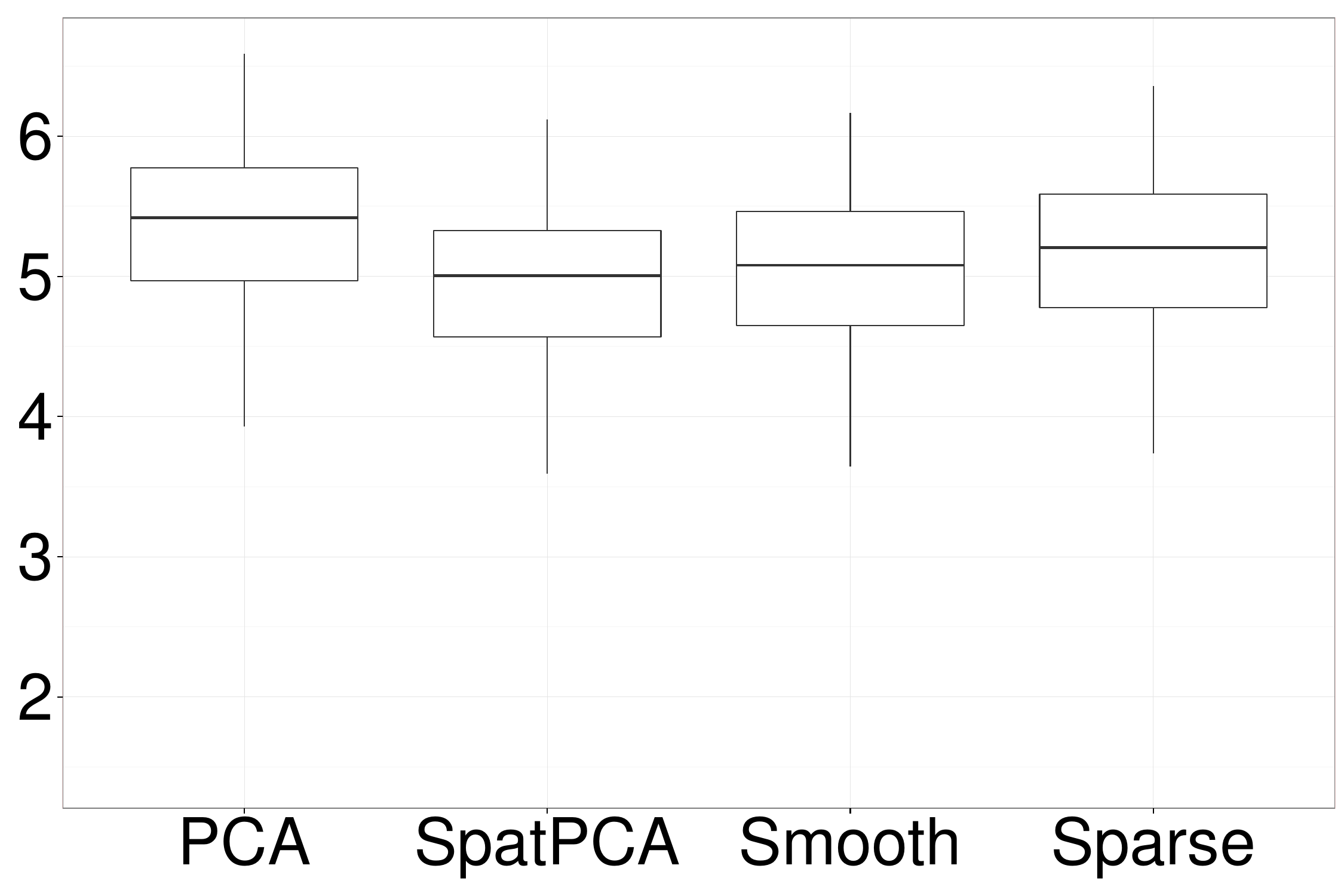}\\
{$(\lambda_1,\lambda_2)=(9,0)$, $K=2$}&{$(\lambda_1,\lambda_2)=(1,0)$, $K=2$}&{$(\lambda_1,\lambda_2)=(9,4)$, $K=2$}\\
\includegraphics[scale=0.12]{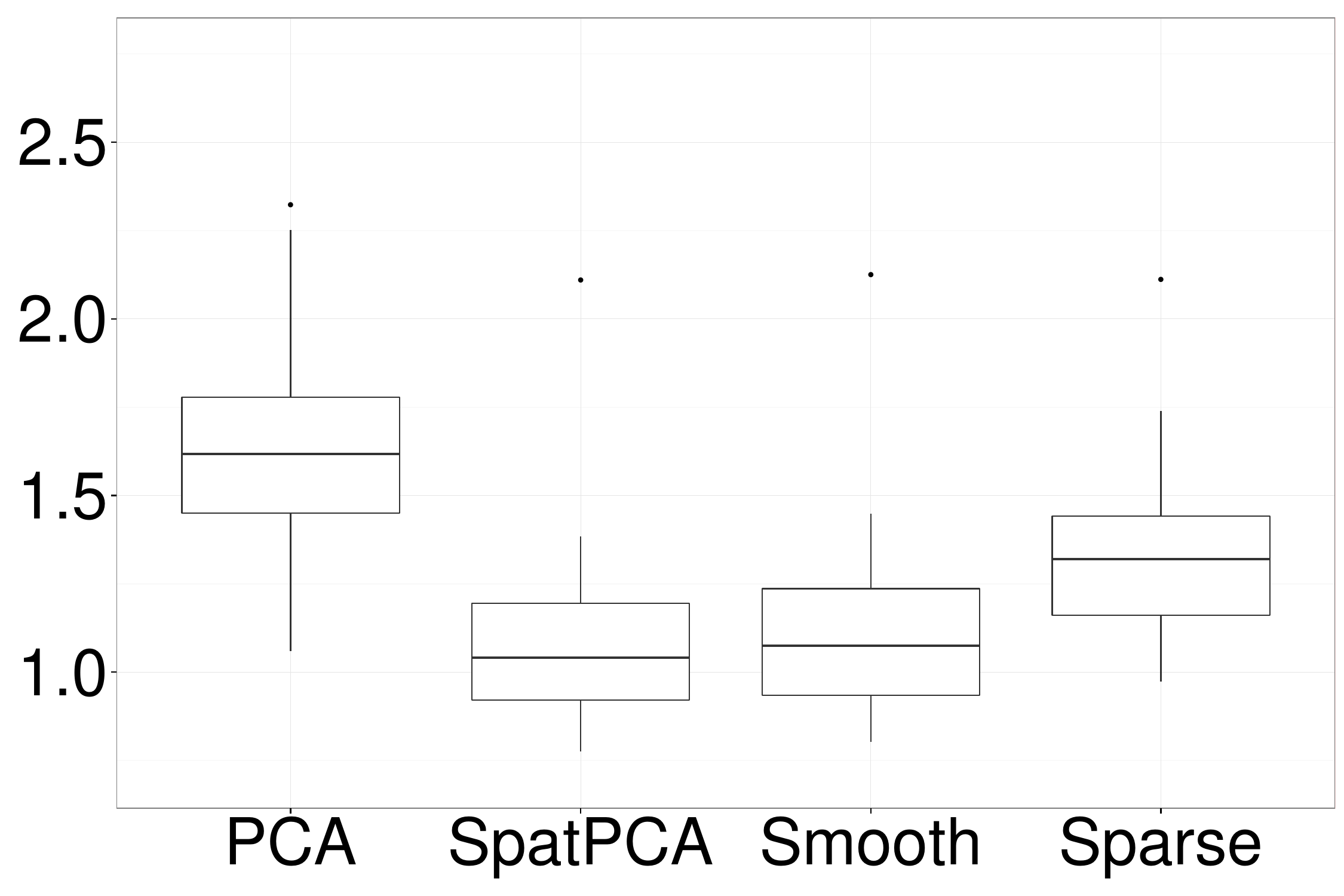}&
\includegraphics[scale=0.12]{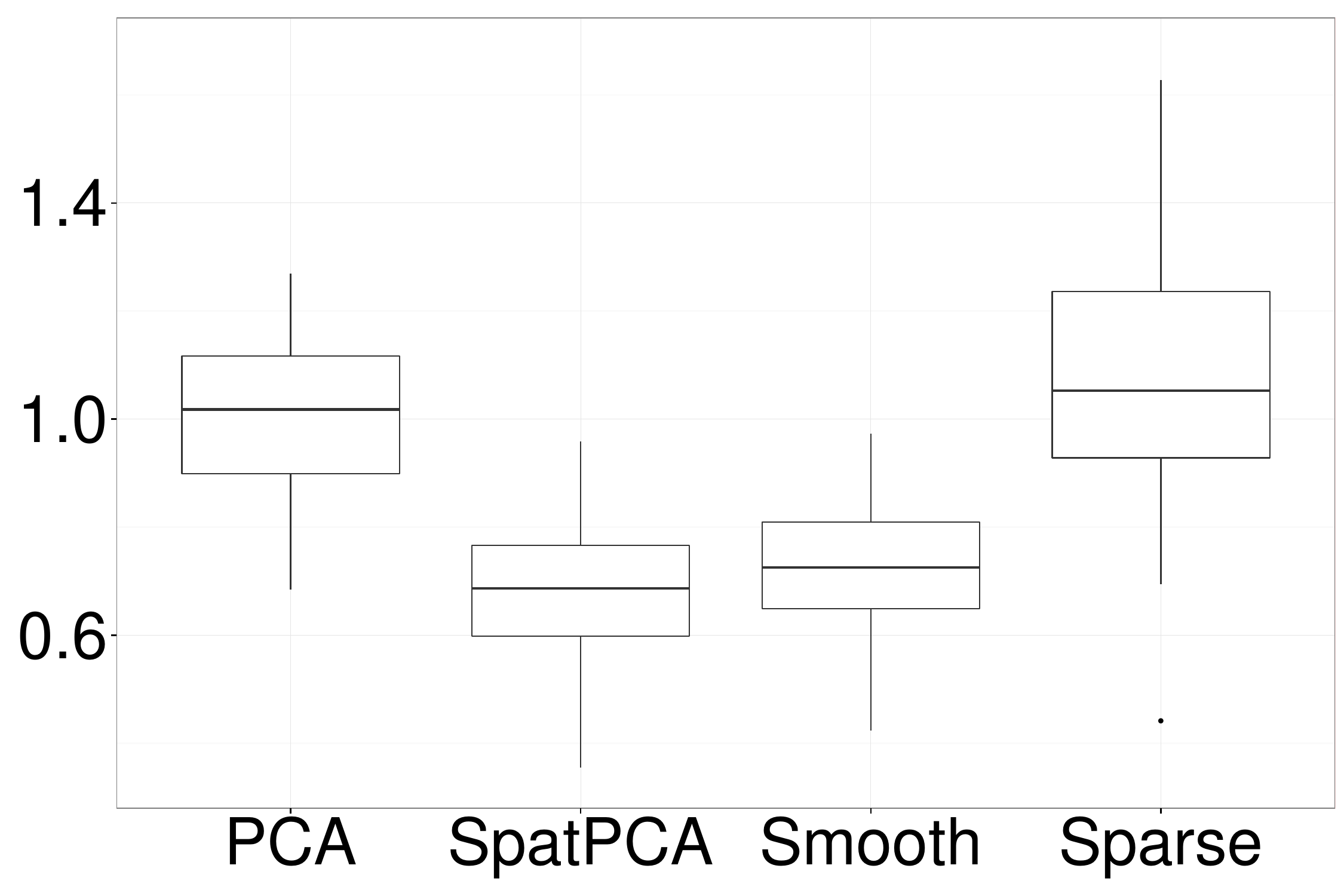}&
\includegraphics[scale=0.12]{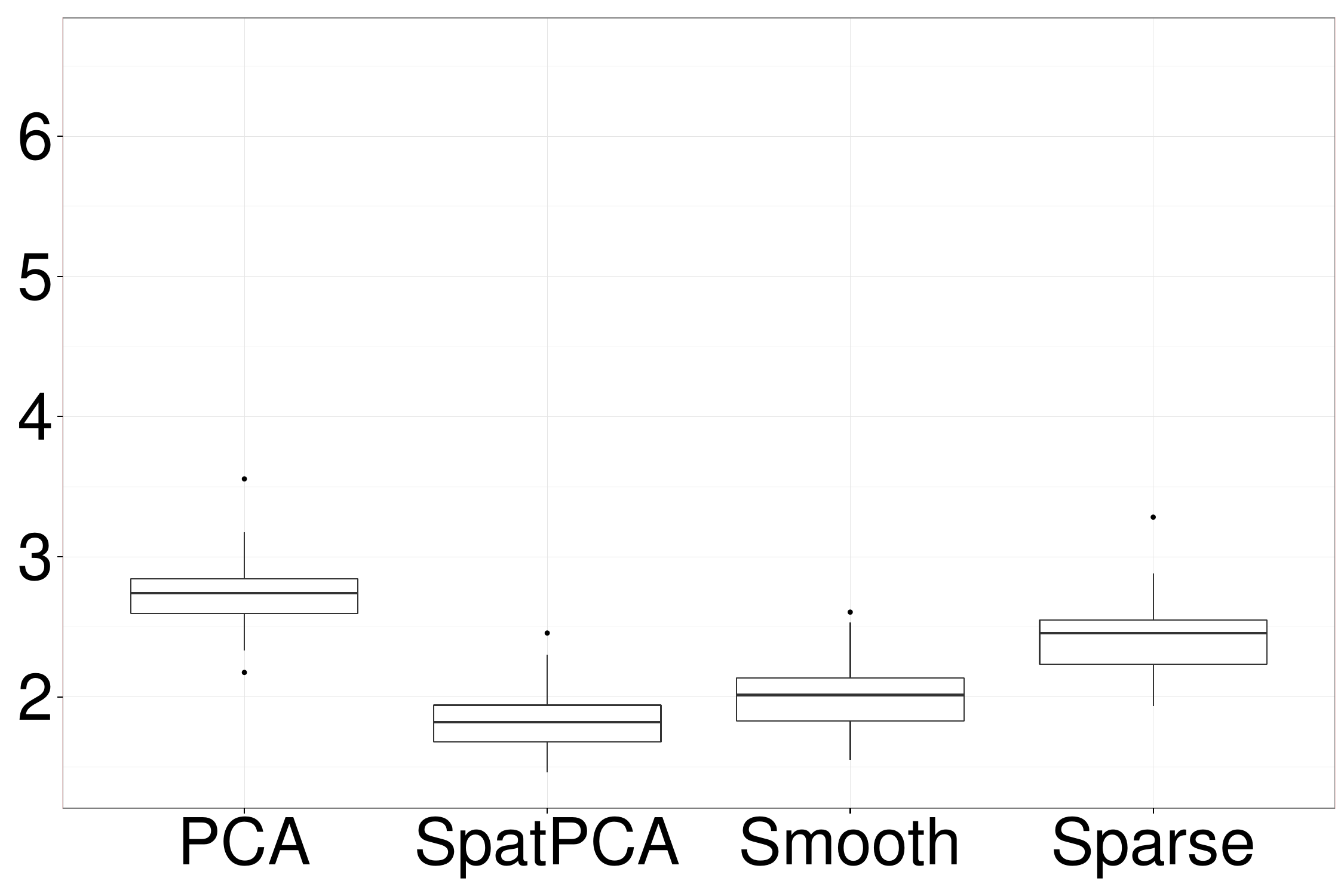}\\
{$(\lambda_1,\lambda_2)=(9,0)$, $K=5$}&{$(\lambda_1,\lambda_2)=(1,0)$, $K=5$}&{$(\lambda_1,\lambda_2)=(9,4)$, $K=5$}\\
\includegraphics[scale=0.12]{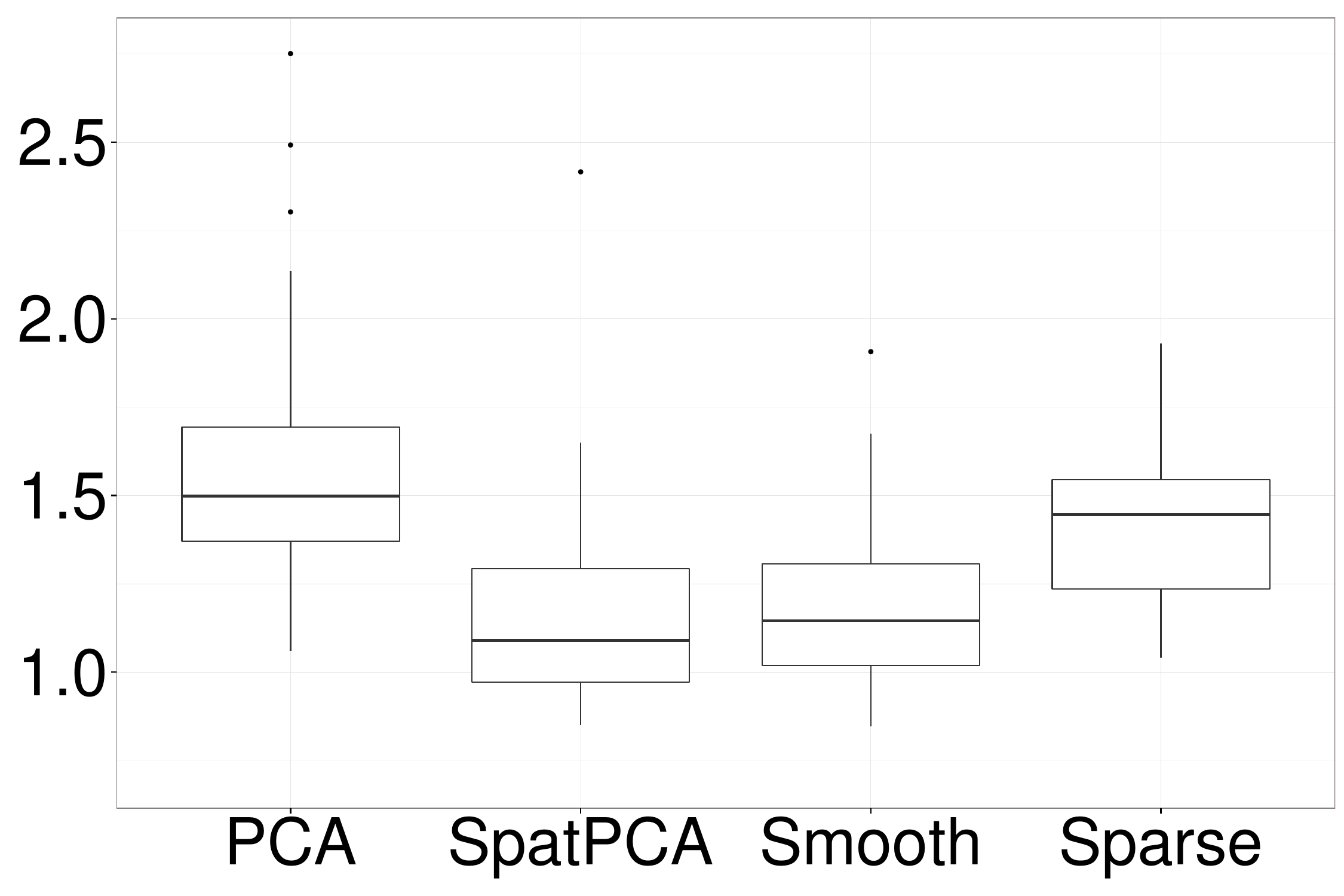}\hspace{4pt}&
\includegraphics[scale=0.12]{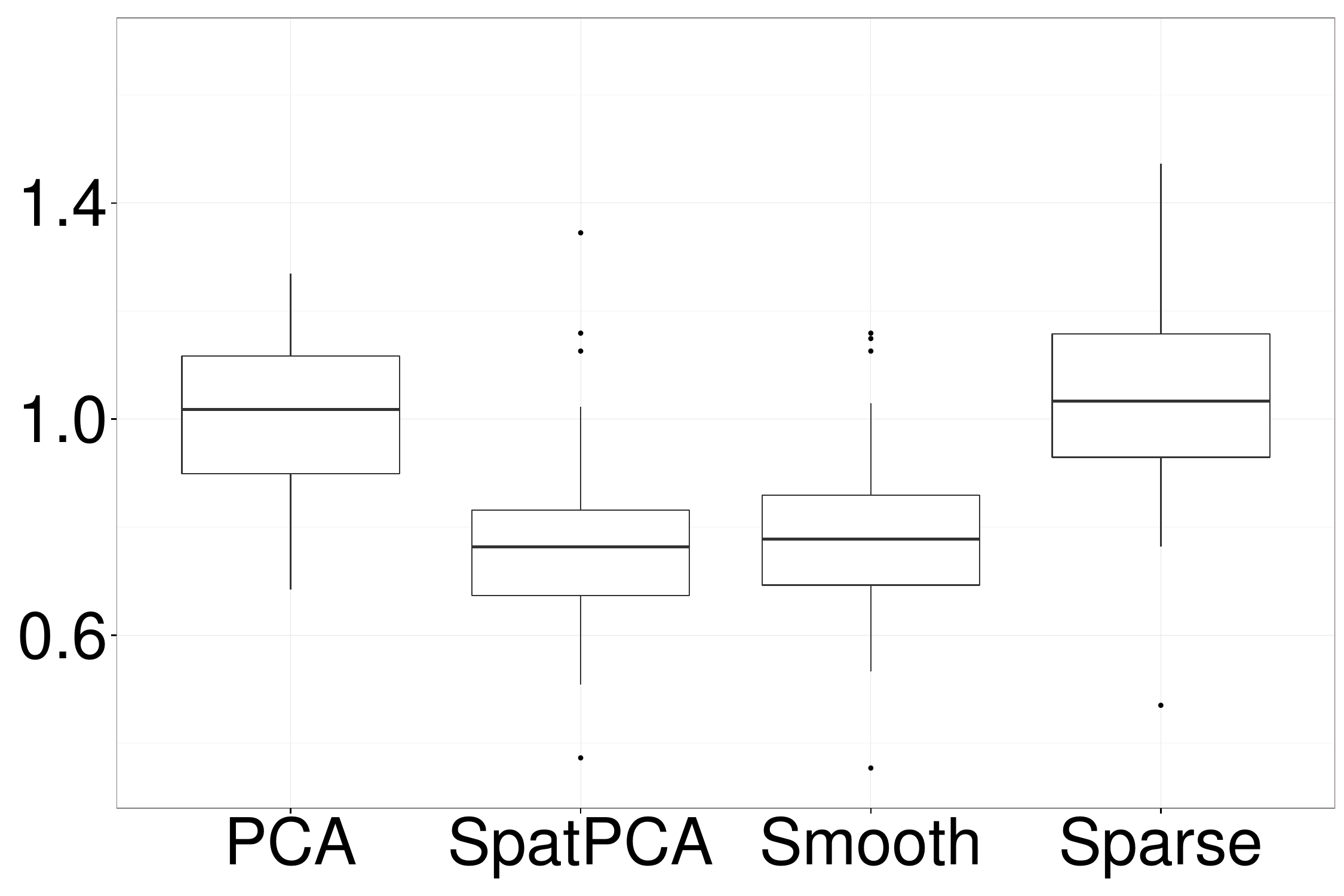}\hspace{4pt}&
\includegraphics[scale=0.12]{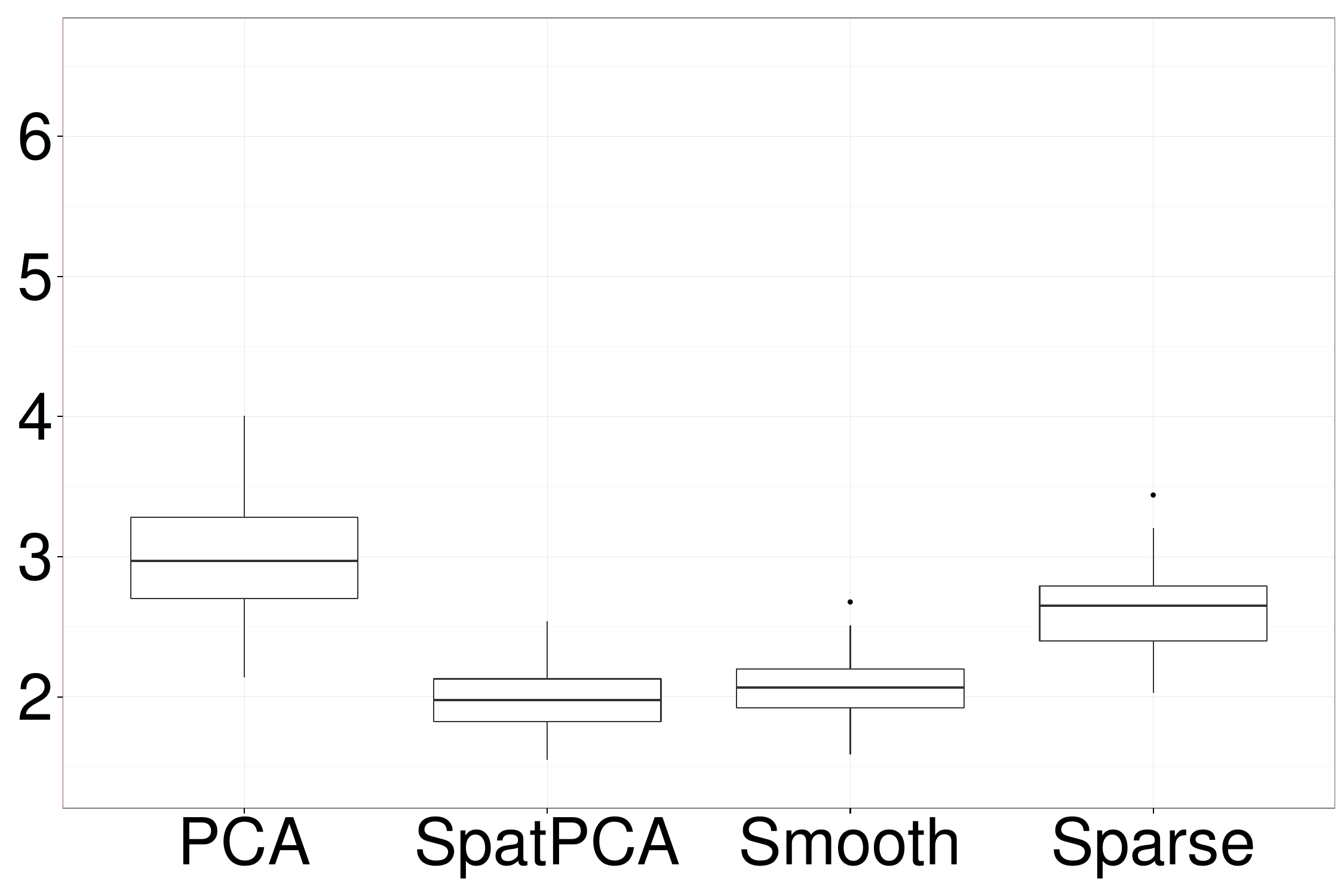}\hspace{4pt}\\
{$(\lambda_1,\lambda_2)=(9,0)$, $K=\hat{K}$}&{$(\lambda_1,\lambda_2)=(1,0)$, $K=\hat{K}$}&{$(\lambda_1,\lambda_2)=(9,4)$, $K=\hat{K}$}\\
\includegraphics[scale=0.12]{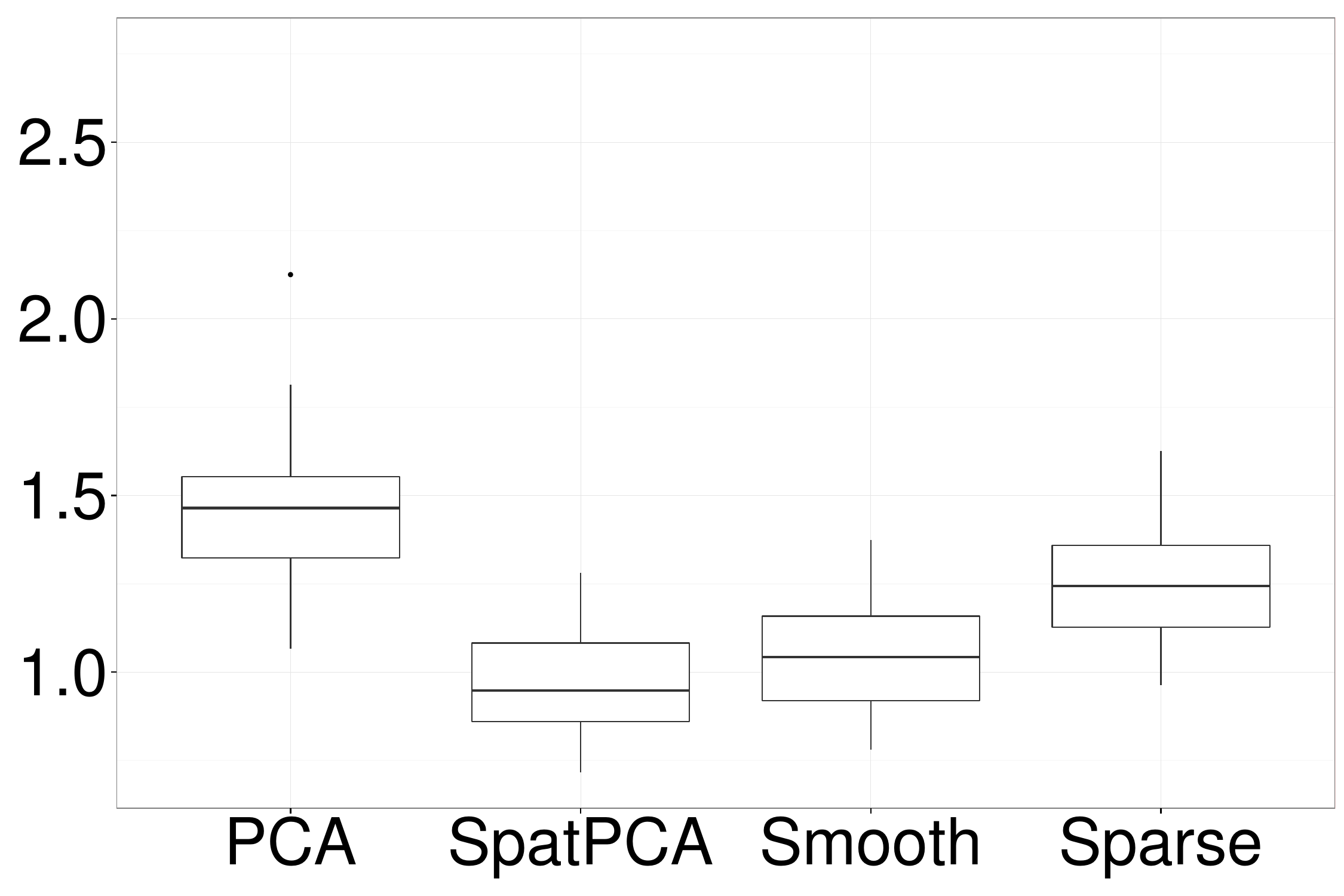}\hspace{4pt}&
\includegraphics[scale=0.12]{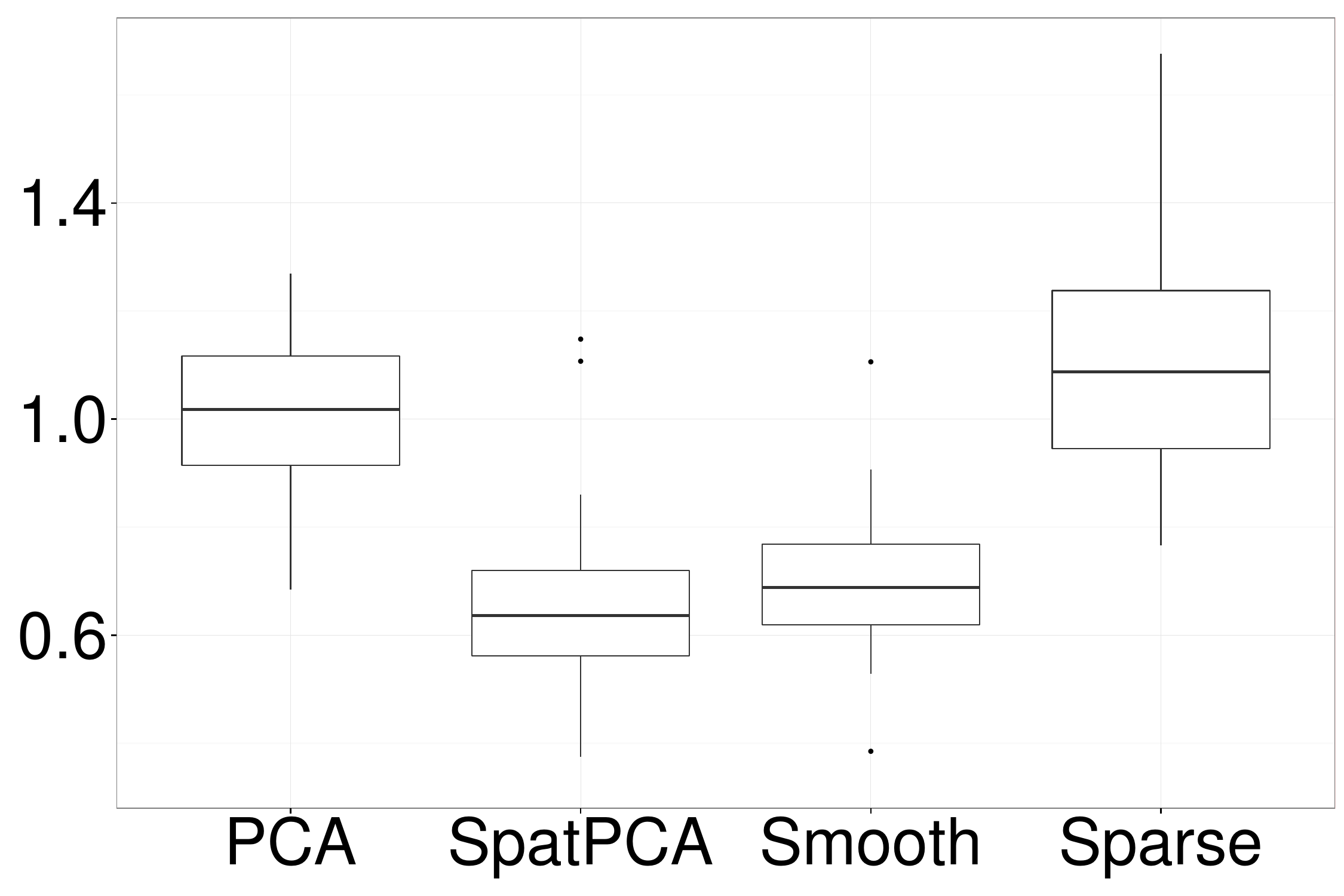}\hspace{4pt}&
\includegraphics[scale=0.12]{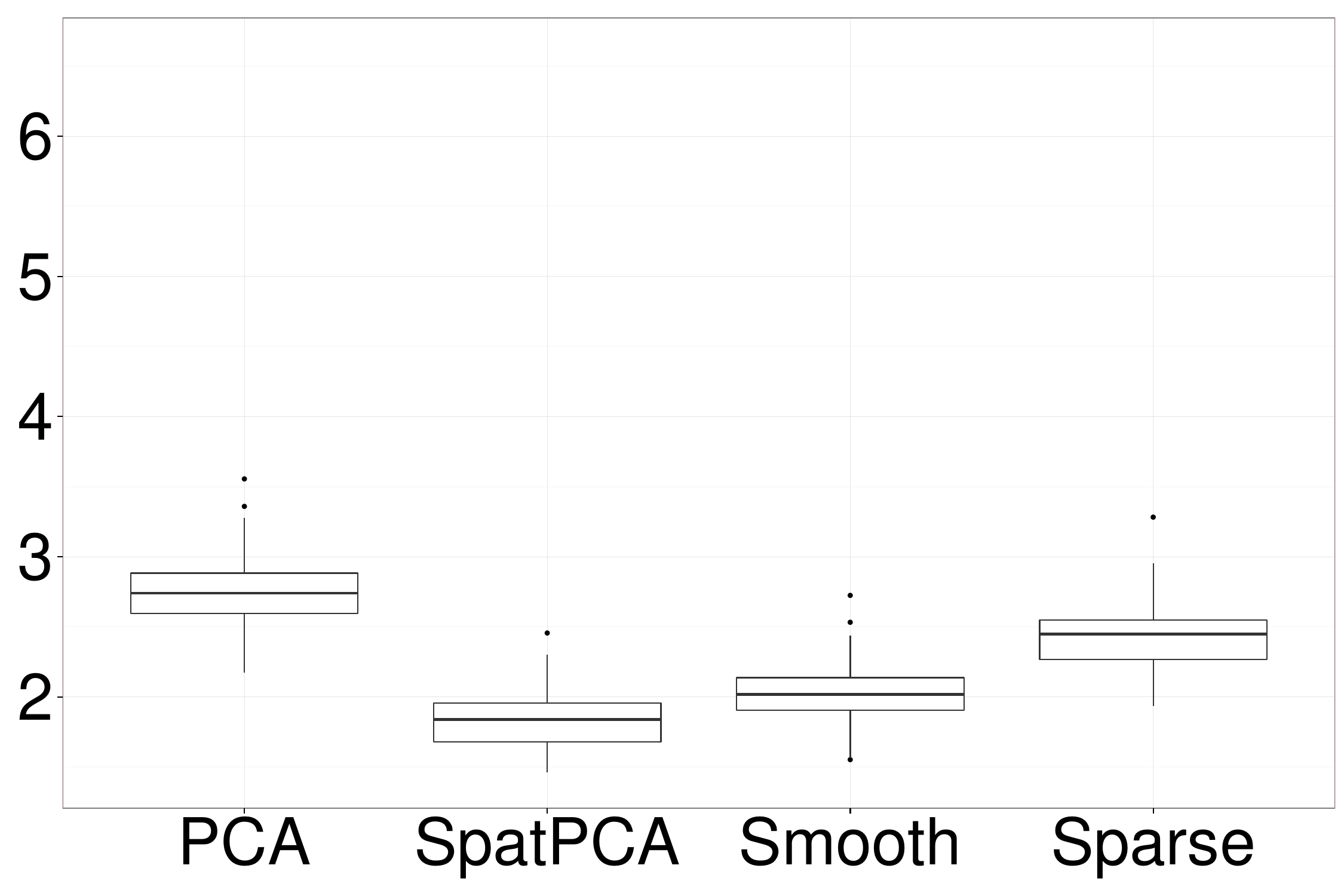}\hspace{4pt}
\end{tabular}
}
\caption{{Boxplots of average squared prediction errors of (\ref{eq:loss_sim}) for various methods in the one-dimensional simulation experiment of
Section \ref{sec:ex1} based on 50 simulation replicates.}}
\label{fig:box_d1_loss1}
\end{figure}

\begin{figure}\centering
 {
 	\begin{tabular}{ccc}
 		{$(\lambda_1,\lambda_2)=(9,0)$, $K=1$}&{$(\lambda_1,\lambda_2)=(1,0)$, $K=1$}&{$(\lambda_1,\lambda_2)=(9,4)$, $K=1$}\\
 		\includegraphics[scale=0.12]{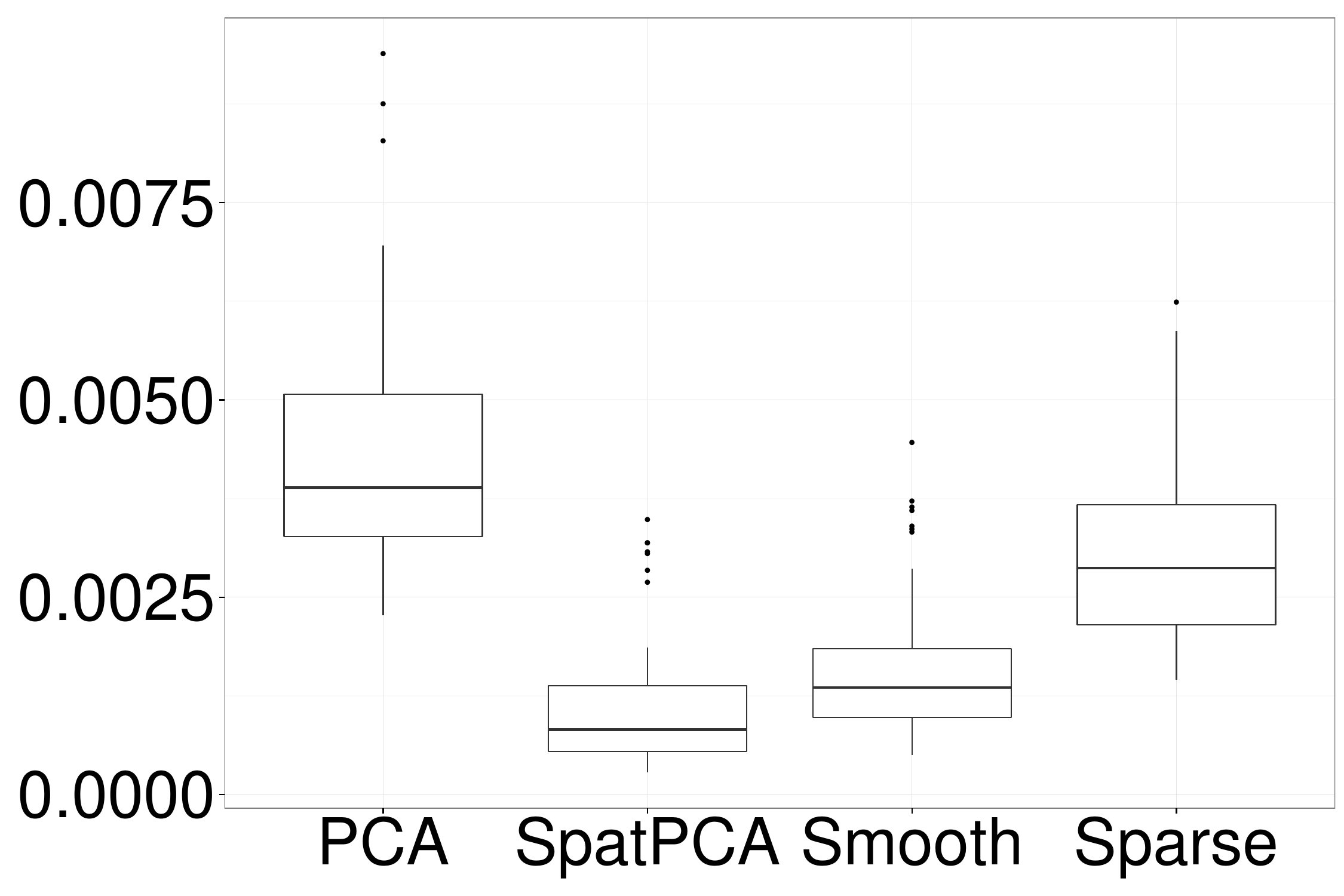}&
 		\includegraphics[scale=0.12]{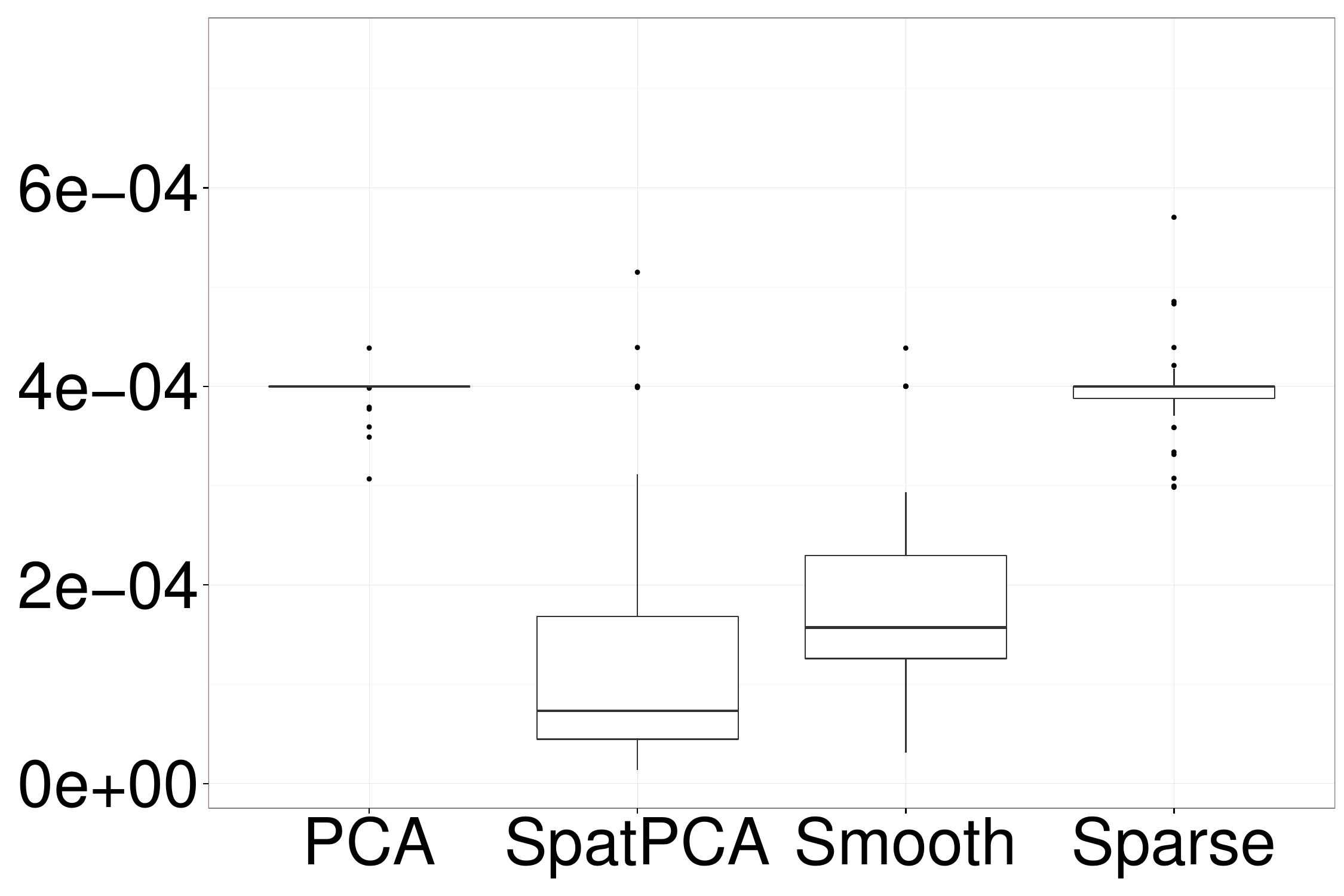}&
 		\includegraphics[scale=0.12]{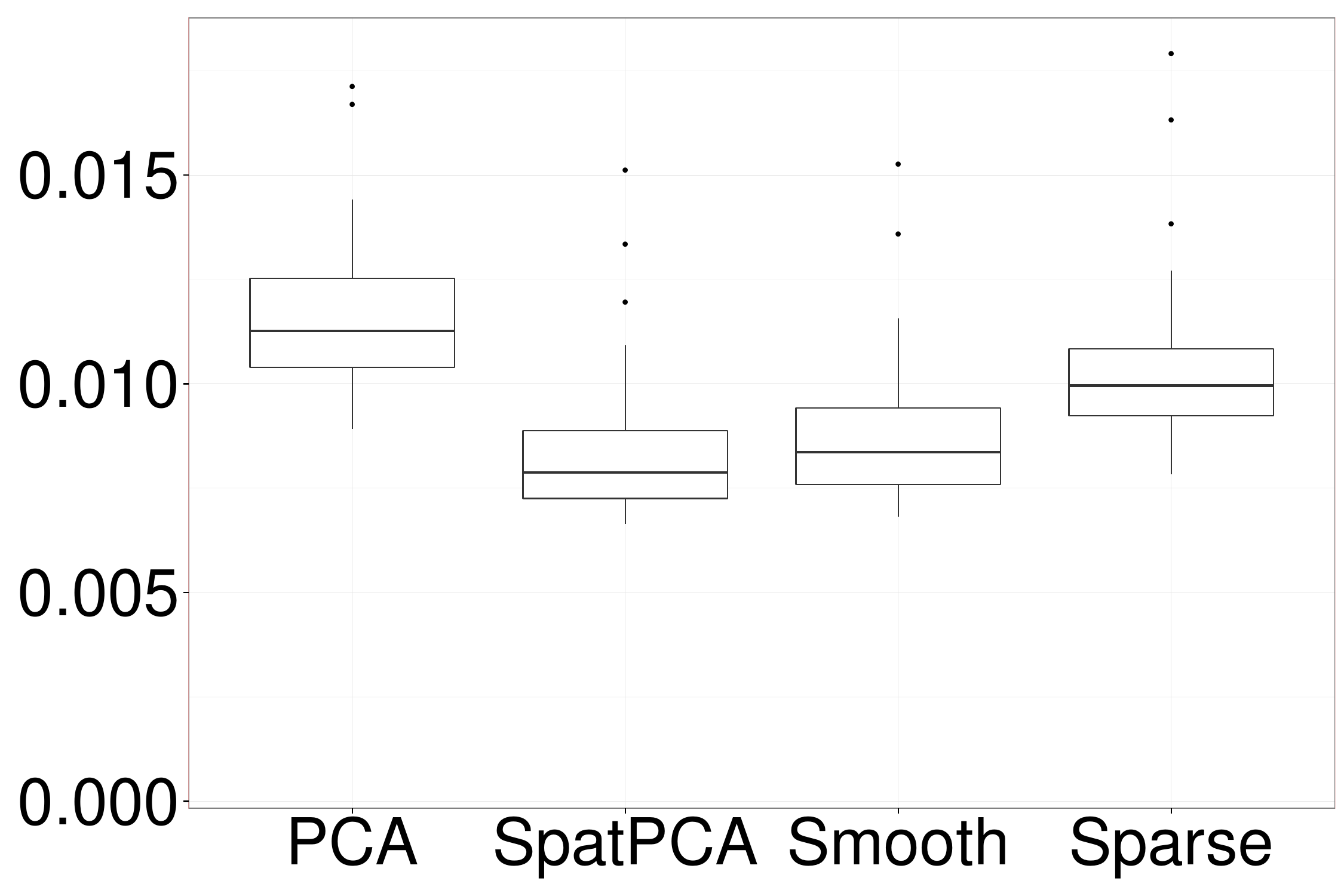}\\
 		{$(\lambda_1,\lambda_2)=(9,0)$, $K=2$}&{$(\lambda_1,\lambda_2)=(1,0)$, $K=2$}&{$(\lambda_1,\lambda_2)=(9,4)$, $K=2$}\\
 		\includegraphics[scale=0.12]{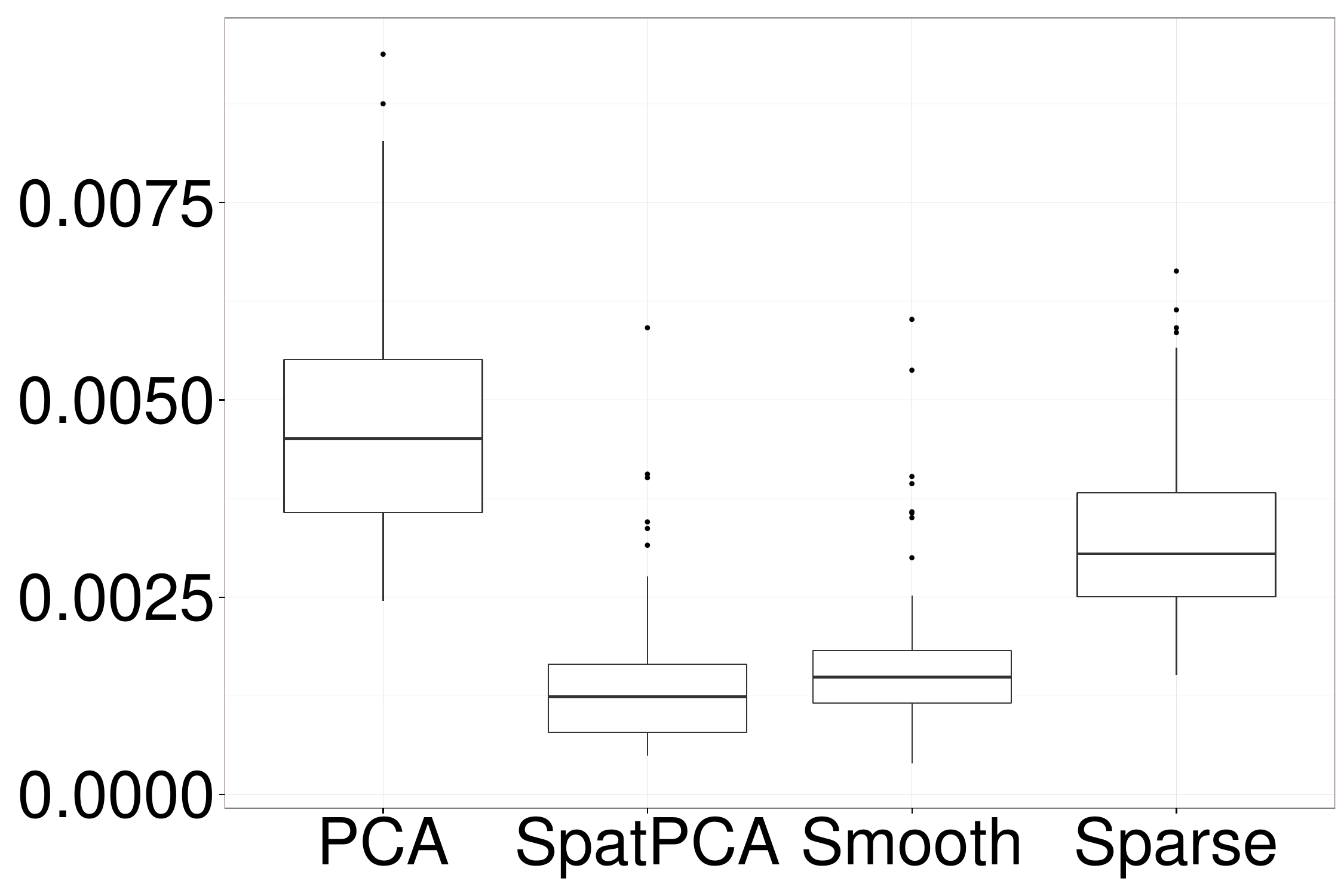}&
 		\includegraphics[scale=0.12]{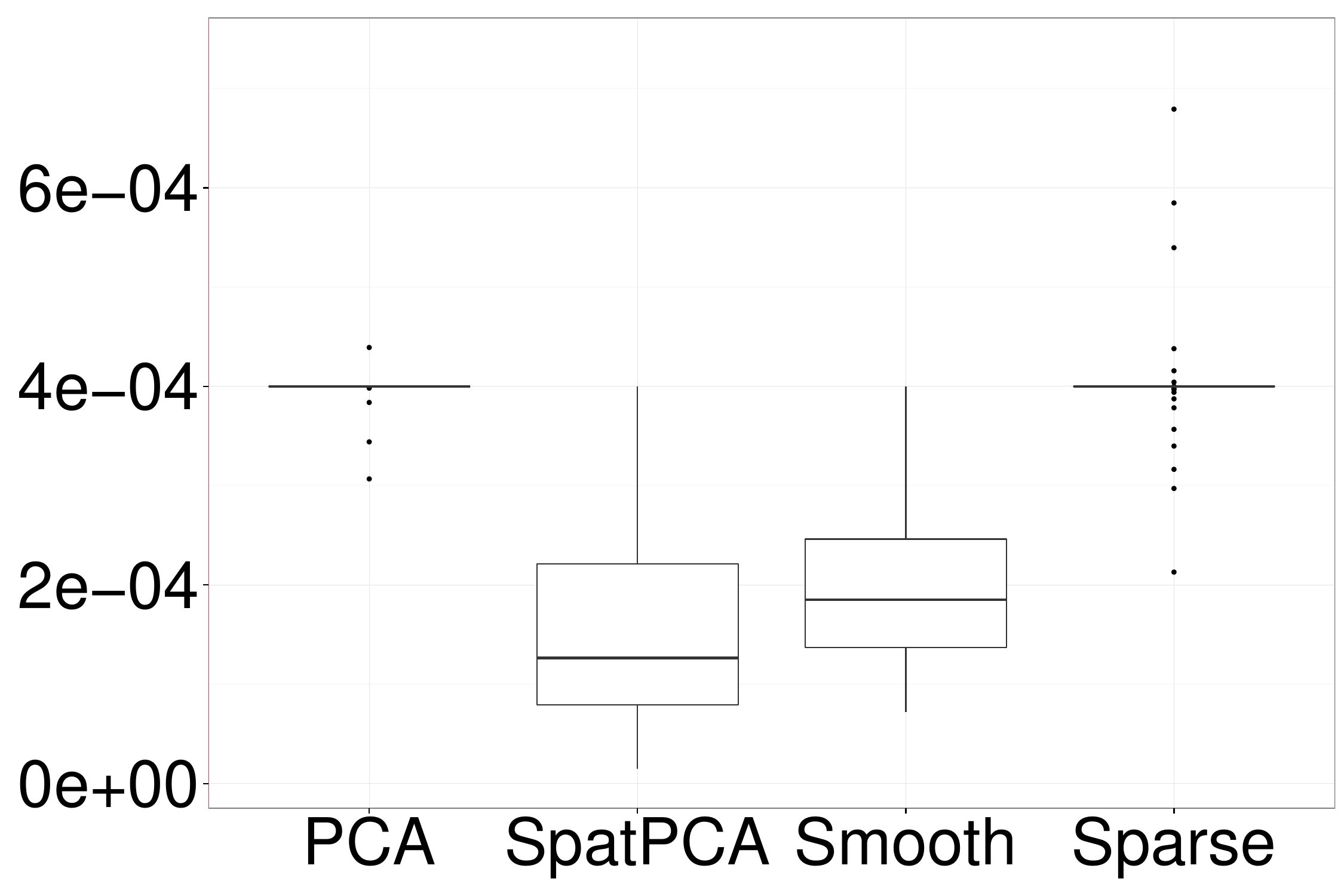}&
 		\includegraphics[scale=0.12]{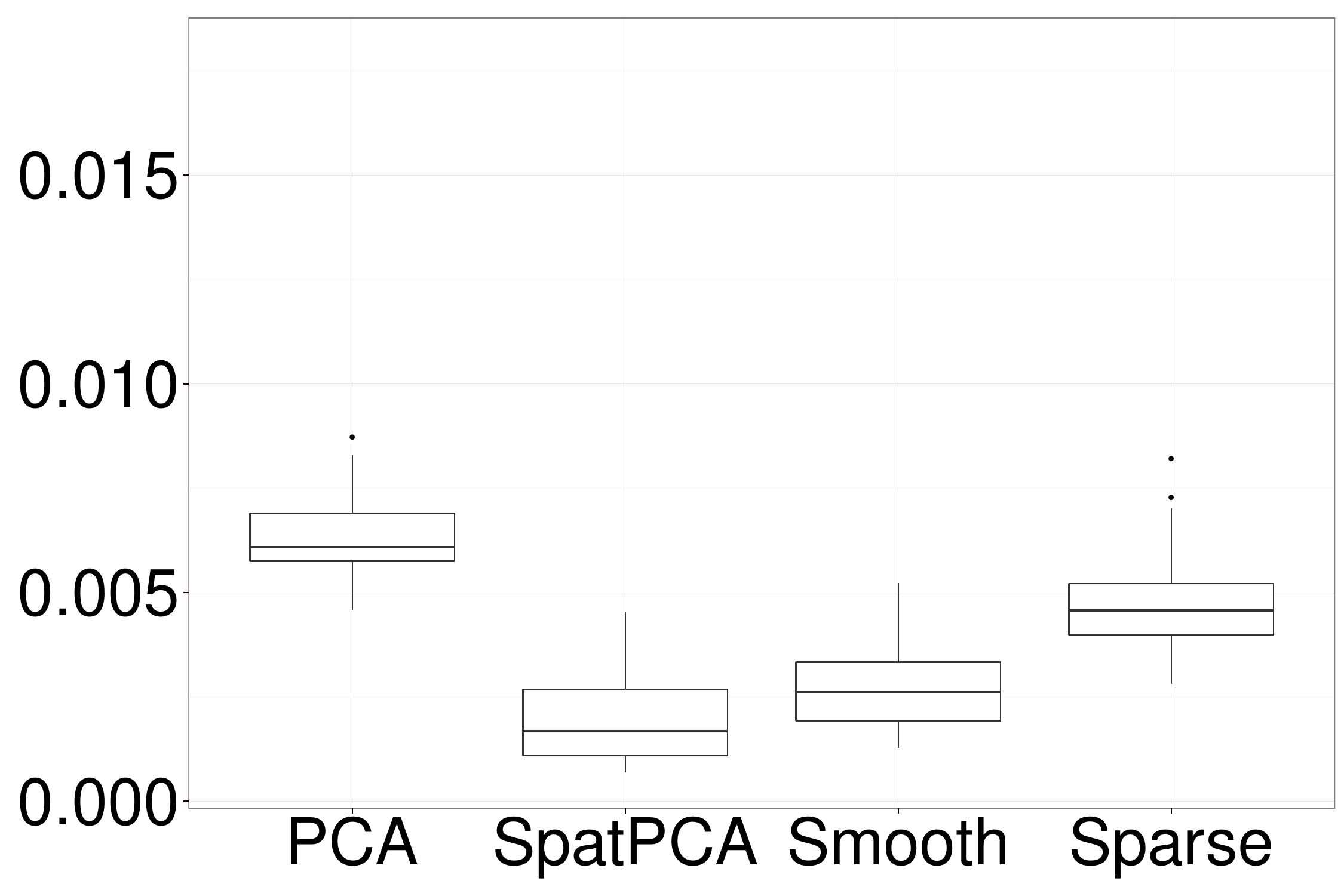}\\
 		{$(\lambda_1,\lambda_2)=(9,0)$, $K=5$}&{$(\lambda_1,\lambda_2)=(1,0)$, $K=5$}&{$(\lambda_1,\lambda_2)=(9,4)$, $K=5$}\\
 		\includegraphics[scale=0.12]{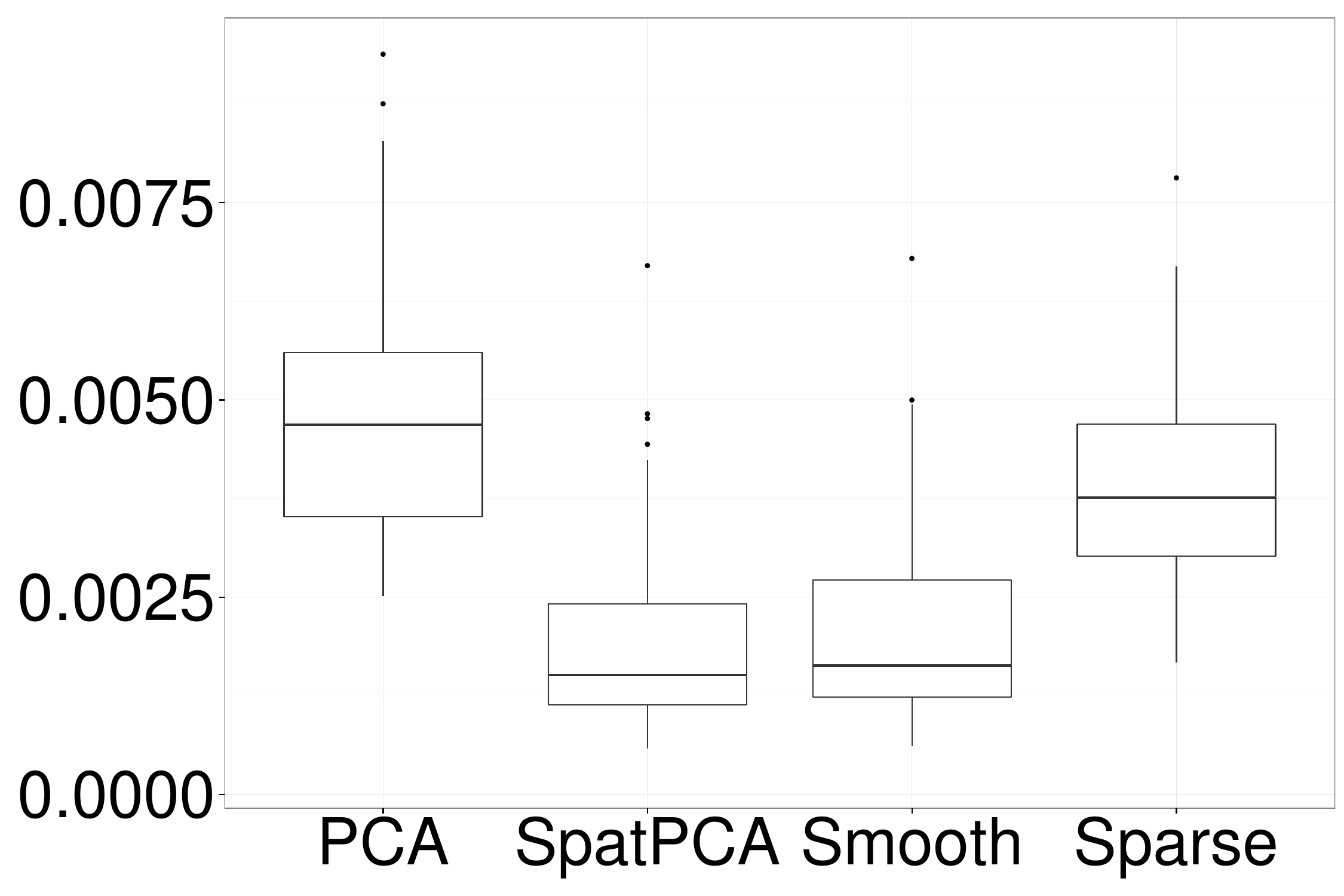}\hspace{4pt}&
 		\includegraphics[scale=0.12]{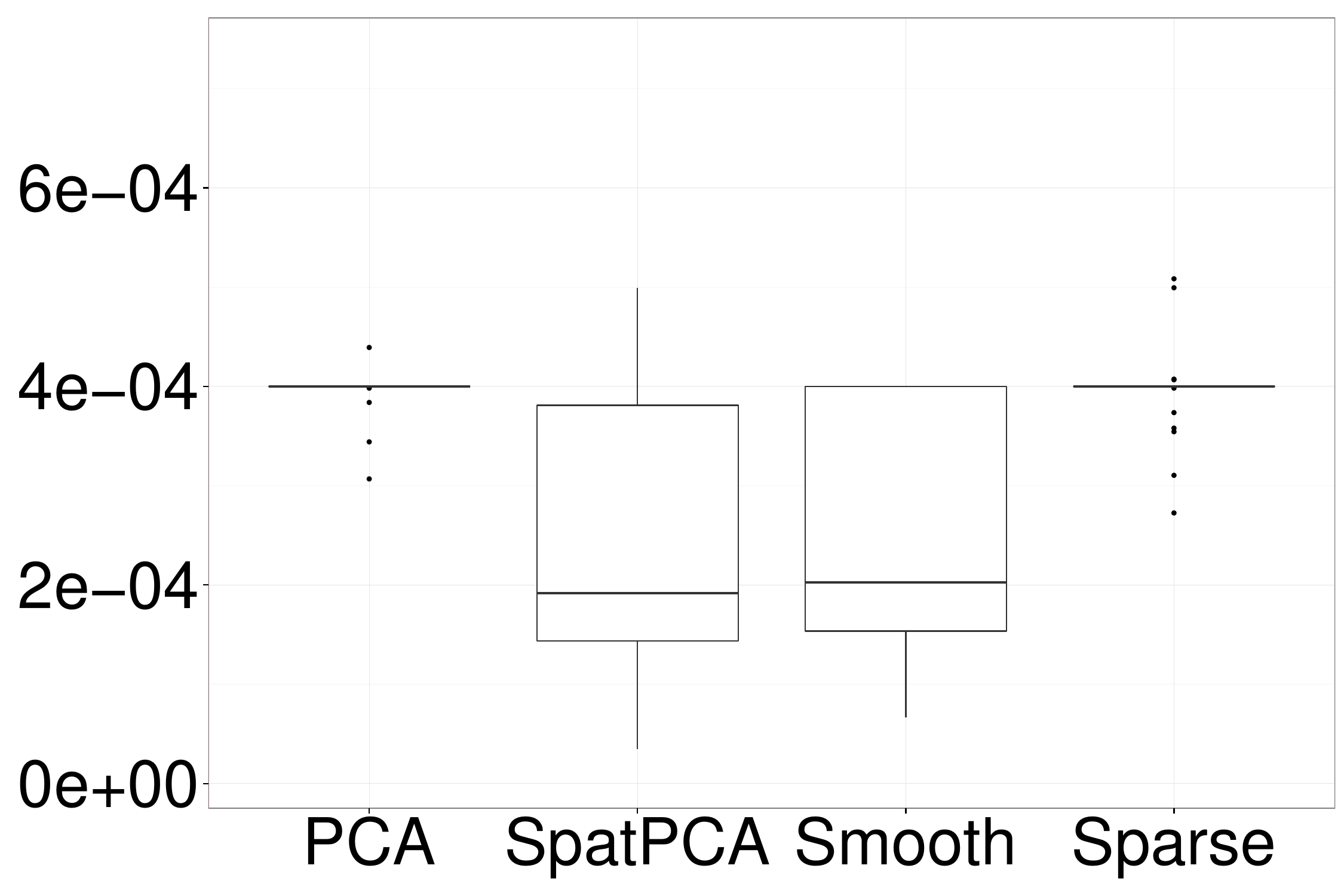}\hspace{4pt}&
 		\includegraphics[scale=0.12]{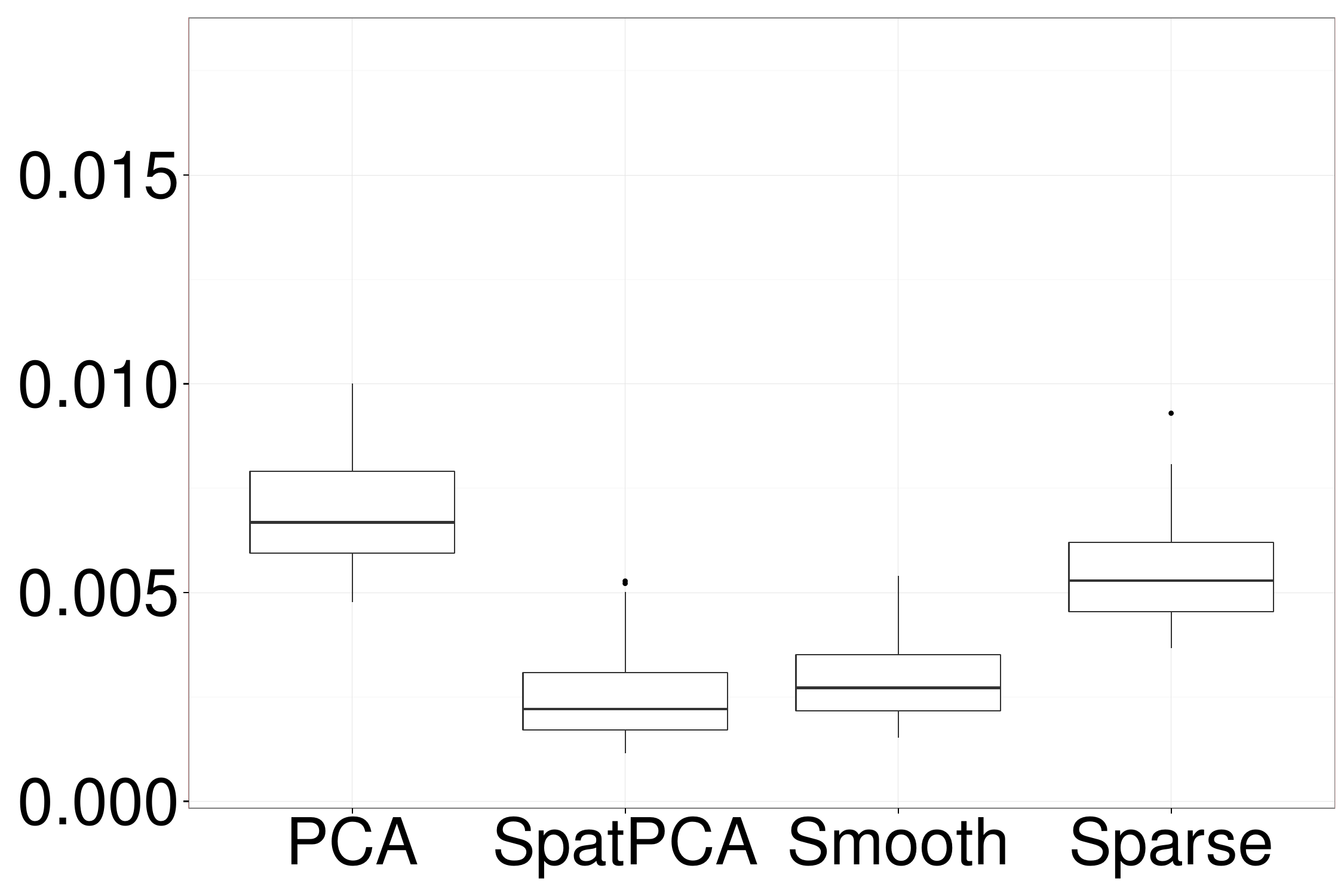}\hspace{4pt}\\
 		{$(\lambda_1,\lambda_2)=(9,0)$, $K=\hat{K}$}&{$(\lambda_1,\lambda_2)=(1,0)$, $K=\hat{K}$}&{$(\lambda_1,\lambda_2)=(9,4)$, $K=\hat{K}$}\\
 		\includegraphics[scale=0.12]{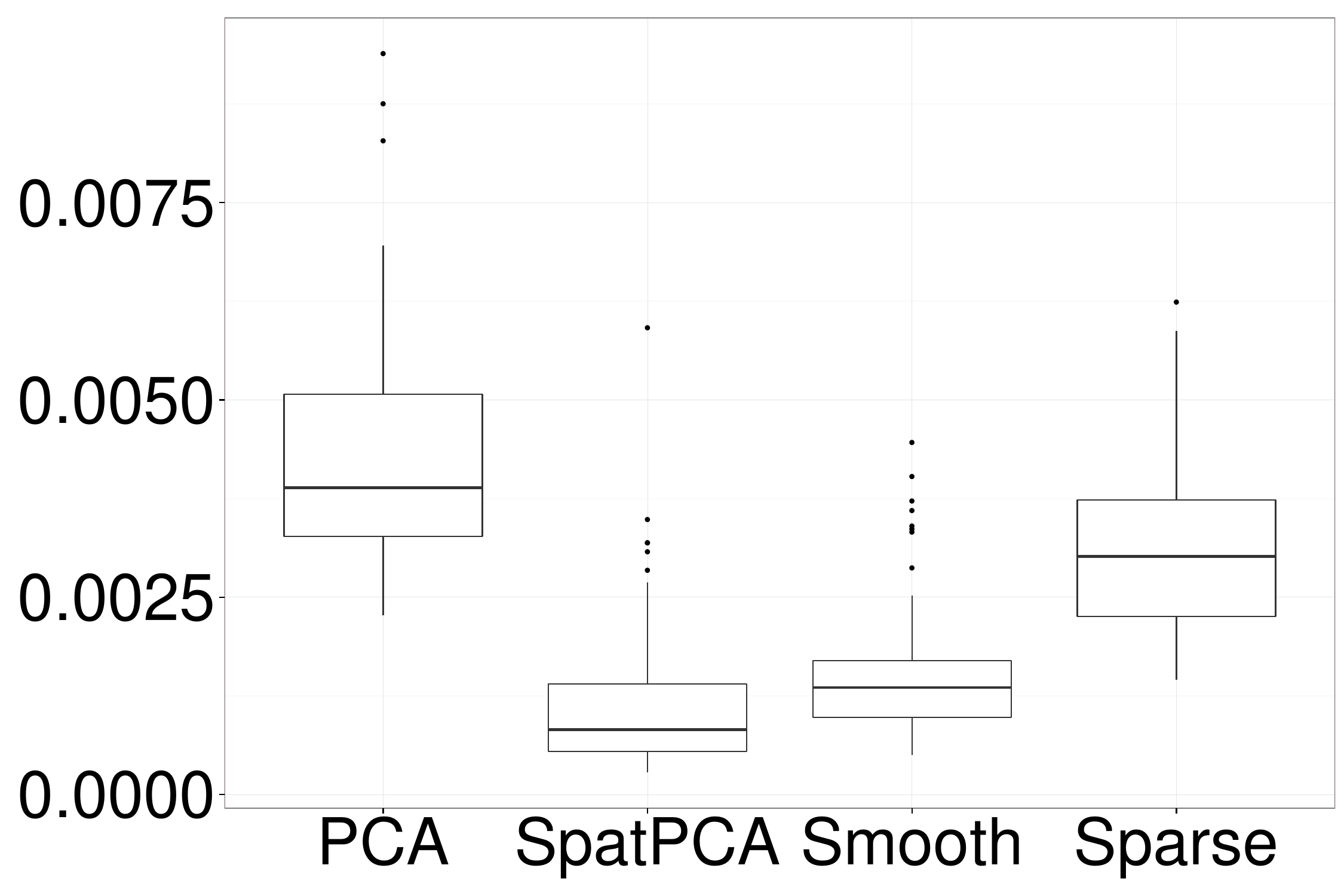}\hspace{4pt}&
 		\includegraphics[scale=0.12]{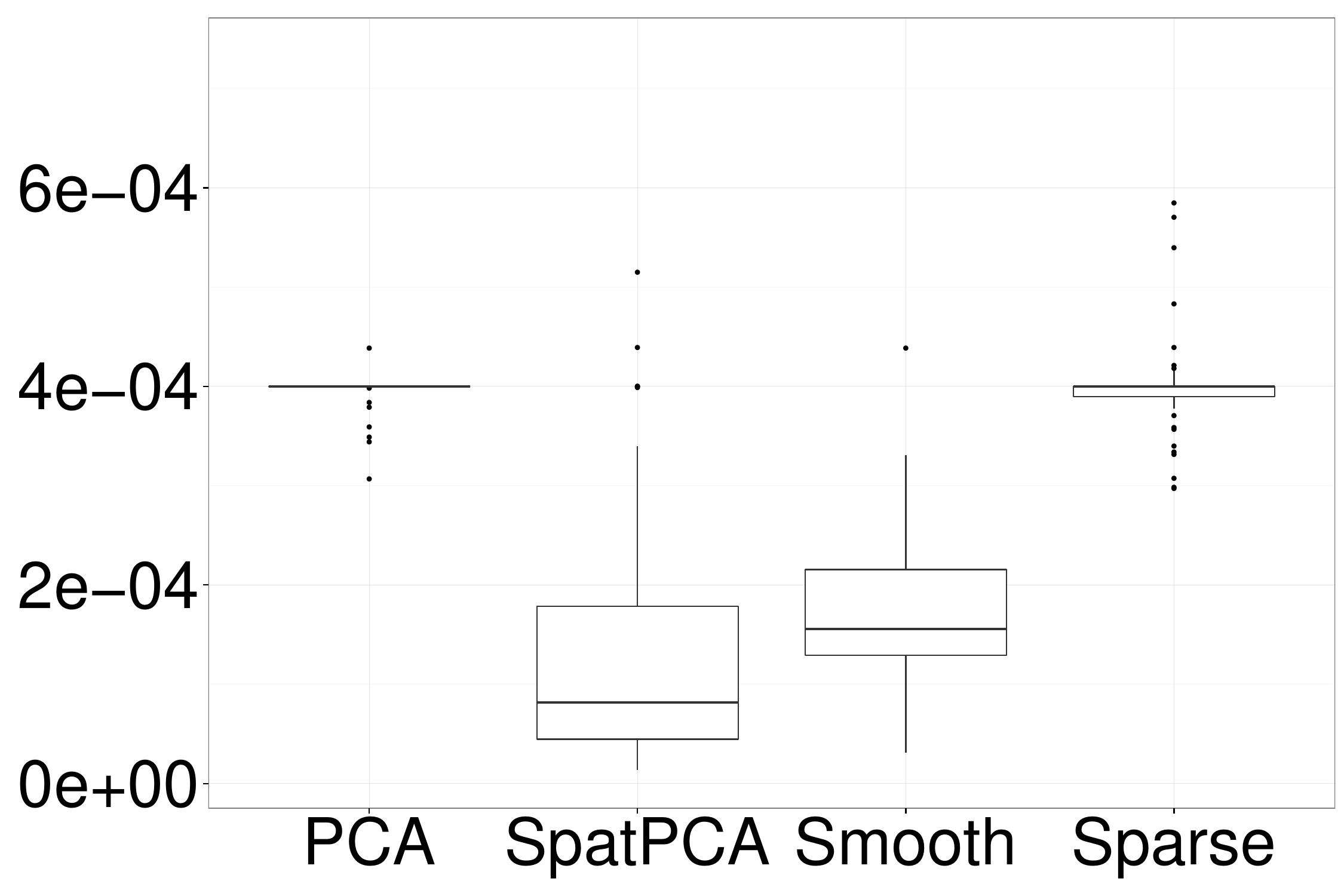}\hspace{4pt}&
 		\includegraphics[scale=0.12]{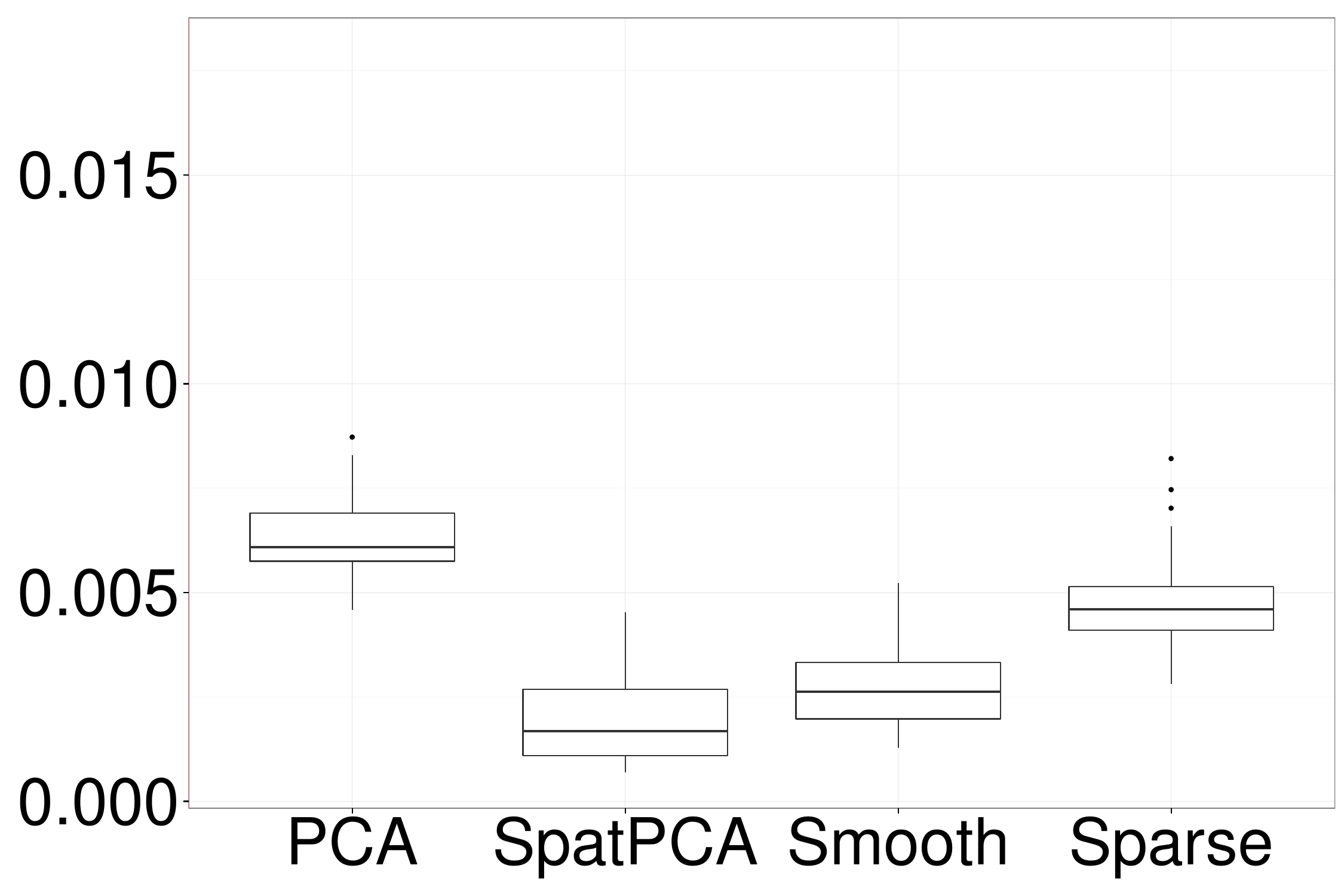}\hspace{4pt}
 	\end{tabular}
 }	
 \caption{{Boxplots of average squared estimation errors of (\ref{eq:loss2_sim}) for various methods in the one-dimensional simulation experiment of
Section \ref{sec:ex1} based on 50 simulation replicates.}}
 \label{fig:box_d1}
 \end{figure}

The performance of the four methods in terms of the loss functions \eqref{eq:loss_sim} and \eqref{eq:loss2_sim} is shown in Figures~\ref{fig:box_d1_loss1} and \ref{fig:box_d1}, respectively, based on $50$ simulation replicates. Once again, SpatPCA outperforms all the other methods in all cases.
 {For $(\lambda_1,\lambda_2)=(9,0)$, the average computation time for SpatPCA
(including selection of $\lambda_1$ and $\lambda_2$ using 5-fold CV) with $K=1,2,5$ are $0.020$, $0.065$ and $0.264$
seconds, respectively, which are larger than $0.002$ seconds required for PCA. The results were conducted using our R package
``SpatPCA" implemented on an iMac PC equipped with a 3.2GHz Intel Core i5 CPU and a 64GB RAM.}
 	 
\subsection{Two-Dimensional Experiment I} 	
\label{sec:2d-1}

\begin{figure}
{
Estimates of $\phi_1(\cdot)$ based on $(\lambda_1,\lambda_2)=(9,0)$ and $K=1$
\includegraphics[scale=0.4]{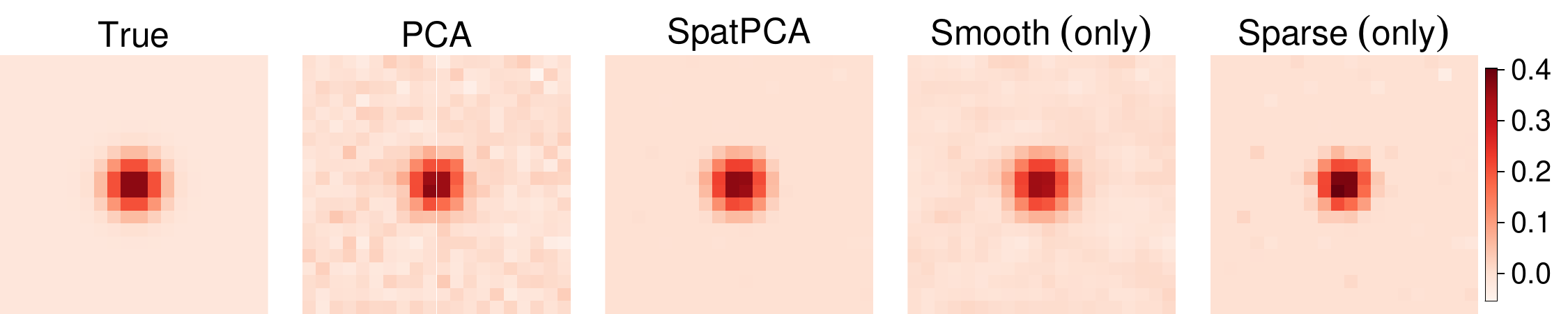}
Estimates $\phi_1(\cdot)$ based on $(\lambda_1,\lambda_2)=(1,0)$ and $K=1$
\includegraphics[scale=0.4]{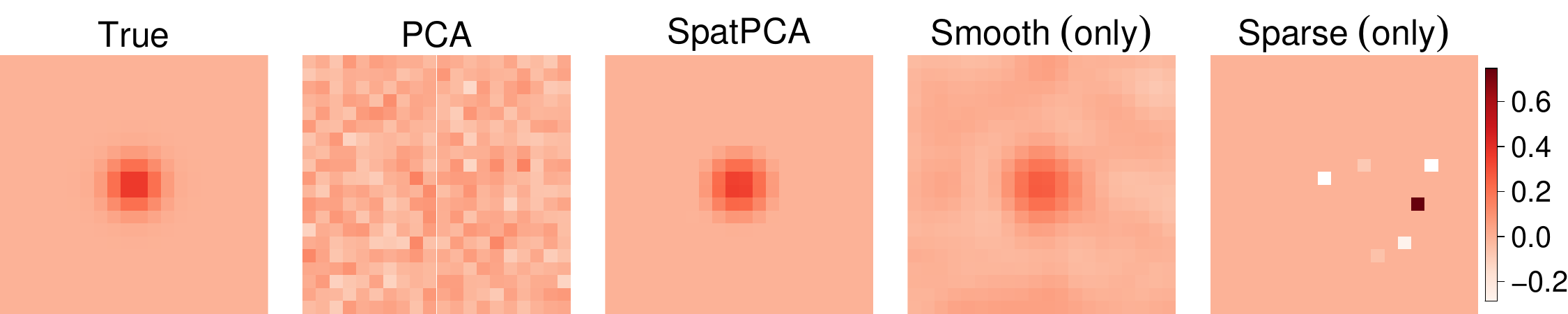}
Estimates of $\phi_1(\cdot)$ based on $(\lambda_1,\lambda_2)=(9,4)$ and $K=2$
\includegraphics[scale=0.4]{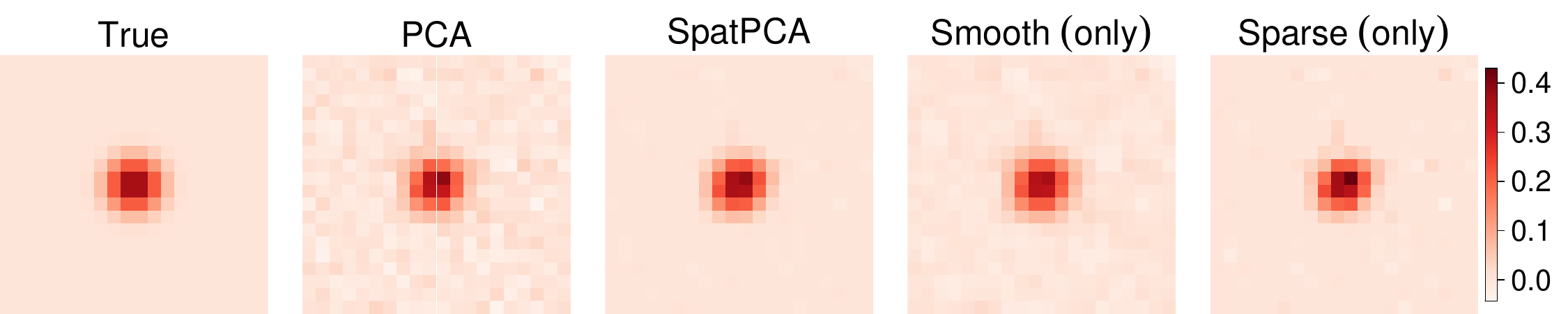}
Estimates of $\phi_2(\cdot)$ based on $(\lambda_1,\lambda_2)=(9,4)$ and $K=2$
\includegraphics[scale=0.4]{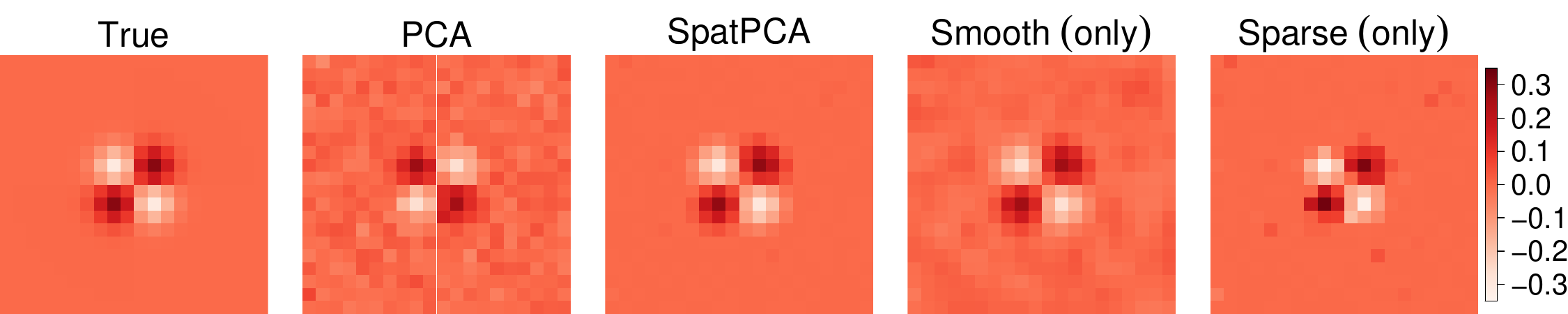}}
\caption{{Estimates of $\phi_1(\cdot)$ and $\phi_2(\cdot)$ obtained from various methods based on three different combinations of eigenvalues.}}
\label{fig:est_d2} 
\end{figure}

We considered a two-dimensional experiment by generating data according to (\ref{eq:measurement}) with $K=2$,
$\bm{\xi}_i\sim N(\bm{0}, \mathrm{diag}(\lambda_1,\lambda_2))$, $\bm{\epsilon}_i\sim N(\bm{0},\bm{I})$, $n=500$, $\bm{s}_1,\dots,\bm{s}_p$
regularly spaced at $p=20^2$ locations in $D=[-5,5]^2$.
Here $\phi_1(\cdot)$ and $\phi_2(\cdot)$ are given by (\ref{eq:phi1_sim}) and (\ref{eq:phi2_sim}) with $d=2$
(see {the images in the first column of Figure~\ref{fig:est_d2}}).

{We considered three pairs of $(\lambda_1,\lambda_2)\in\{(9,0),(1,0),(9,4)\}$, and applied the proposed
SpatPCA {with $K\in\{1,2,5\}$ and $\hat{K}$ selected from (\ref{eq:khat})}. As in the one-dimensional experiment,
we used 5-fold CV of \eqref{eq:cv} and a two-step procedure to select among the same 11 values of $\tau_1$ and $31$ values of $\tau_2$.
Similarly, we used 5-fold CV of \eqref{eq:cv.gamma} to select among the same 11 values of $\gamma$ for covariance function estimation.} 

\begin{figure}\centering
{
		\begin{tabular}{ccc}
			{$(\lambda_1,\lambda_2)=(9,0)$, $K=1$}&{$(\lambda_1,\lambda_2)=(1,0)$, $K=1$}&{$(\lambda_1,\lambda_2)=(9,4)$, $K=1$}\\
			\includegraphics[scale=0.12]{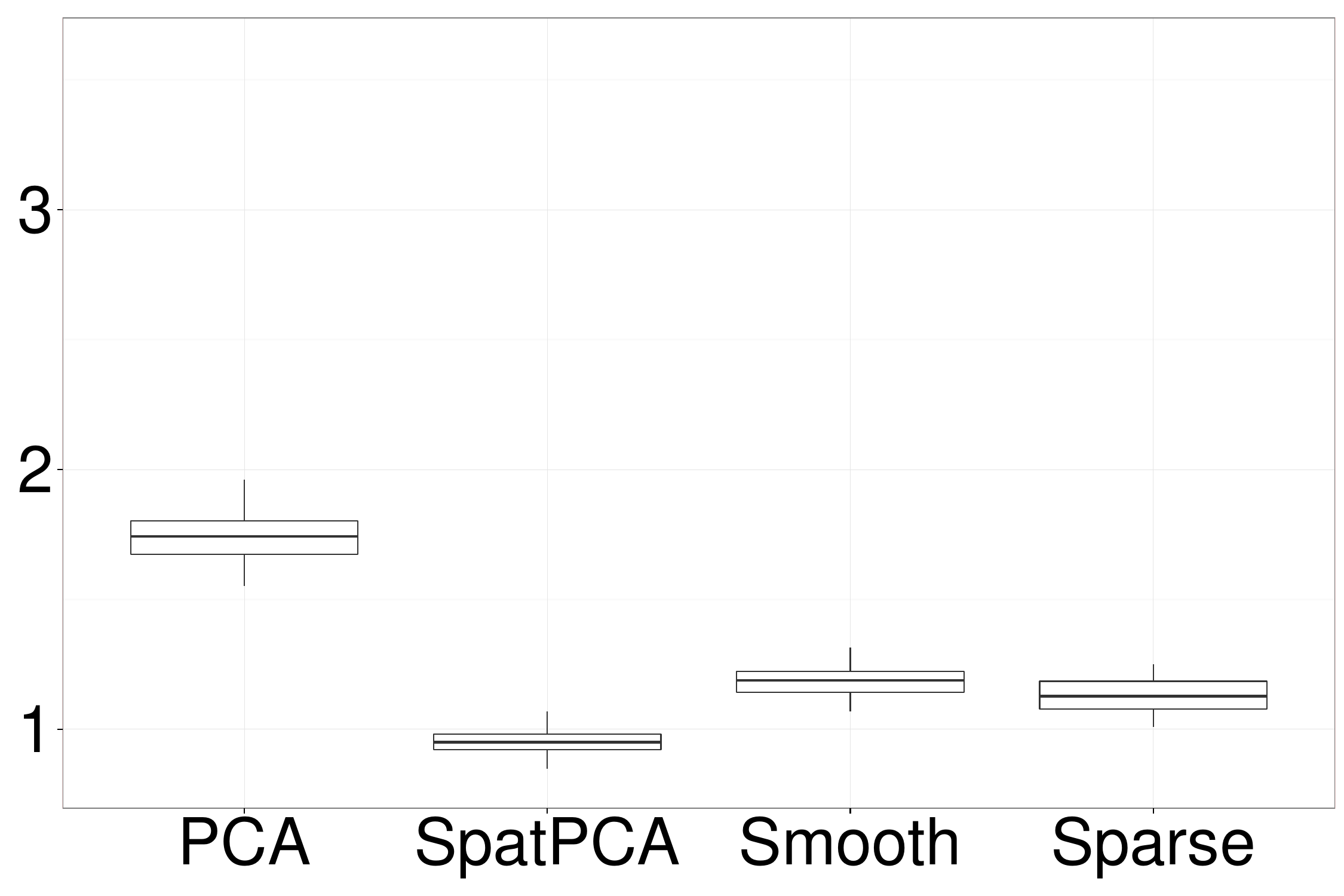}&
			\includegraphics[scale=0.12]{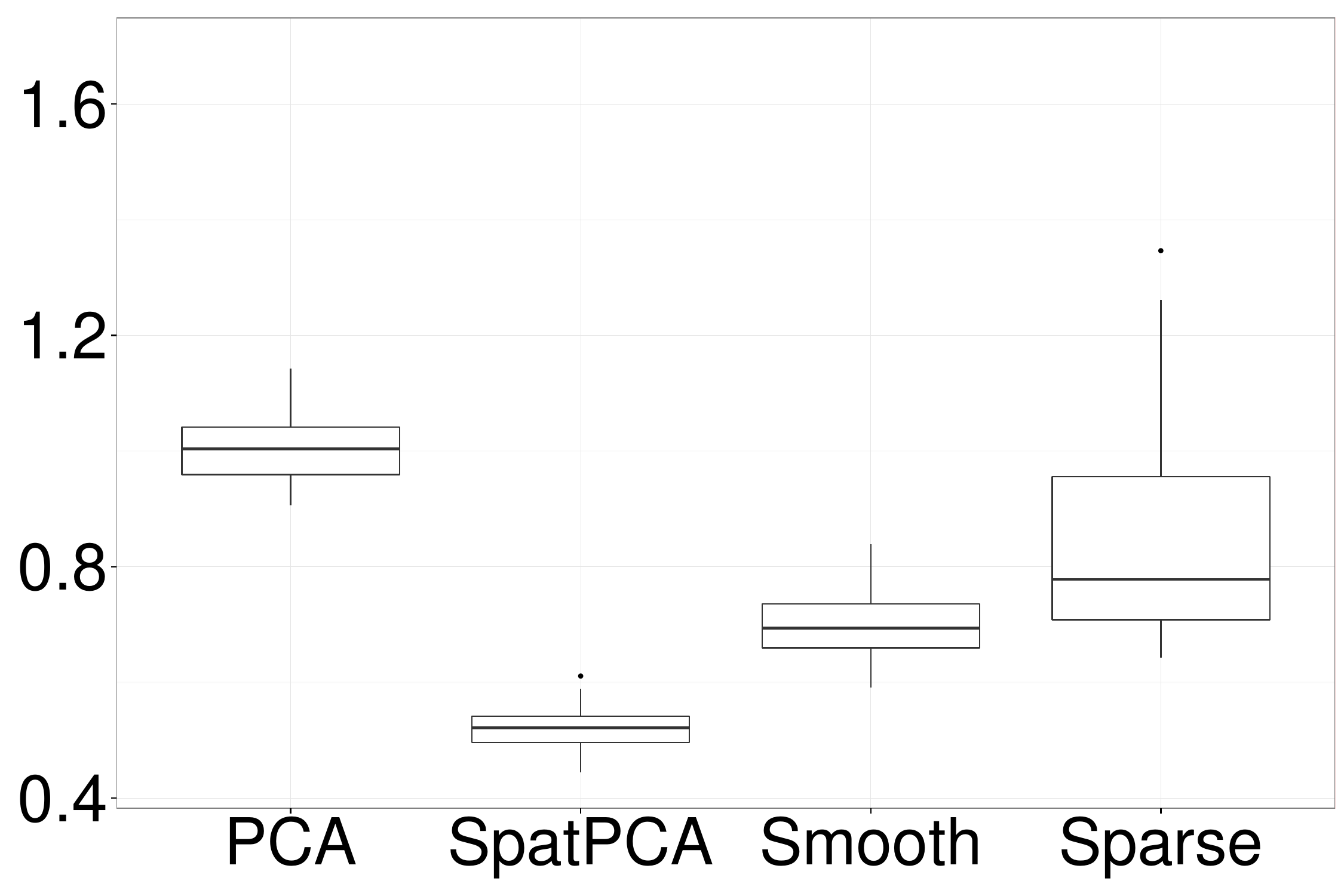}&
			\includegraphics[scale=0.12]{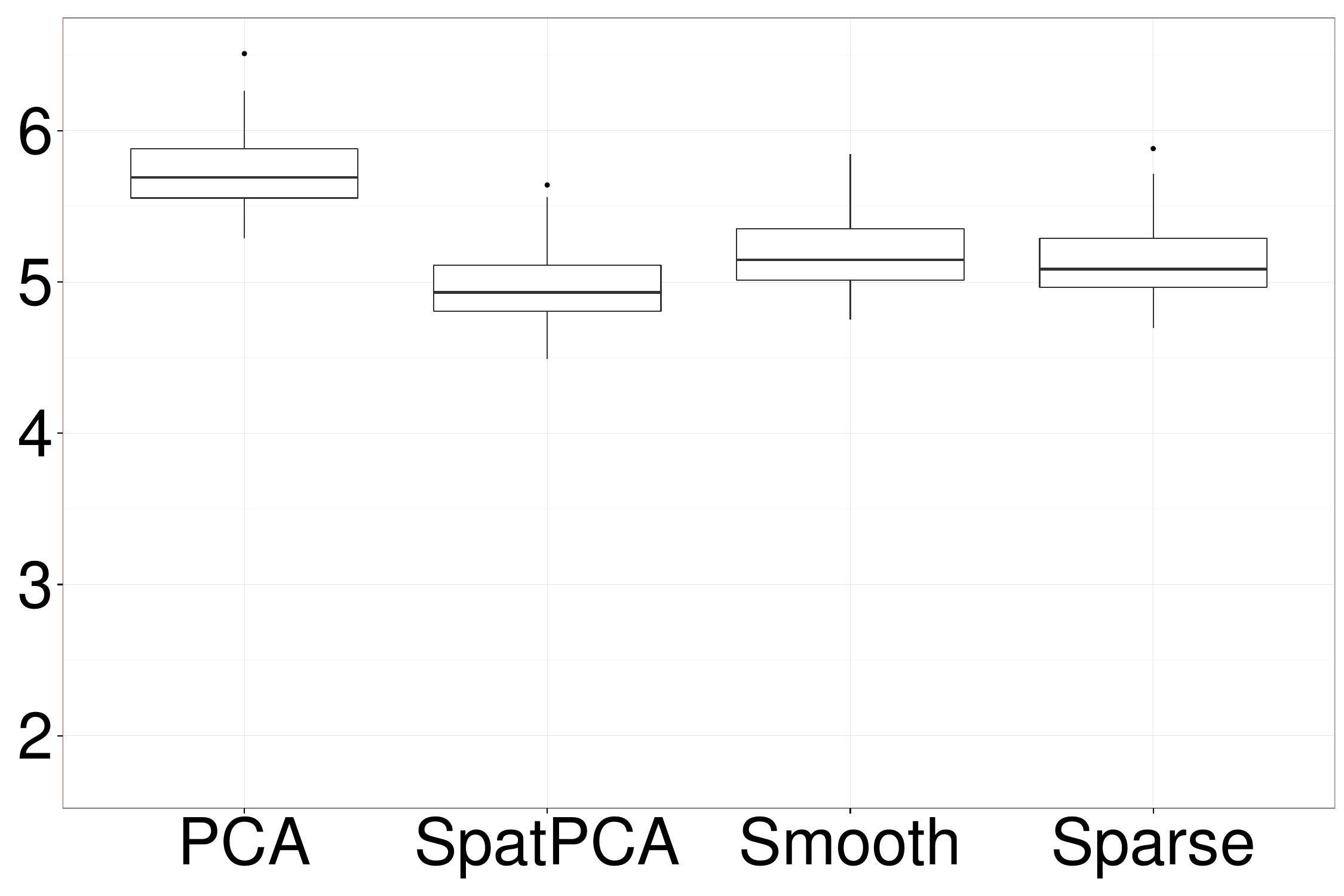}\\
			{$(\lambda_1,\lambda_2)=(9,0)$, $K=2$}&{$(\lambda_1,\lambda_2)=(1,0)$, $K=2$}&{$(\lambda_1,\lambda_2)=(9,4)$, $K=2$}\\
			\includegraphics[scale=0.12]{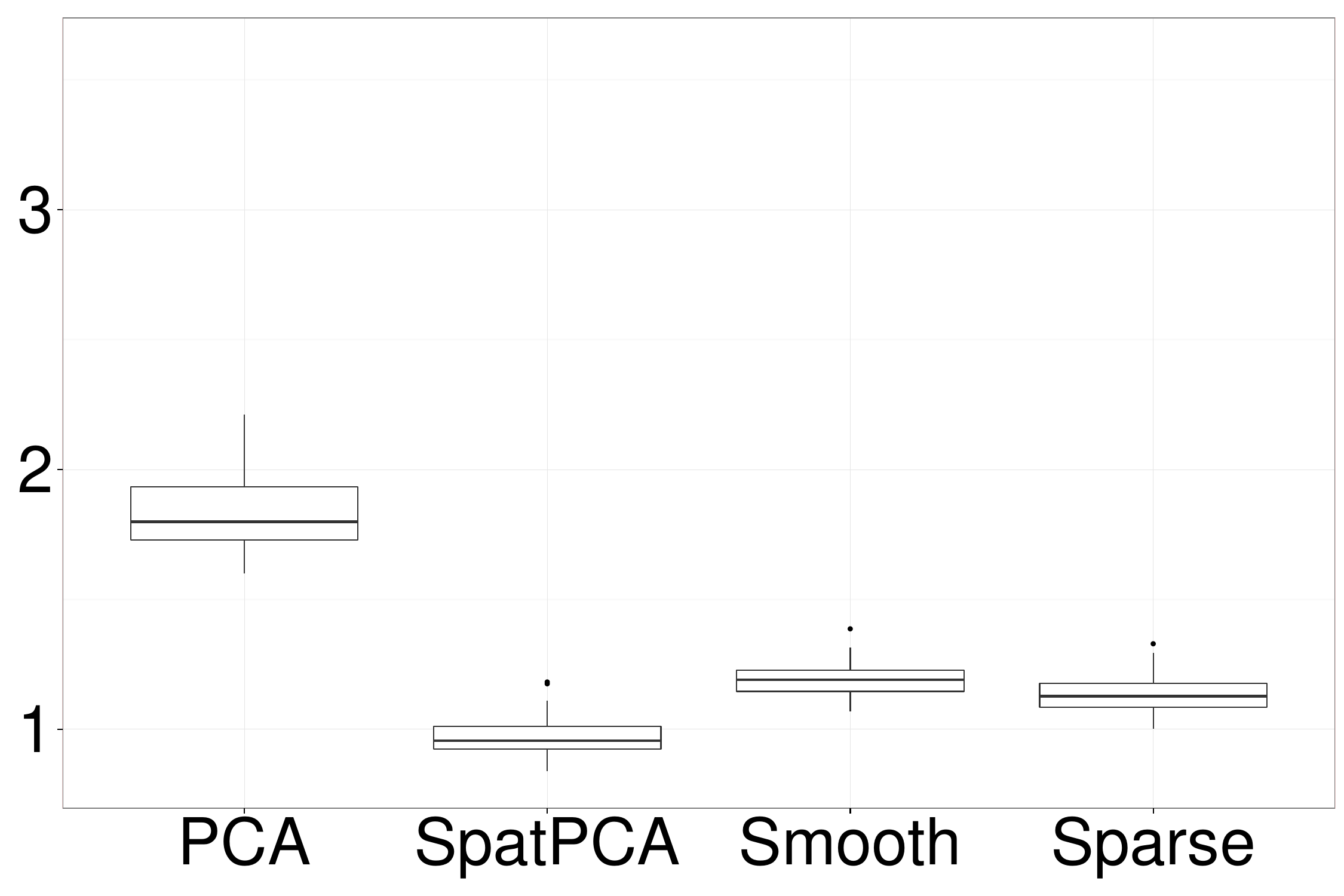}&
			\includegraphics[scale=0.12]{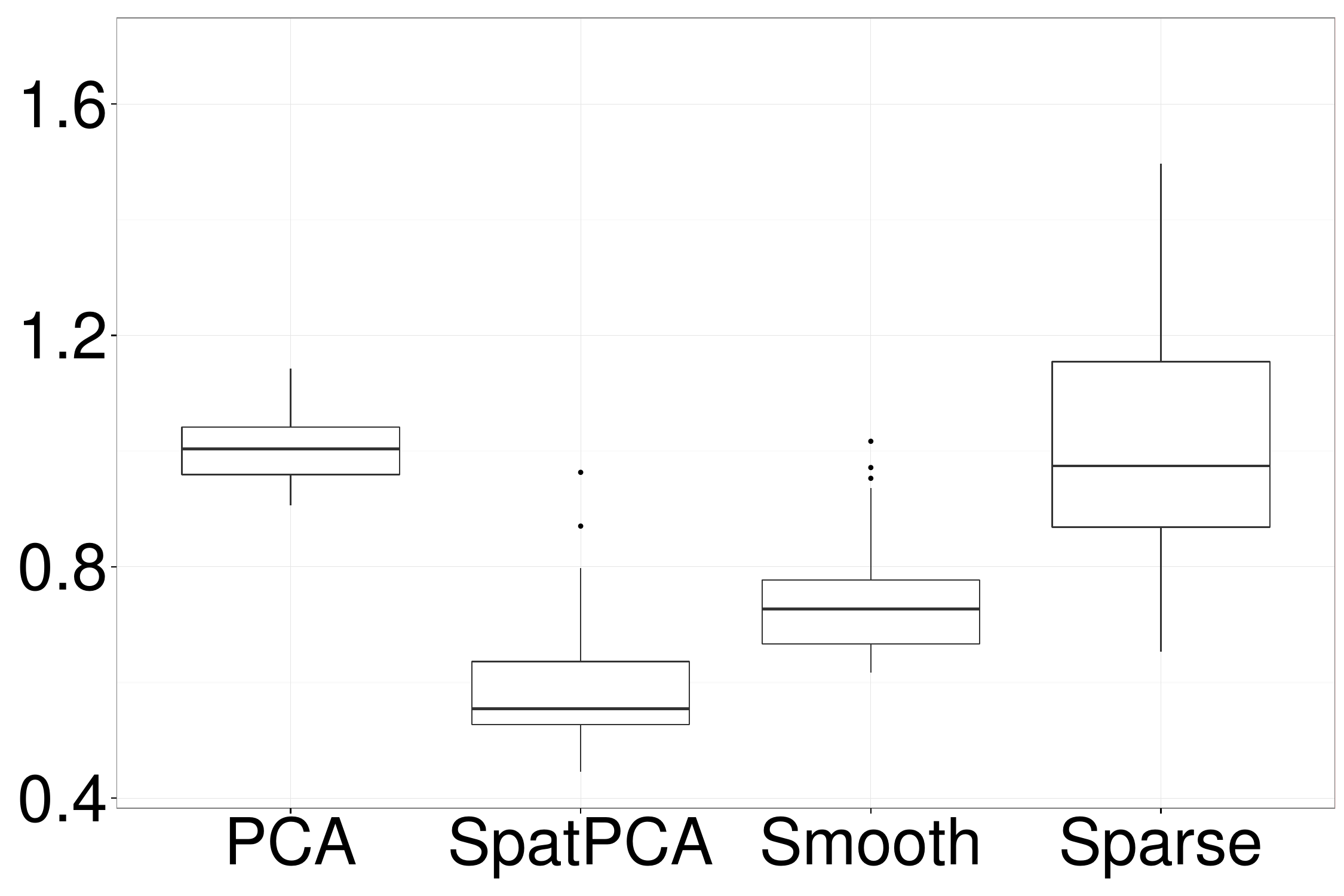}&
			\includegraphics[scale=0.12]{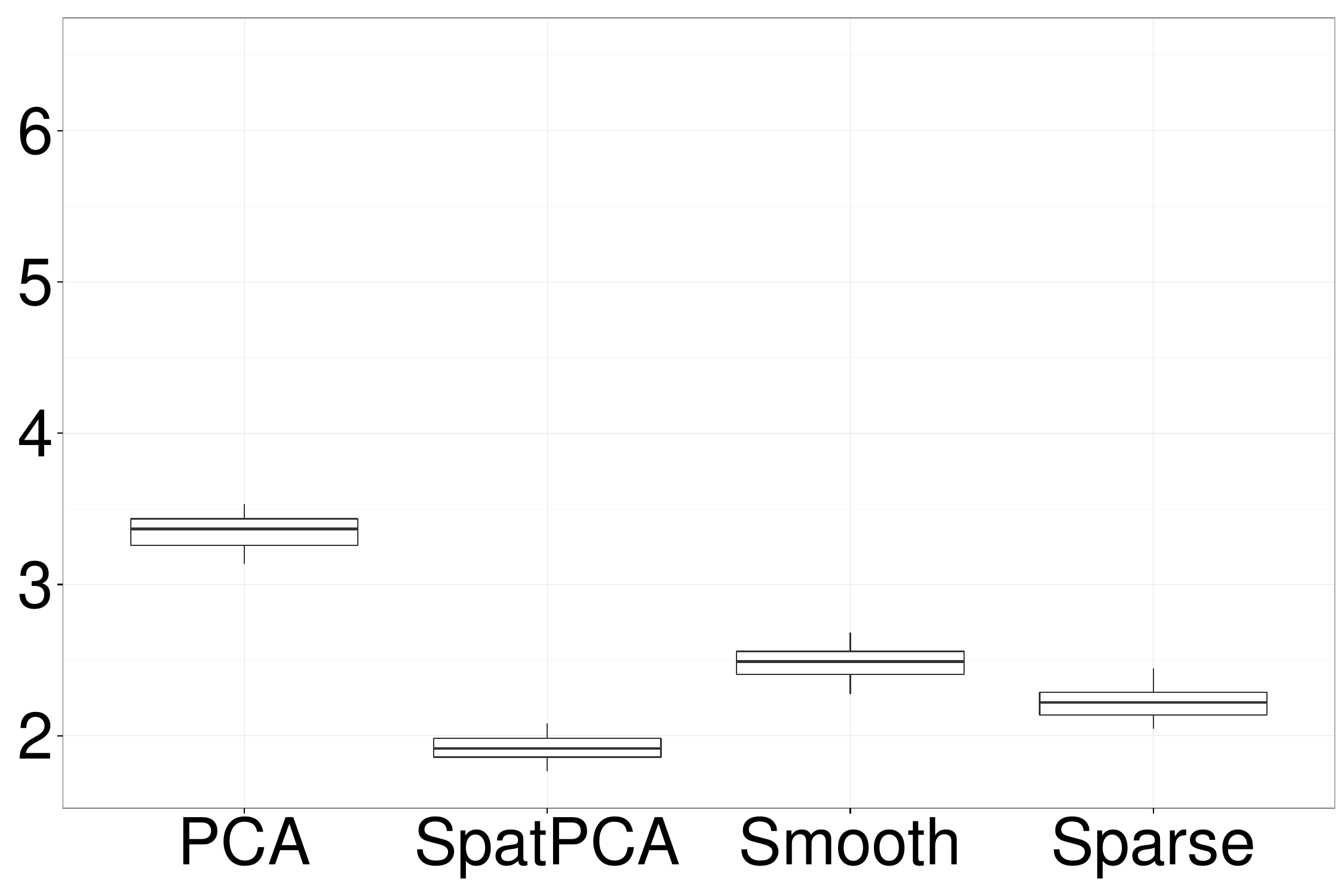}\\
			{$(\lambda_1,\lambda_2)=(9,0)$, $K=5$}&{$(\lambda_1,\lambda_2)=(1,0)$, $K=5$}&{$(\lambda_1,\lambda_2)=(9,4)$, $K=5$}\\
			\includegraphics[scale=0.12]{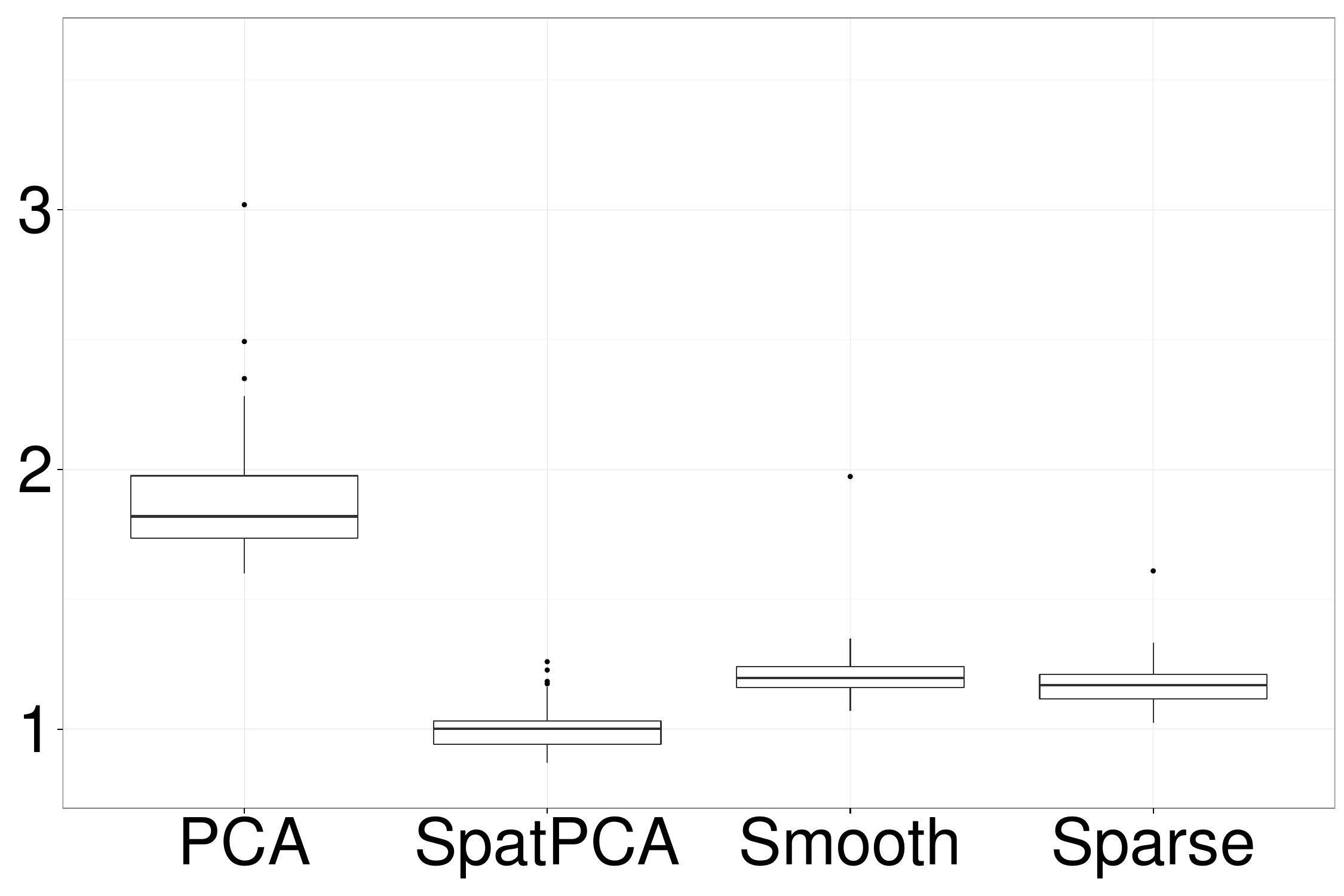}\hspace{4pt}&
			\includegraphics[scale=0.12]{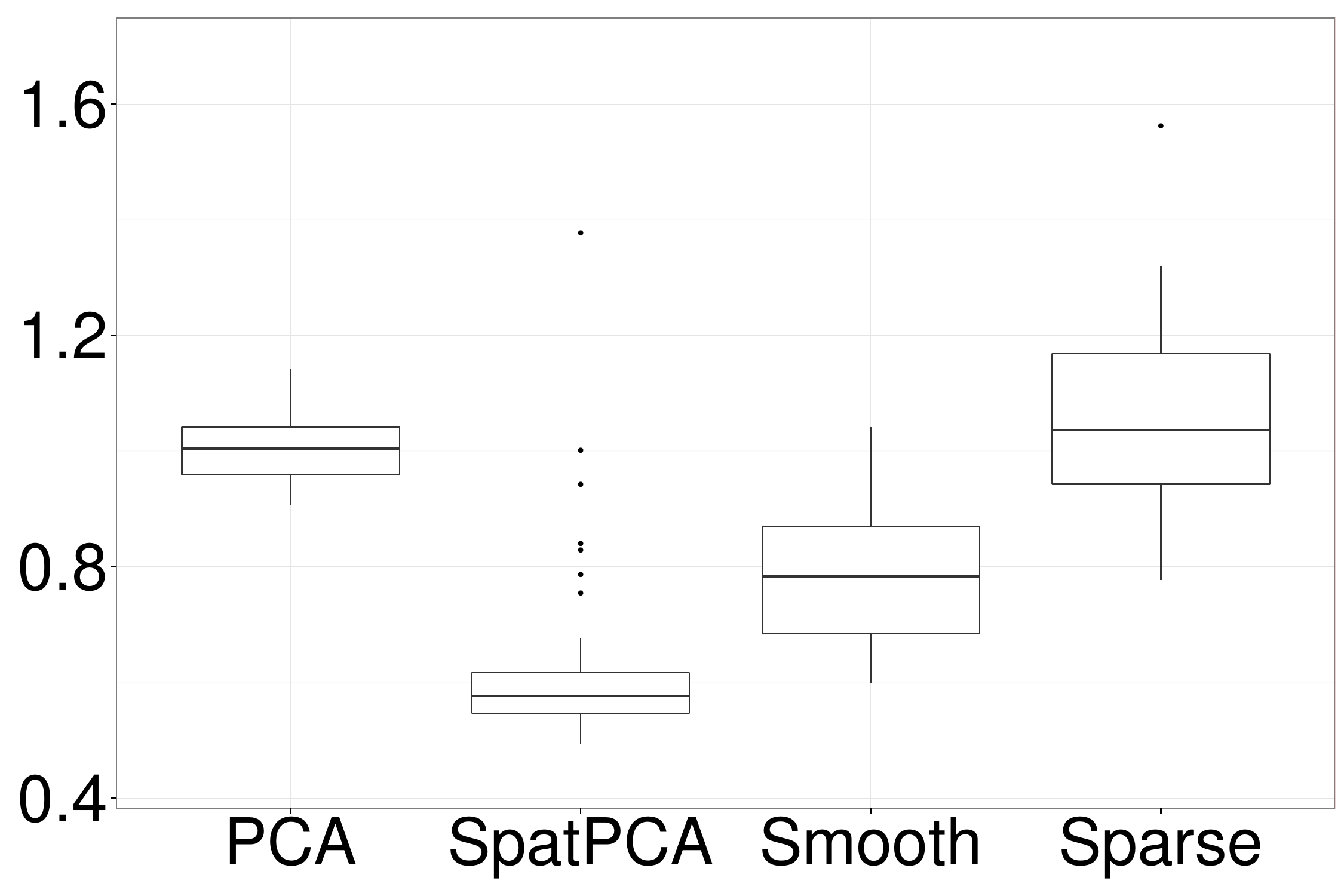}\hspace{4pt}&
			\includegraphics[scale=0.12]{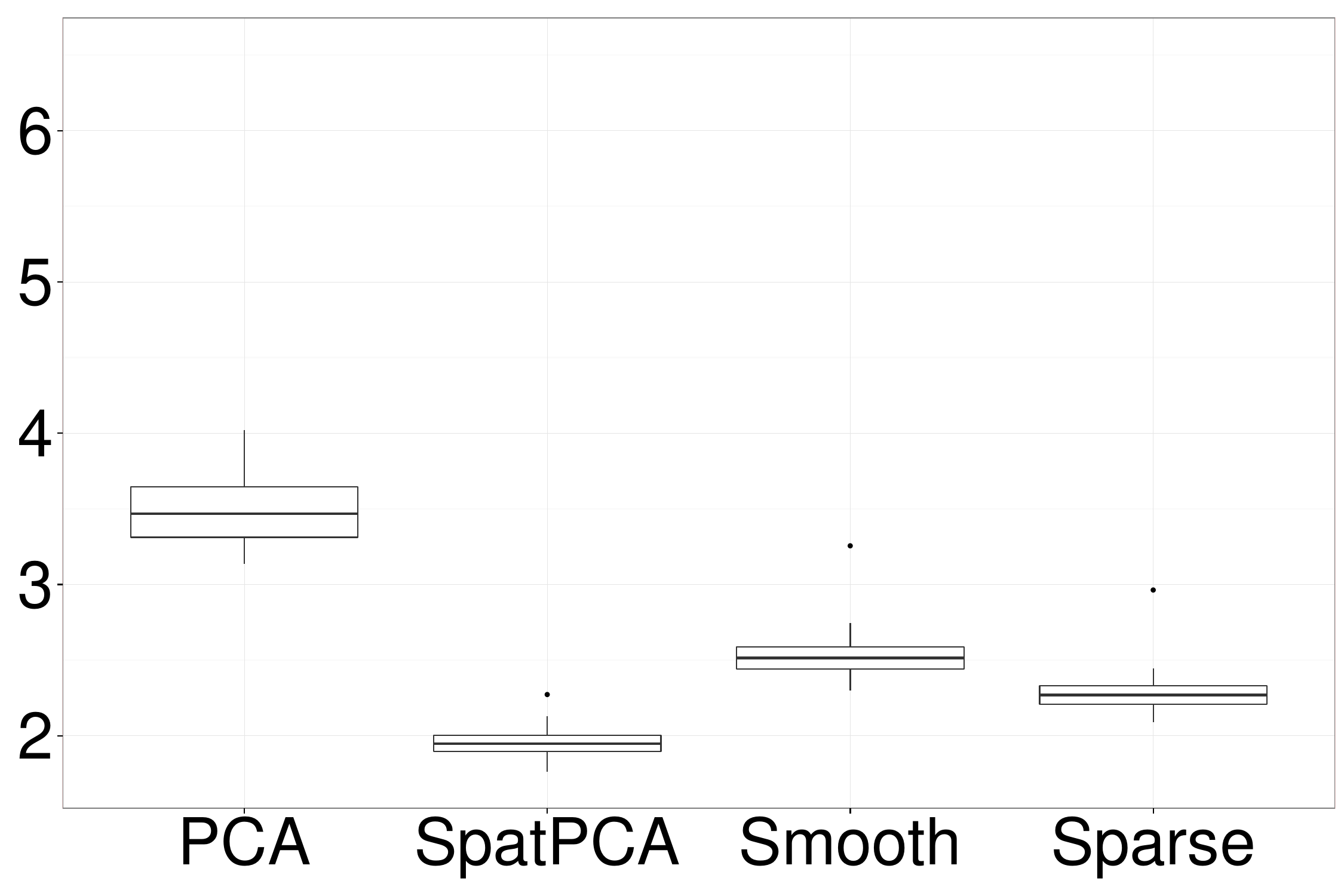}\hspace{4pt}\\
			{$(\lambda_1,\lambda_2)=(9,0)$, $K=\hat{K}$}&{$(\lambda_1,\lambda_2)=(1,0)$, $K=\hat{K}$}&{$(\lambda_1,\lambda_2)=(9,4)$, $K=\hat{K}$}\\
			\includegraphics[scale=0.12]{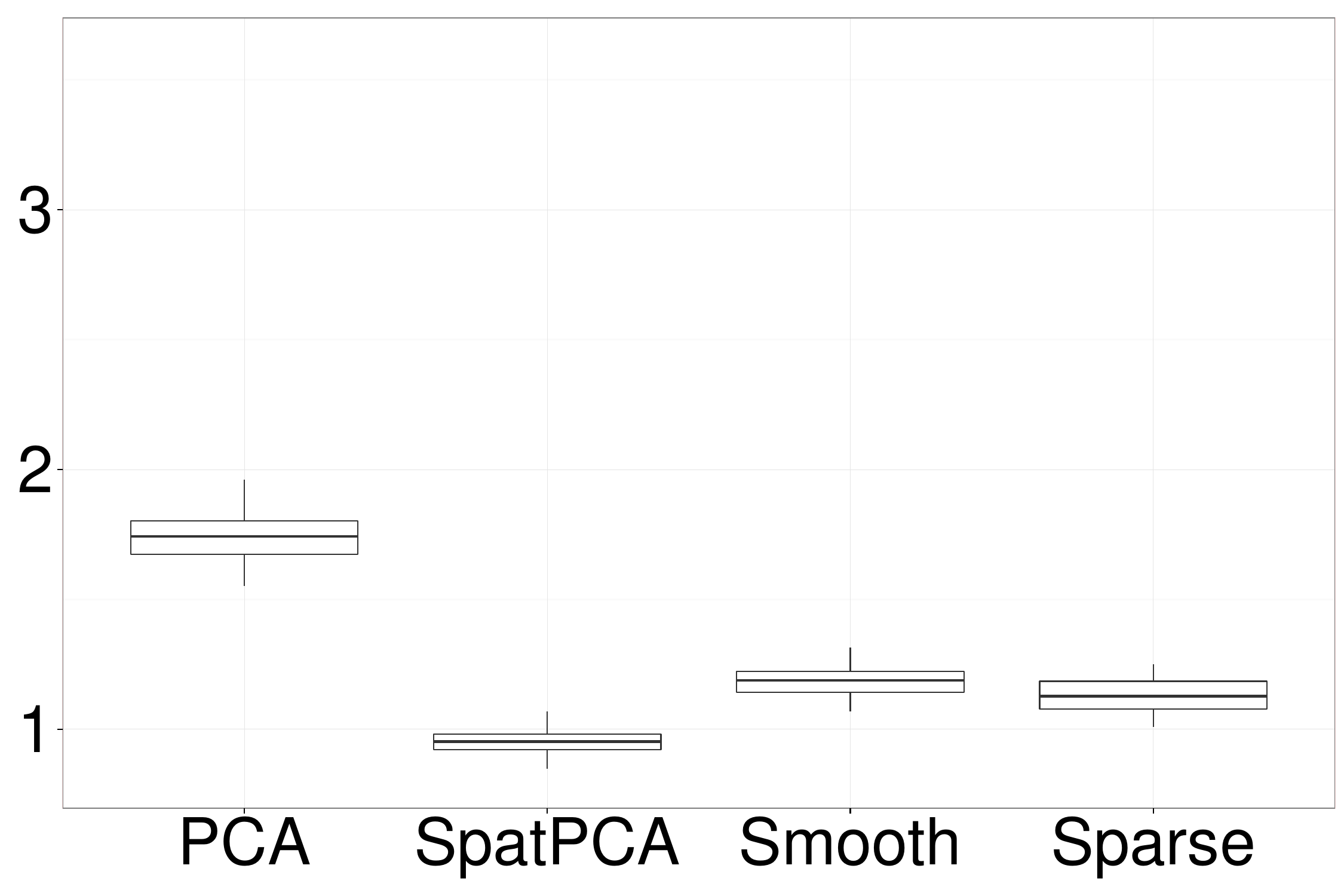}\hspace{4pt}&
			\includegraphics[scale=0.12]{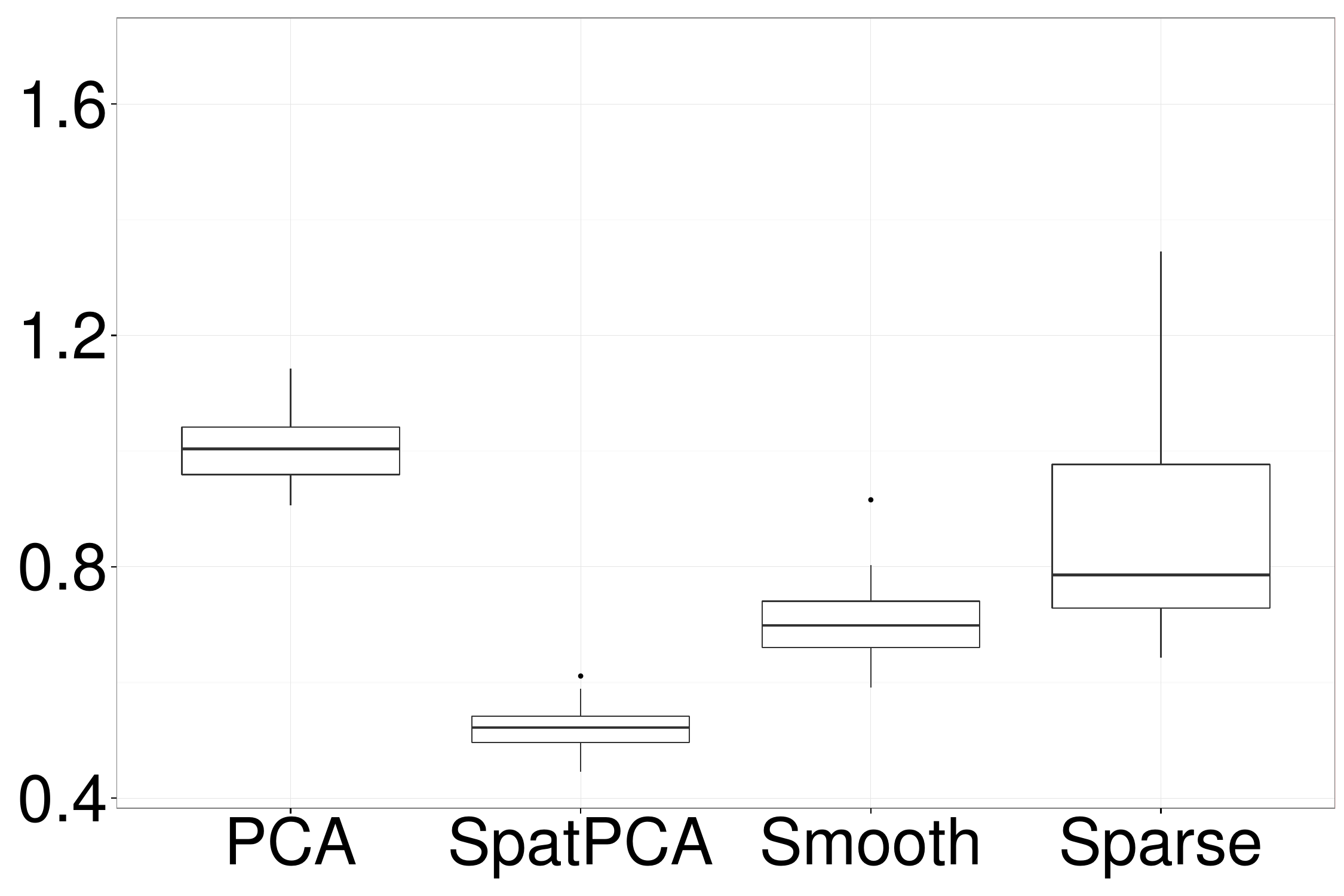}\hspace{4pt}&
			\includegraphics[scale=0.12]{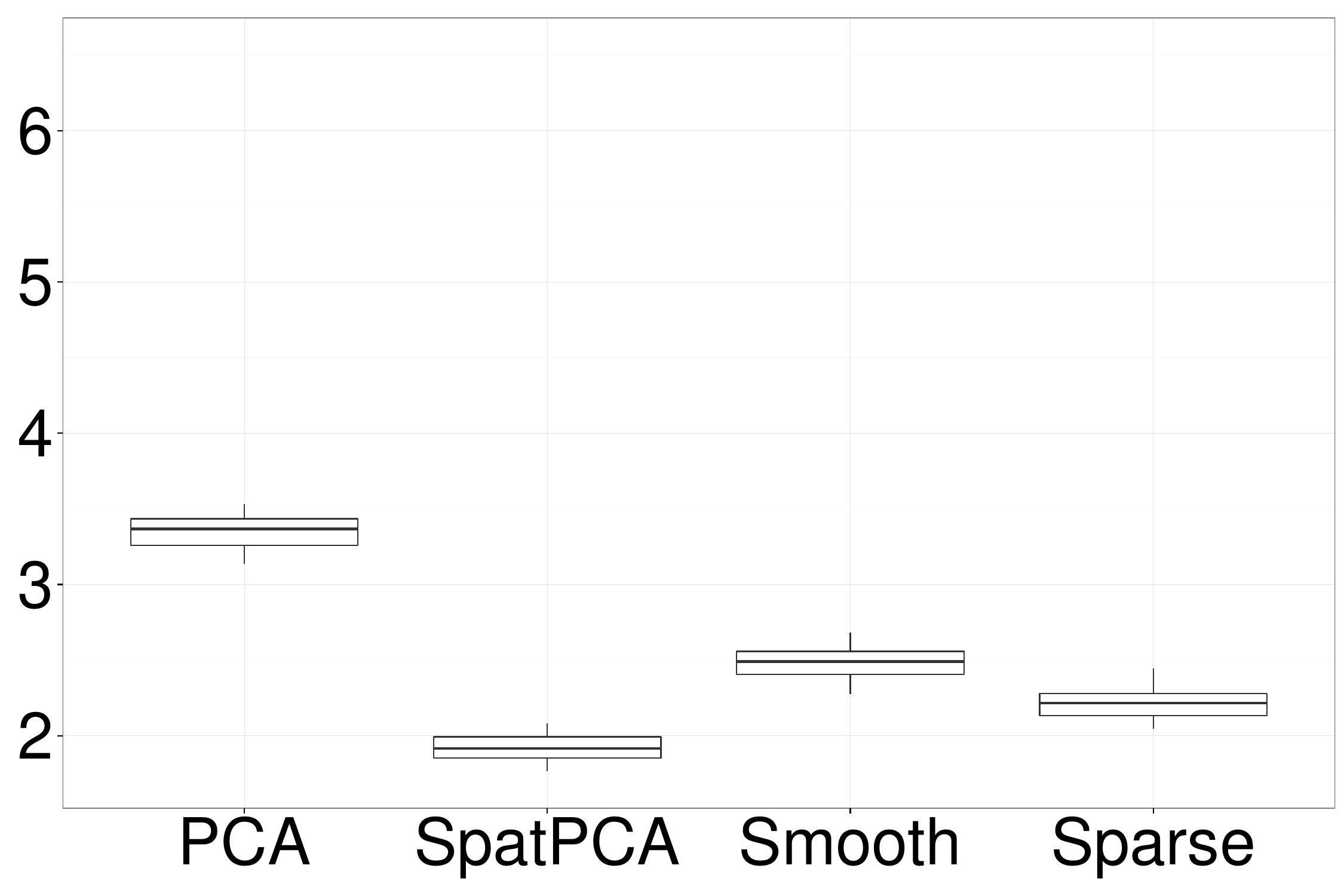}\hspace{4pt}
		\end{tabular}
	}
\caption{{Boxplots of average squared prediction errors of (\ref{eq:loss_sim}) for various methods in the two-dimensional simulation experiment of
Section \ref{sec:2d-1} based on 50 simulation replicates.}}
 \label{fig:box_d2_loss1}
 \end{figure}

\begin{figure}\centering
	{
		\begin{tabular}{ccc}
			{$(\lambda_1,\lambda_2)=(9,0)$, $K=1$}&{$(\lambda_1,\lambda_2)=(1,0)$, $K=1$}&{$(\lambda_1,\lambda_2)=(9,4)$, $K=1$}\\
			\includegraphics[scale=0.12]{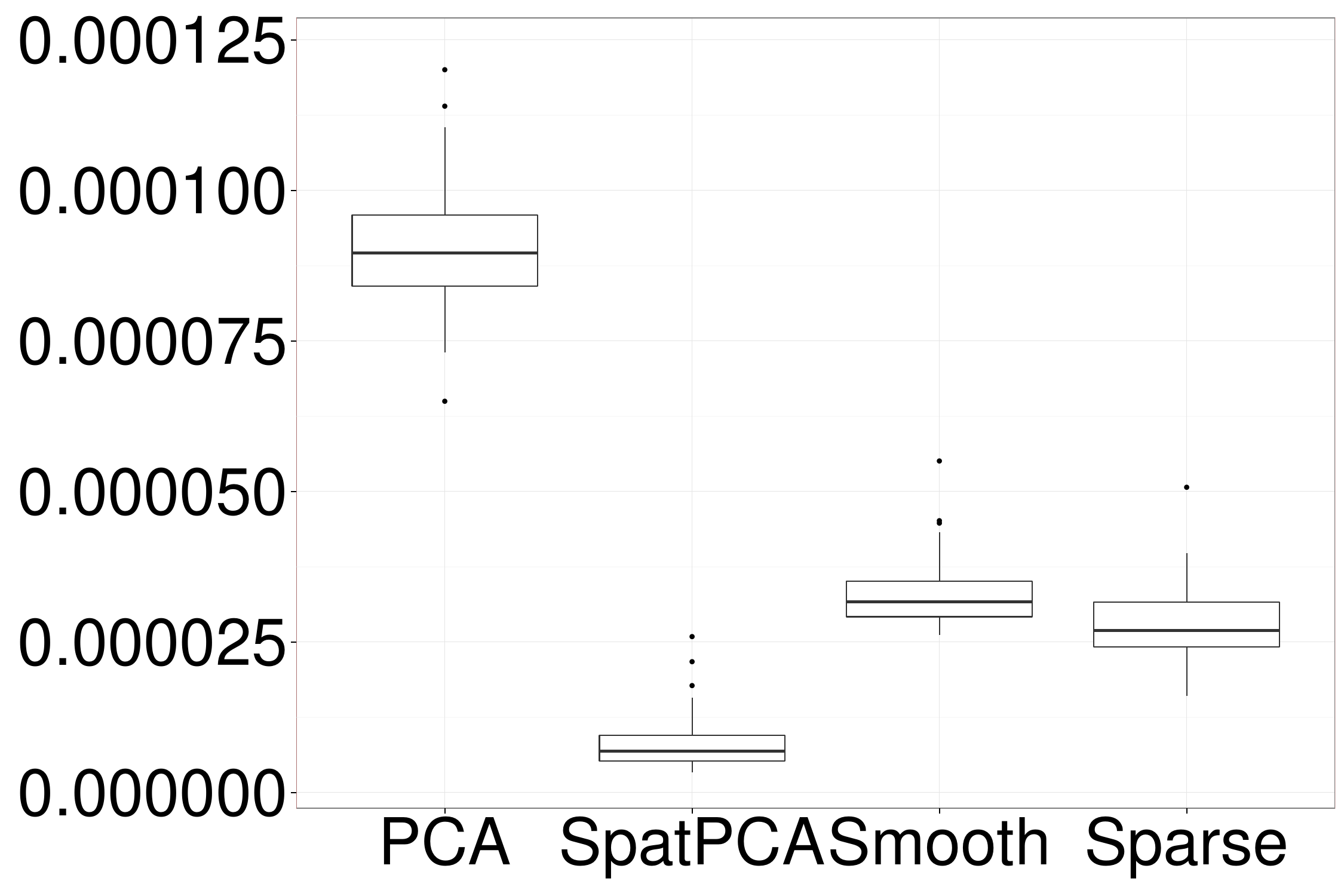}&
			\includegraphics[scale=0.12]{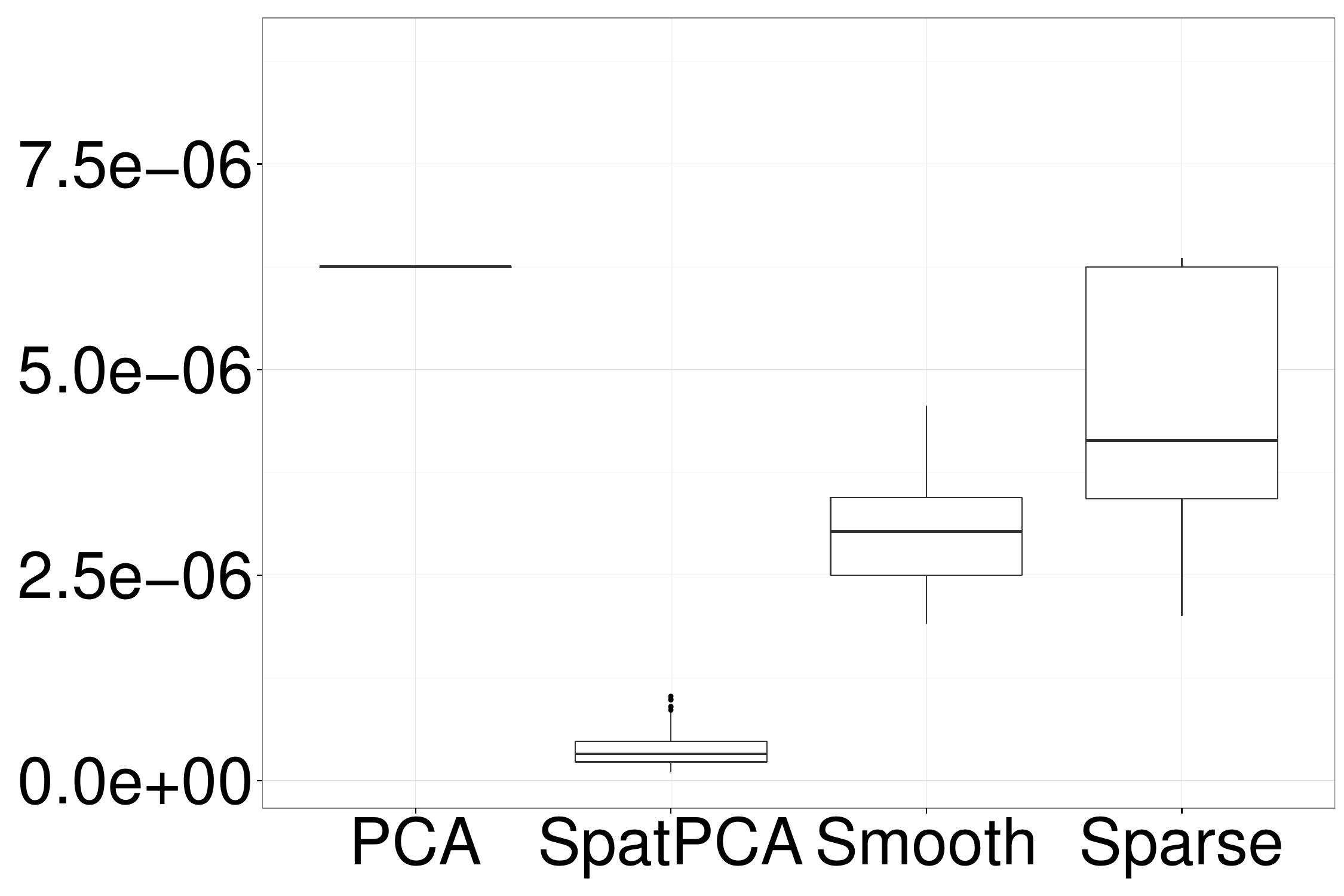}&
			\includegraphics[scale=0.12]{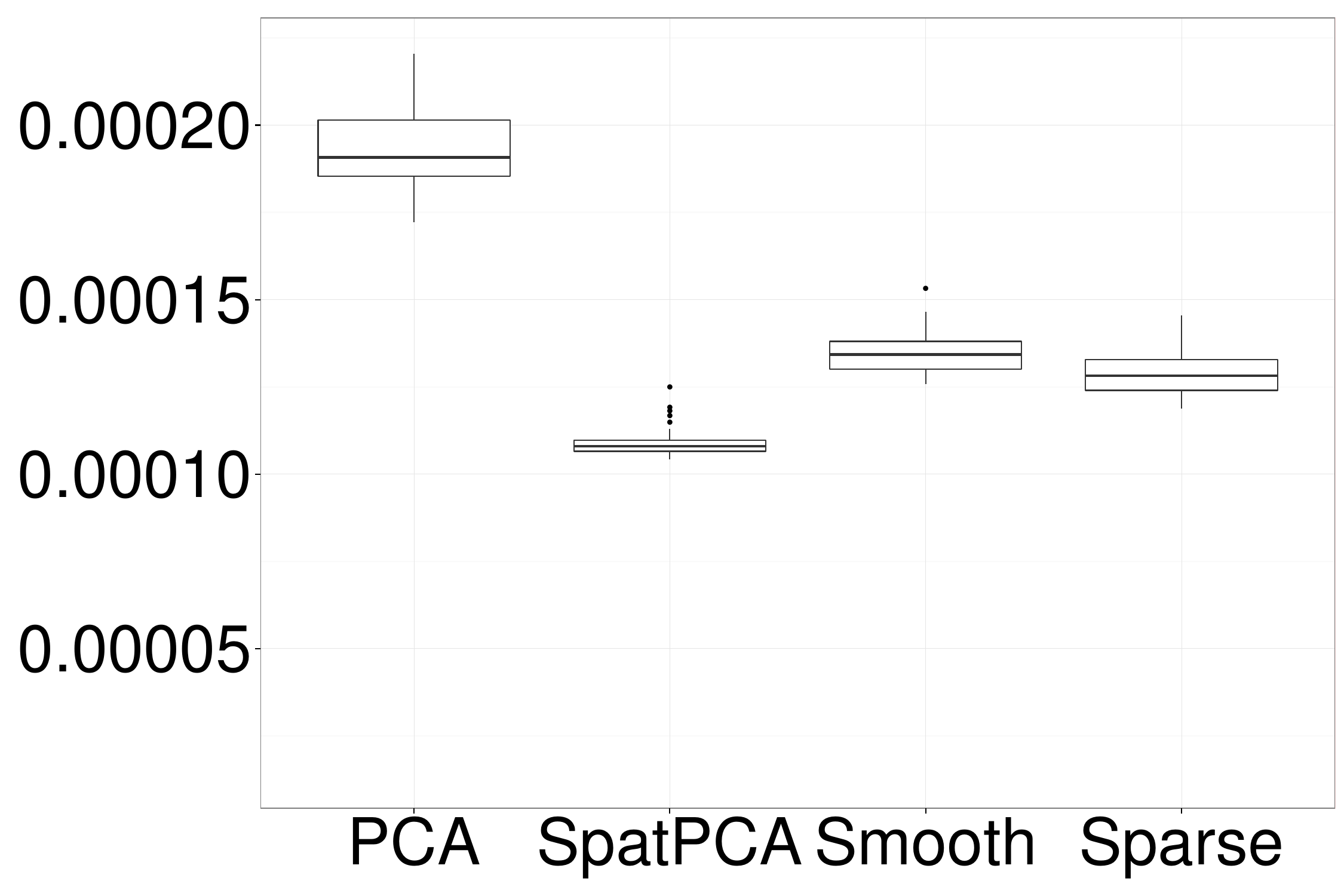}\\
			{$(\lambda_1,\lambda_2)=(9,0)$, $K=2$}&{$(\lambda_1,\lambda_2)=(1,0)$, $K=2$}&{$(\lambda_1,\lambda_2)=(9,4)$, $K=2$}\\
			\includegraphics[scale=0.12]{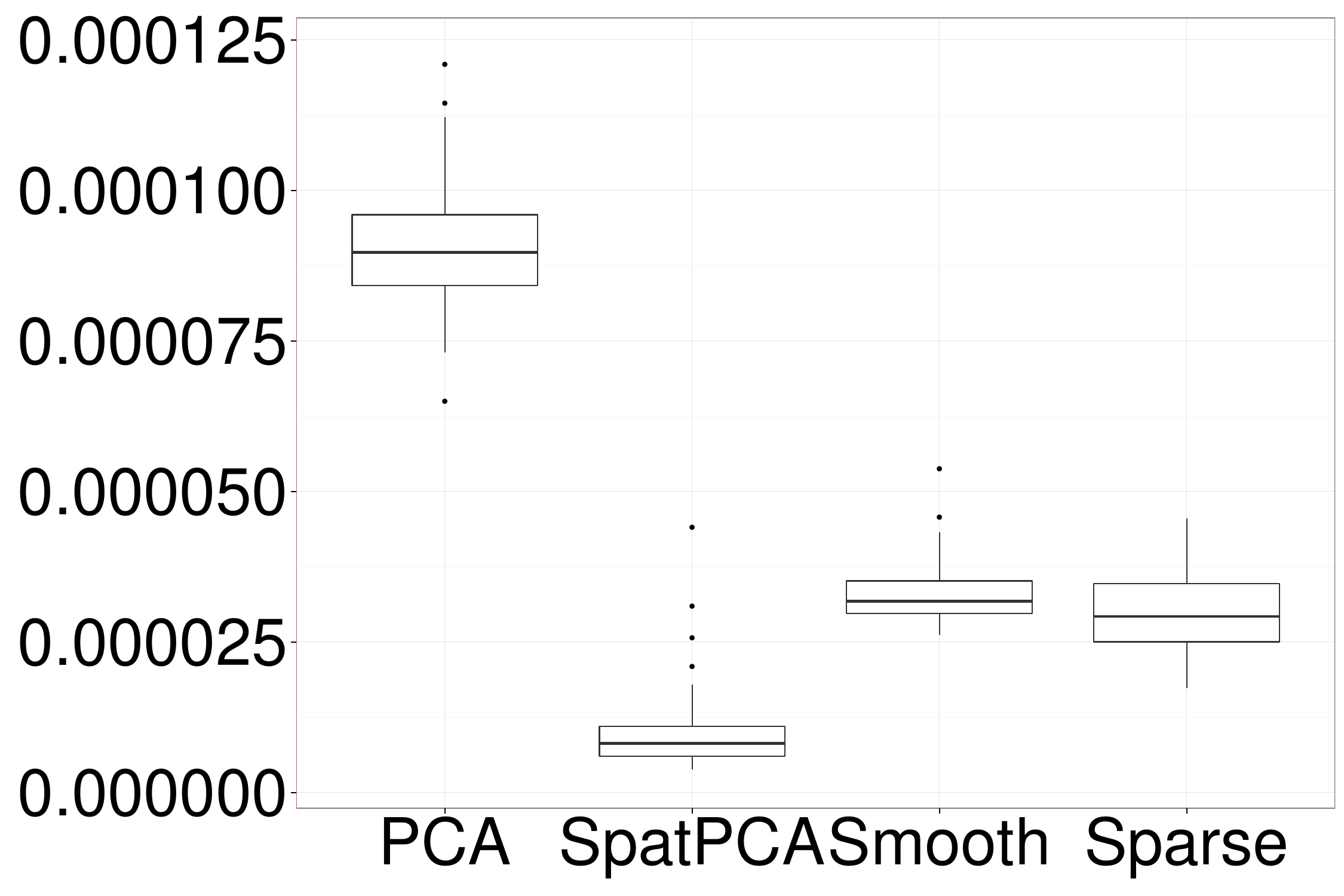}&
			\includegraphics[scale=0.12]{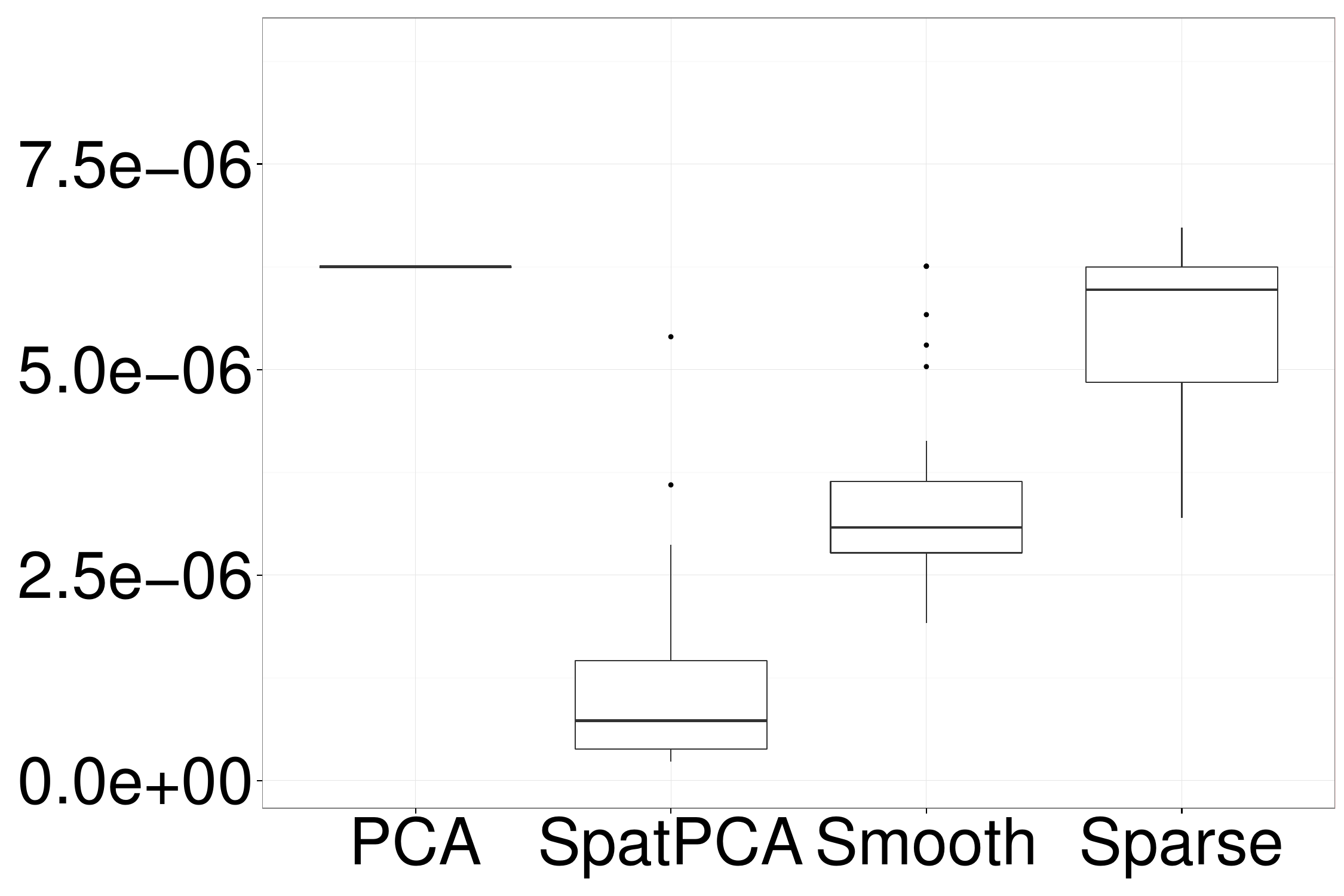}&
			\includegraphics[scale=0.12]{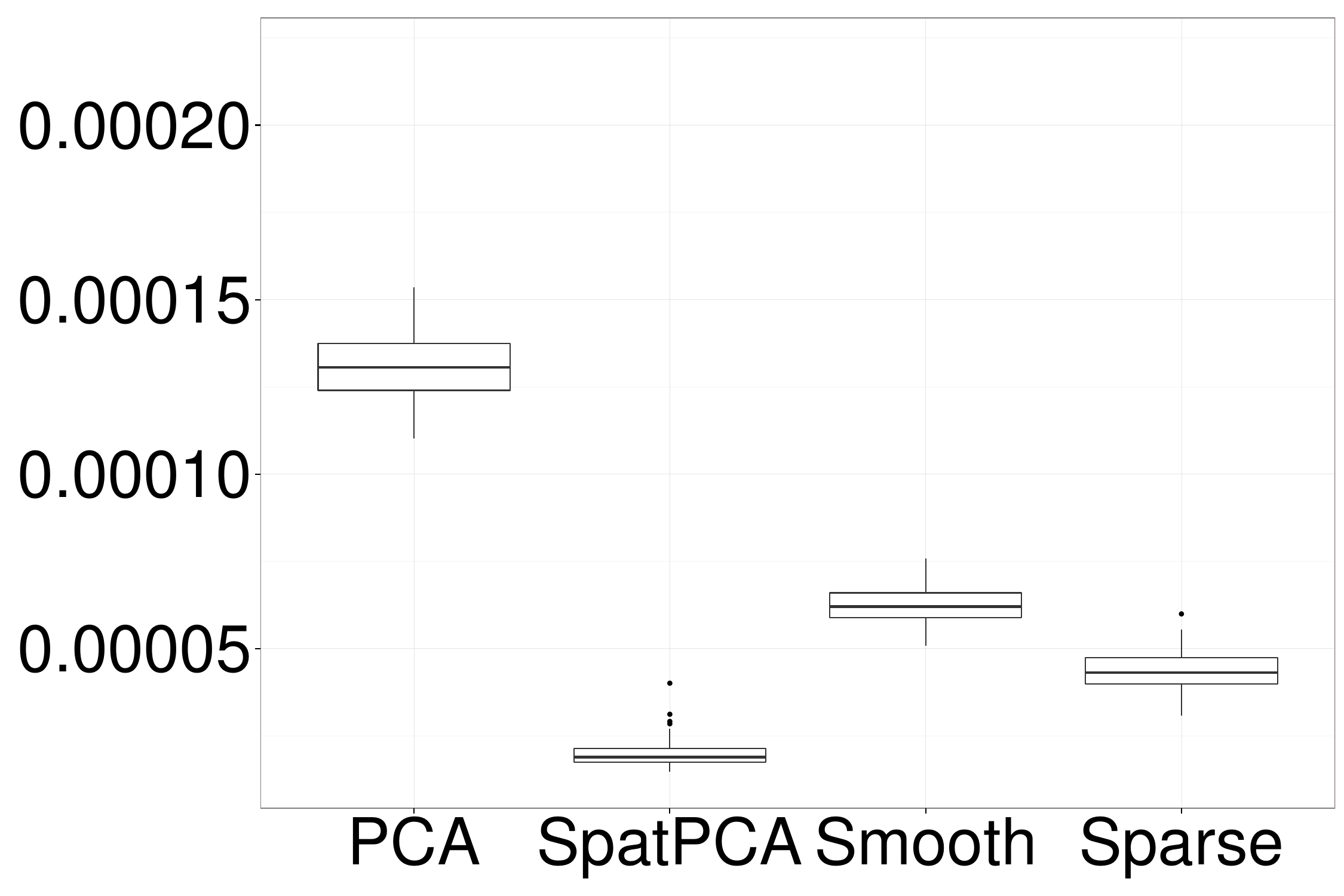}\\
			{$(\lambda_1,\lambda_2)=(9,0)$, $K=5$}&{$(\lambda_1,\lambda_2)=(1,0)$, $K=5$}&{$(\lambda_1,\lambda_2)=(9,4)$, $K=5$}\\
			\includegraphics[scale=0.12]{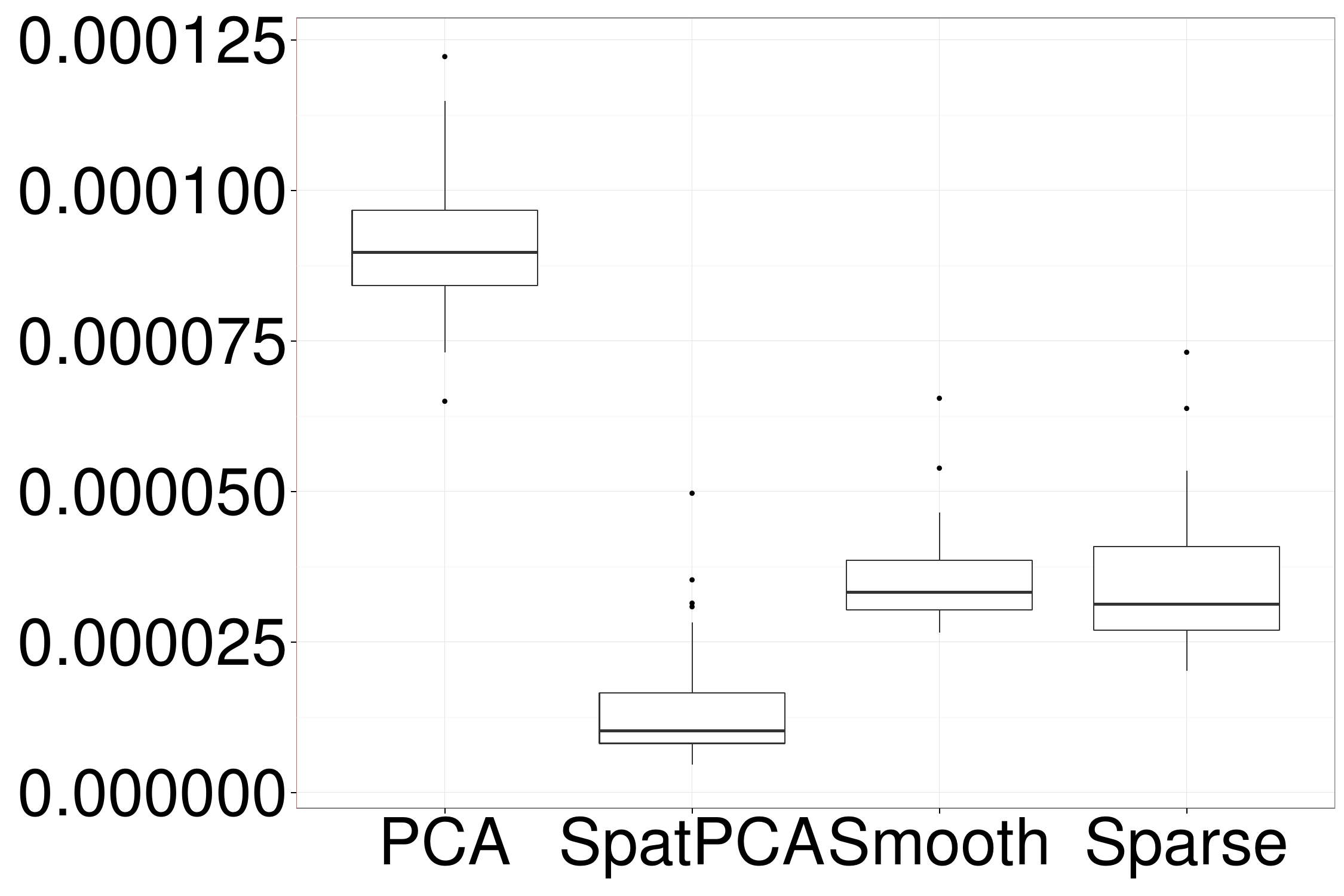}\hspace{4pt}&
			\includegraphics[scale=0.12]{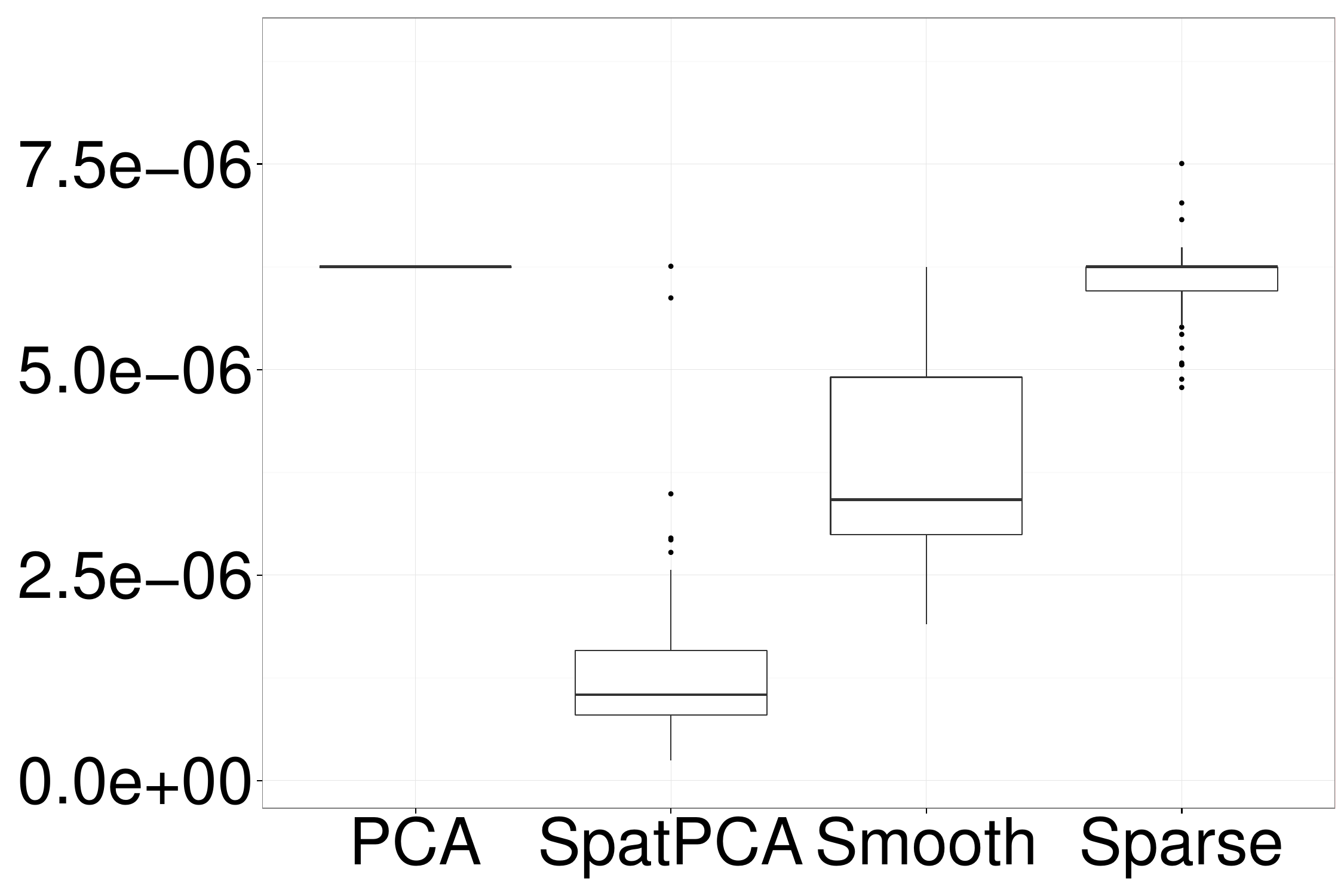}\hspace{4pt}&
			\includegraphics[scale=0.12]{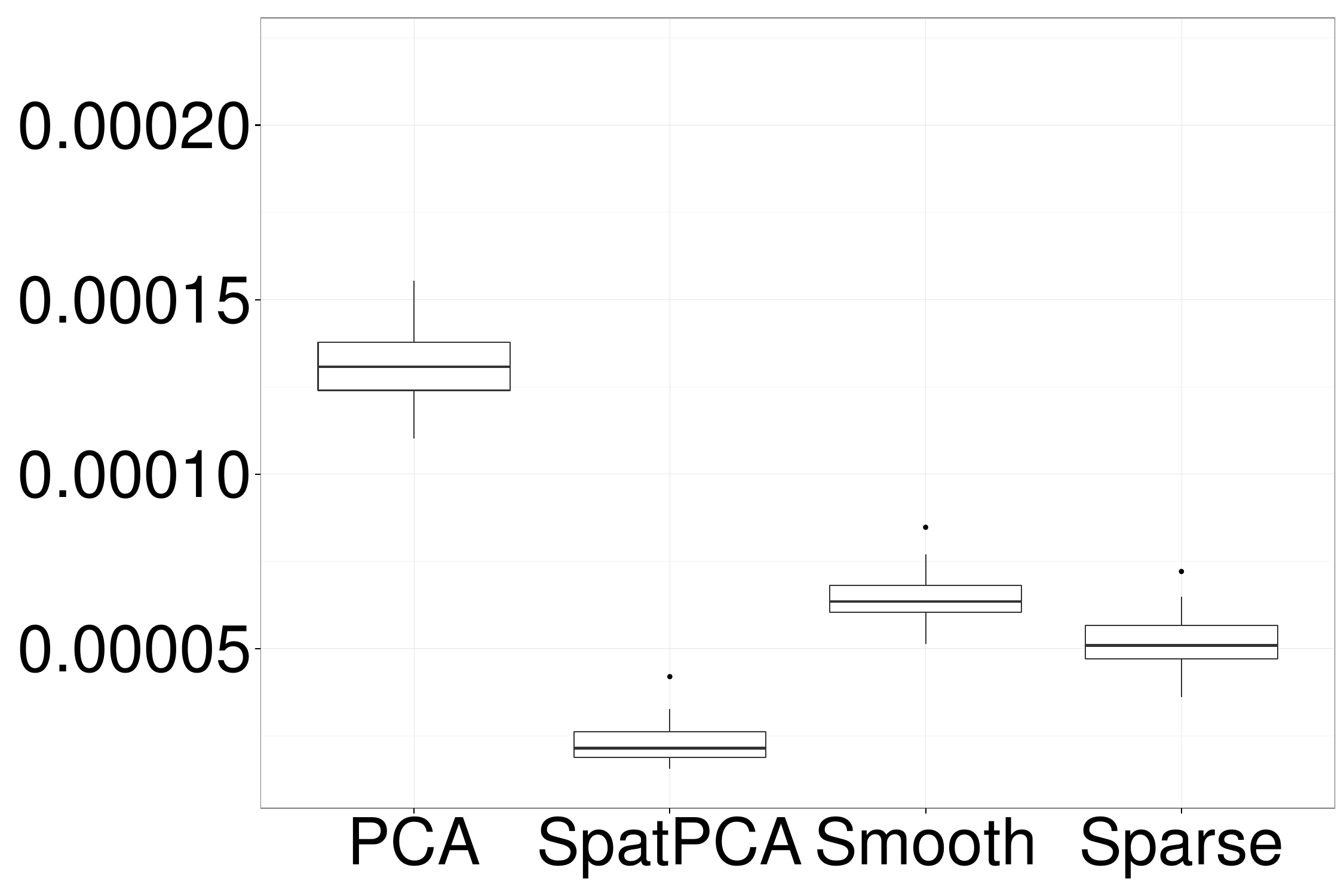}\hspace{4pt}\\
			{$(\lambda_1,\lambda_2)=(9,0)$, $K=\hat{K}$}&{$(\lambda_1,\lambda_2)=(1,0)$, $K=\hat{K}$}&{$(\lambda_1,\lambda_2)=(9,4)$, $K=\hat{K}$}\\
			\includegraphics[scale=0.12]{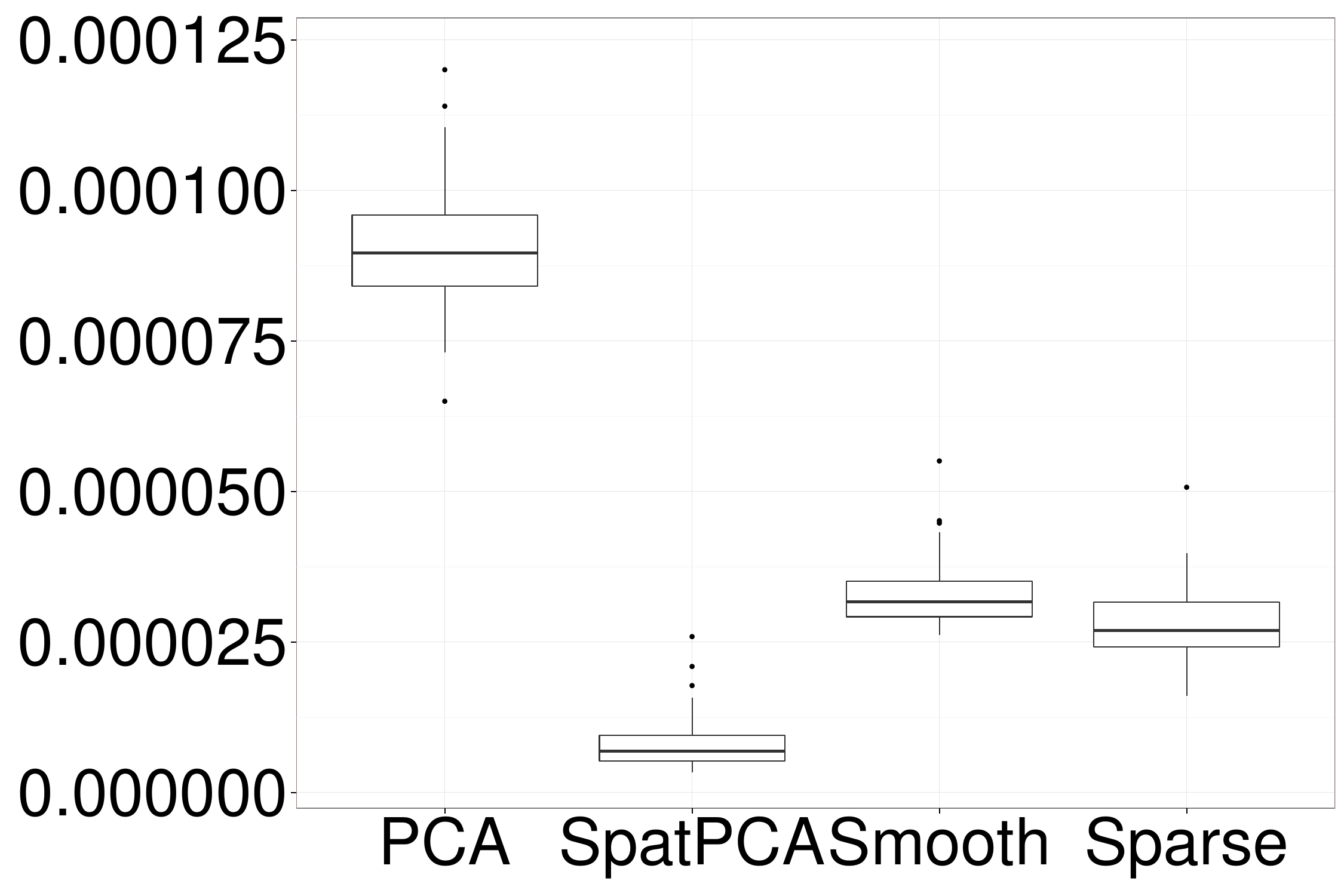}\hspace{4pt}&
			\includegraphics[scale=0.12]{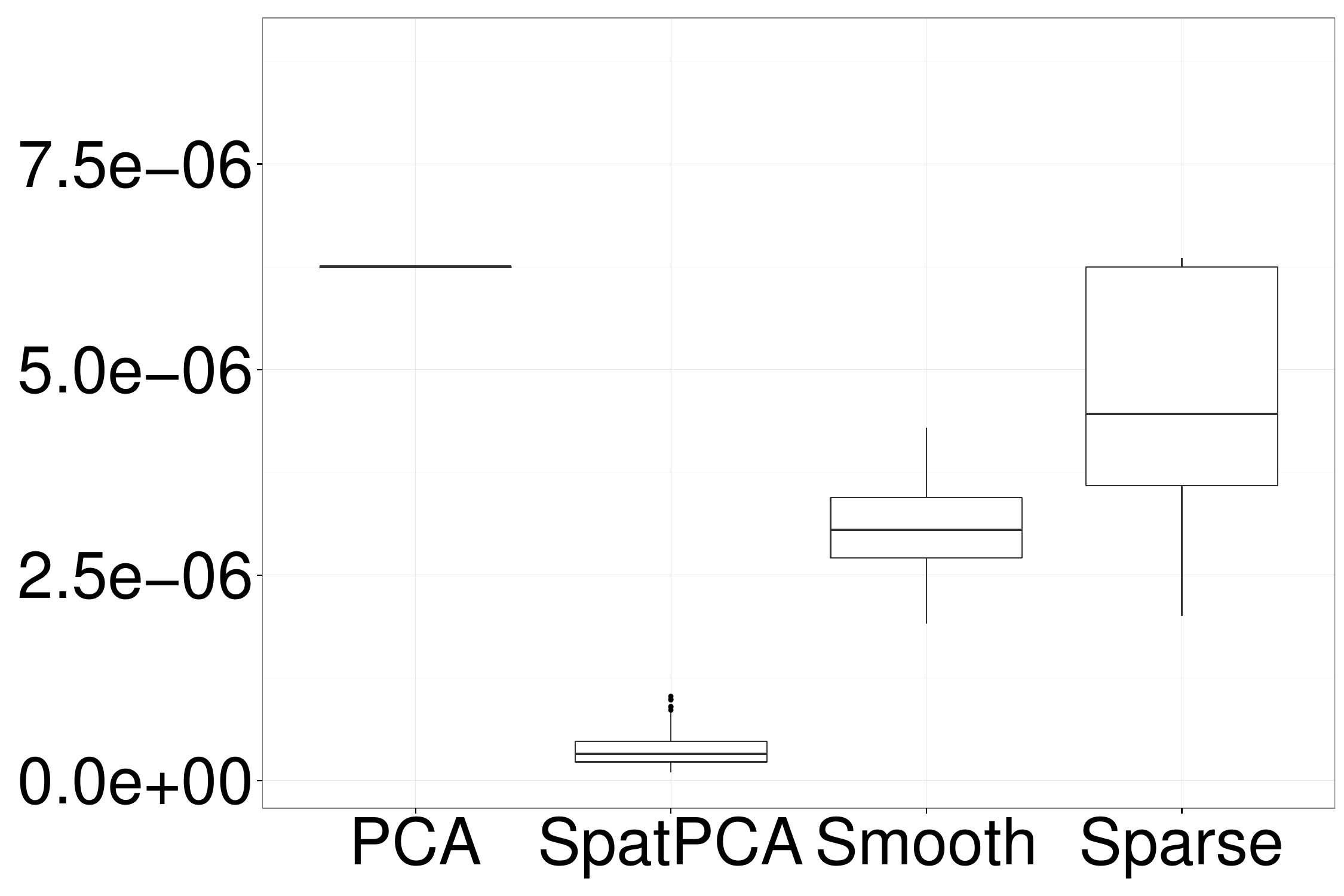}\hspace{4pt}&
			\includegraphics[scale=0.12]{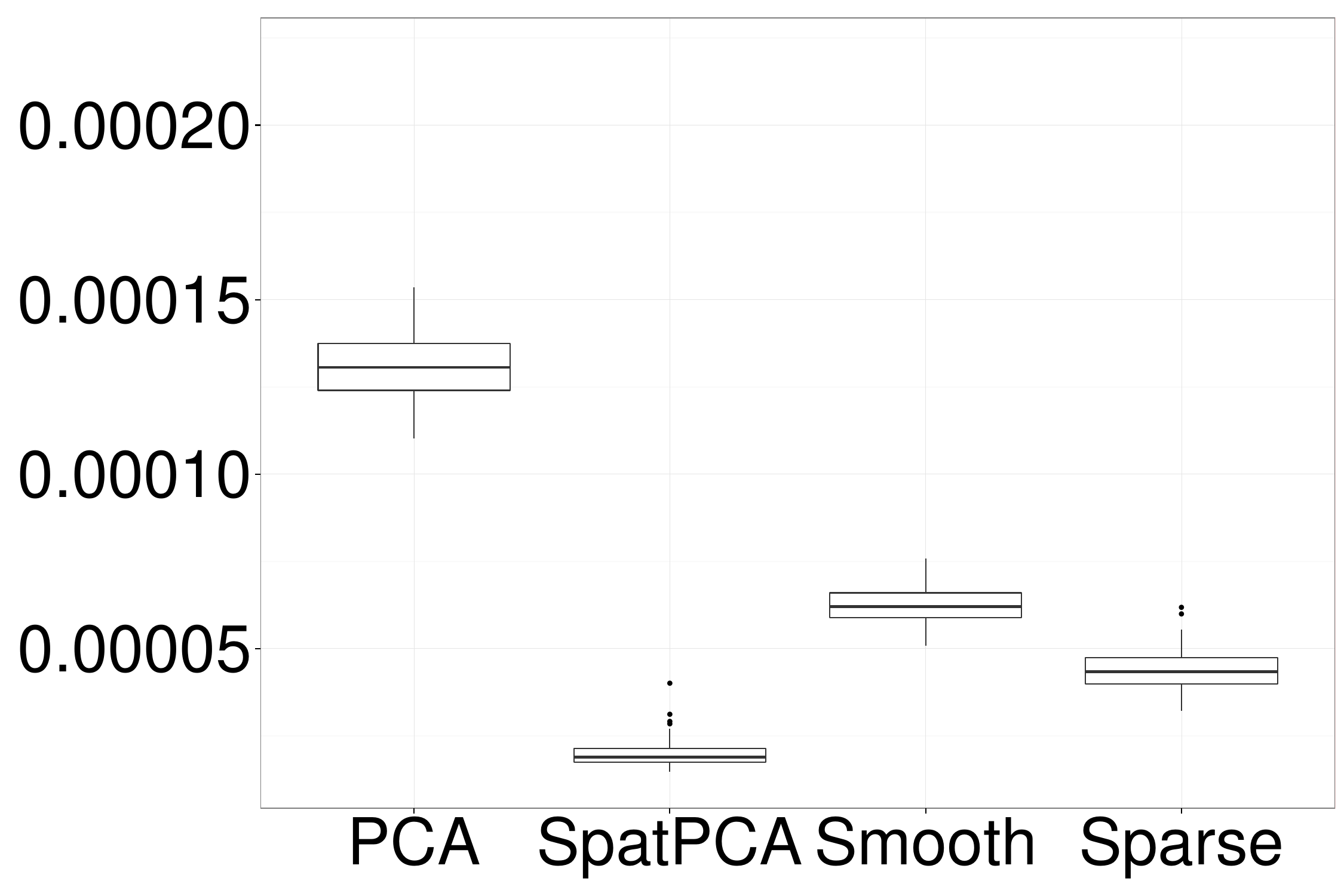}\hspace{4pt}
		\end{tabular}
	}
 \caption{{Boxplots of average squared estimation errors of (\ref{eq:loss2_sim}) for various methods in the two-dimensional simulation experiment of
Section \ref{sec:2d-1} based on 50 simulation replicates.}}
\label{fig:box_d2}
\end{figure}

{Figure~\ref{fig:est_d2} shows the estimates of $\phi_1(\cdot)$ and $\phi_2(\cdot)$ obtained from the four methods
for various cases based on a randomly generated dataset.
The performance of the four methods in terms of the loss functions \eqref{eq:loss_sim} and \eqref{eq:loss2_sim} is summarized in
Figure~\ref{fig:box_d2_loss1} and Figure~\ref{fig:box_d2}, respectively, based on 50 simulation replicates.
Similarly to the one-dimensional examples, SpatPCA performs significantly better than all the other methods in all cases.
For $(\lambda_1,\lambda_2)=(9,0)$, the average computation time for SpatPCA
(including selection of $\lambda_1$ and $\lambda_2$ using 5-fold CV) with $K=1,2,5$ are $3.105$, $4.242$ and $16.160$
seconds, respectively, using the R package ``SpatPCA" implemented on an iMac PC with a 3.2GHz Intel Core i5 CPU and a 64GB RAM.
While SpatPCA is slower than PCA (requiring only $0.267$ seconds), it is reasonably fast and provides much improved results. }

\subsection{An Application to a Sea Surface Temperature Dataset}
\label{sec:application}

{Since the proposed SpatPCA works better when both smoothness and sparseness penalties are involved according to the simulation experiments
in Sections \ref{sec:ex1} and \ref{sec:2d-1},
we applied the proposed SpatPCA with both penalty terms to a sea surface temperature (SST) dataset observed over a region in the Indian Ocean,
and only compared it with PCA.
The data are monthly averages of SST obtained from the Met Office Marine Data Bank
(available at \url{http://www.metoffice.gov.uk/hadobs/hadisst/)} on $1$ degree latitude by 1 degree longitude ($1 ^{\circ}\times 1 ^{\circ}$)
equiangular grid cells
from January 2001 to December 2010 in the region between latitudes $20^\circ N$ and $20^\circ S$ and between longitudes $39^\circ E$ and $120^\circ E$.
Out of $40\times 81=3,240$ grid cells, there are $460$ cells on the land where no data are available.
Hence the data we used are observed at $p=2,780$ cells and $120$ time points.
We first detrended the SST data by subtracting the SST for a given cell and a given month by the average SST for that cell and that month over the whole period.
We decomposed the data into two parts with one part consisting of $60$ time points of $\{1,3,\dots,119\}$ for training data,
and the other part, consisting of $60$ time points of even numbers, for validation purpose.}

{We applied SpatPCA on the training data with $K$ selected by $\hat{K}$ of $(\ref{eq:khat})$.
Similar to the two-step method described in Section \ref{sec:ex1}, we selected among 11 values of $\tau_1$ (including $0$, and the other 10 values from $10^3$
to $10^8$ equally spaced on the log scale) and 31 values of $\tau_2$ (including $0$, and the other $30$ values from $1$ to $10^3$
equally spaced on the log scale) using 5-fold CV of $\eqref{eq:cv}$.
For both PCA and SpatPCA, we applied $5$-fold CV of $\eqref{eq:cv.gamma}$ to select among
11 values of $\gamma$ (including $0$ and other 10 values from $\hat{d}_1/10^3$ to $\hat{d}_1$ equally spaced on the log scale), where $\hat{d}_1$ is the largest eigenvalue of
$\hat{\bm{\Phi}}'\bm{S}\hat{\bm{\Phi}}$.}


\begin{figure}\centering
{$\hat{\phi}_1(\cdot)$ from PCA\hspace{120pt}$\hat{\phi}_1(\cdot)$ from SpatPCA}
\includegraphics[scale=0.2]{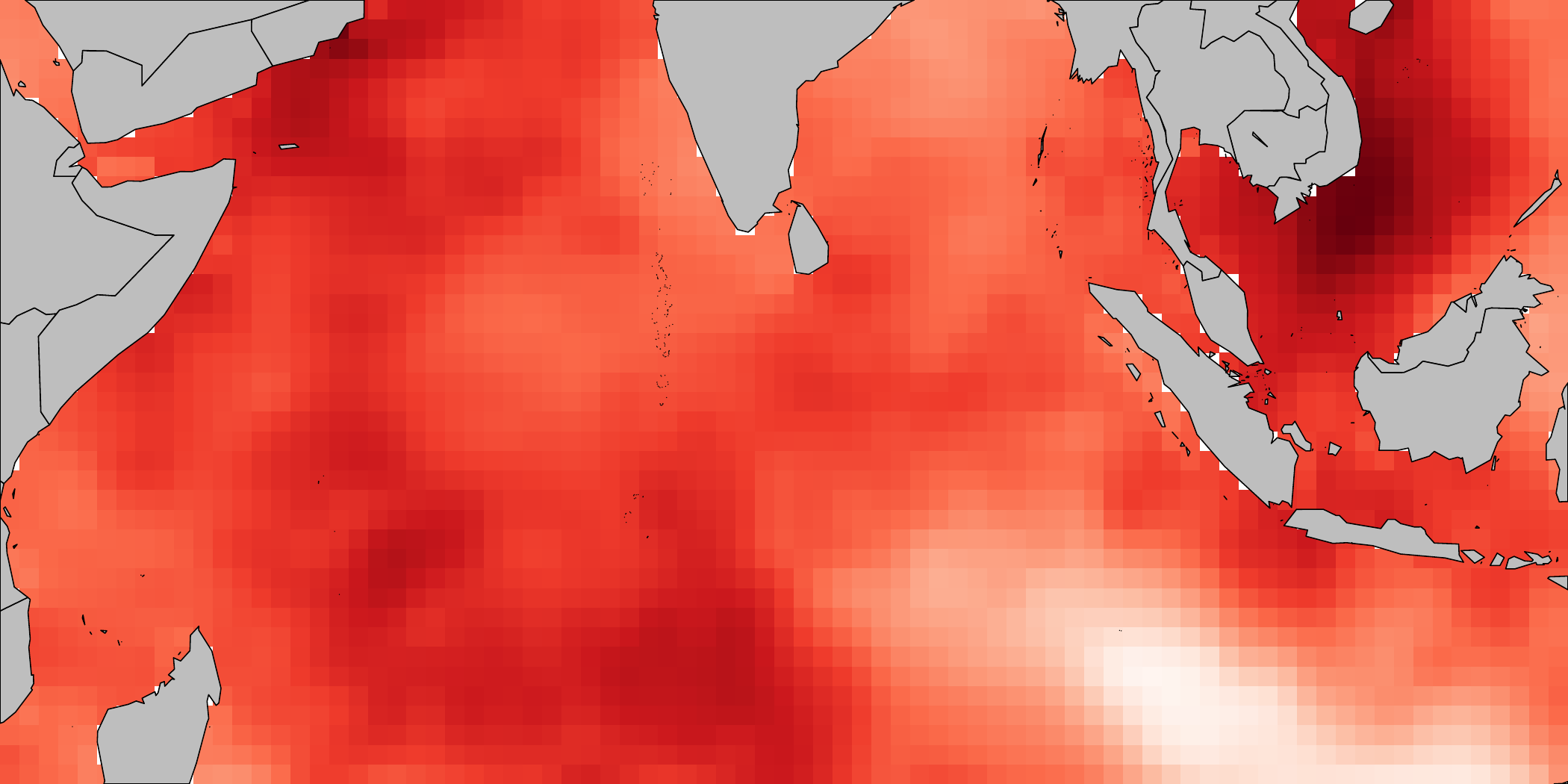}
\includegraphics[scale=0.2]{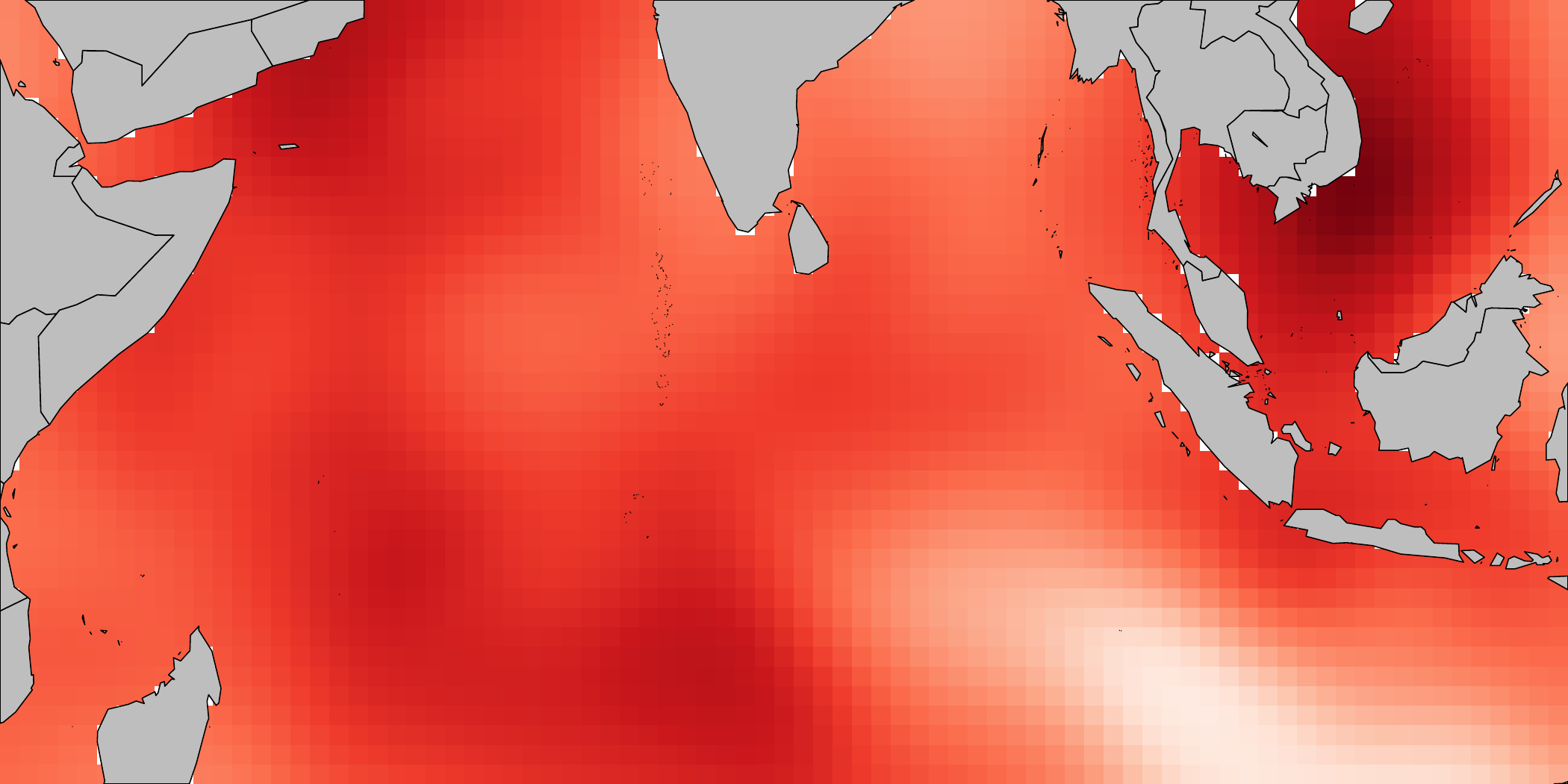}\\
\includegraphics[scale=0.2]{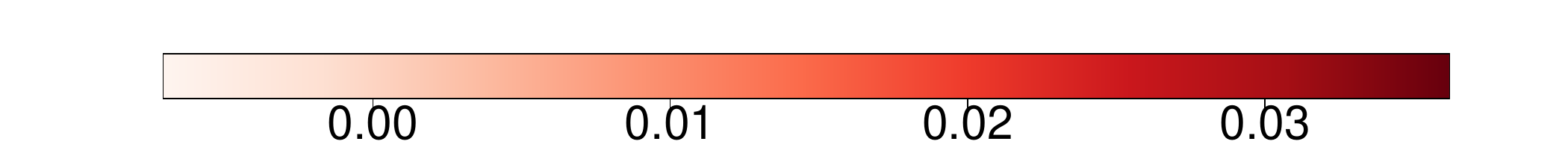}\\
{$\hat{\phi}_2(\cdot)$ from PCA\hspace{120pt}$\hat{\phi}_2(\cdot)$ from SpatPCA}
\includegraphics[scale=0.2]{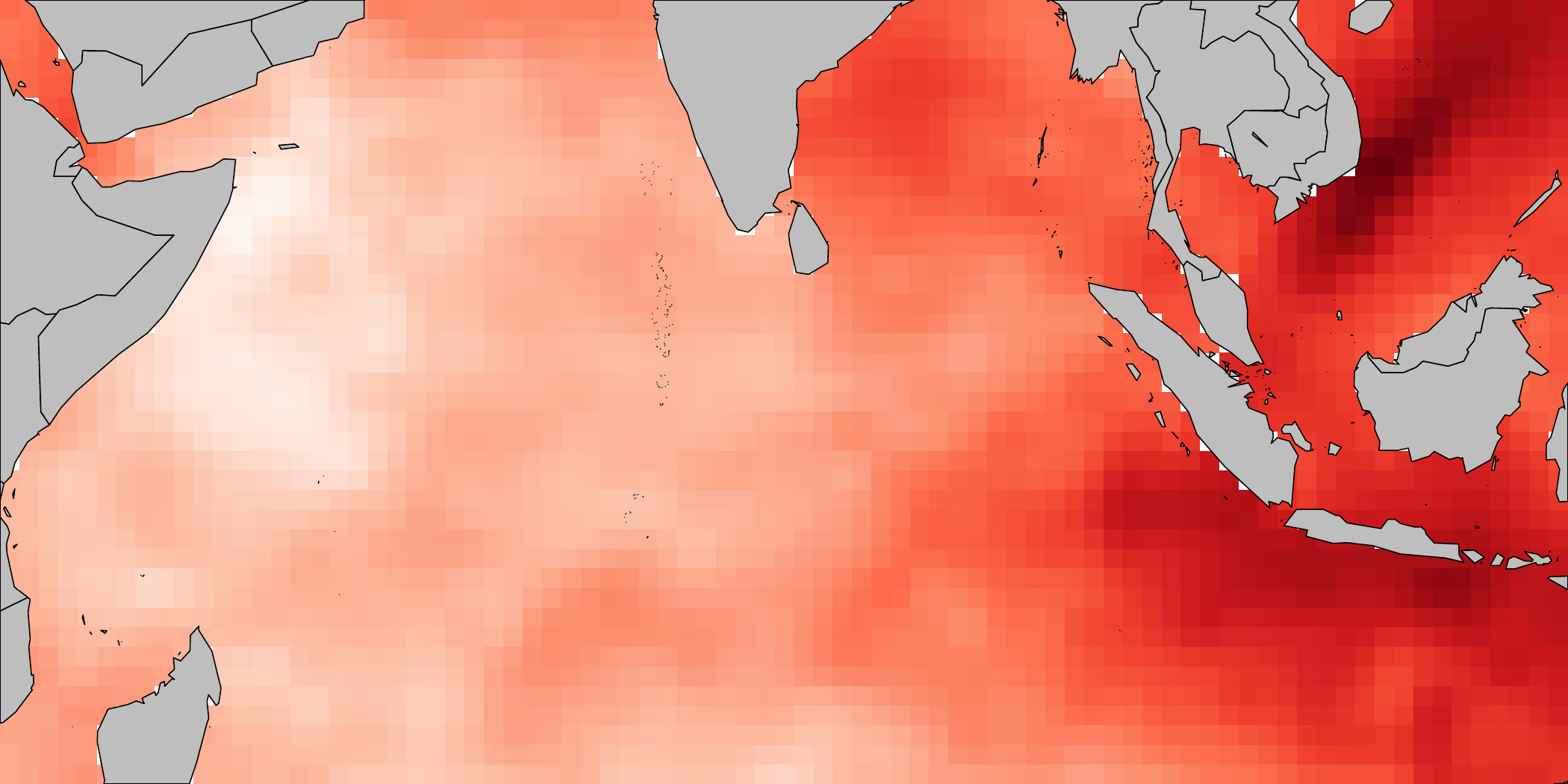}
\includegraphics[scale=0.2]{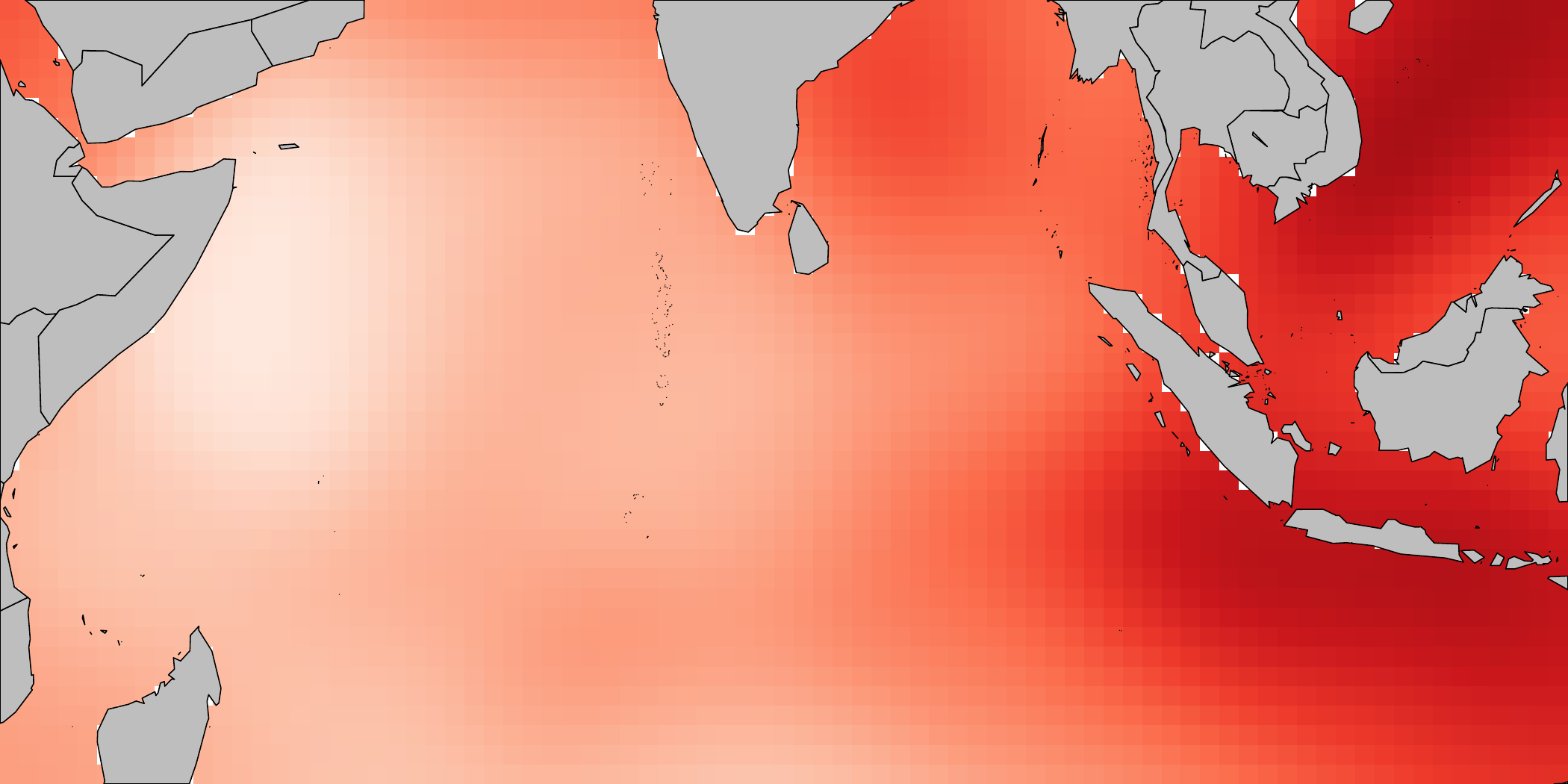}\\
\includegraphics[scale=0.2]{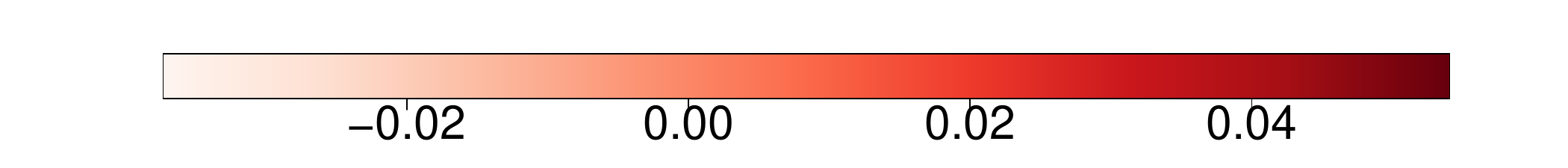}
\caption{{Estimated eigenimages obtained form PCA and SpatPCA over a region in the Indian Ocean,} where the gray regions correspond to the land.}
\label{fig:sst_real}
\end{figure}

The first {two dominant} patterns estimated from PCA and SpatPCA are shown in Figure~\ref{fig:sst_real}.
Both methods identify similar patterns with the ones estimated from SpatPCA
being a bit smoother than those estimated from PCA. The first pattern
is a basin-wide mode and the second one corresponds to the east-west dipole mode (\citet{sst}).

\begin{figure}\centering
{
\includegraphics[height=5cm, width=6 cm]{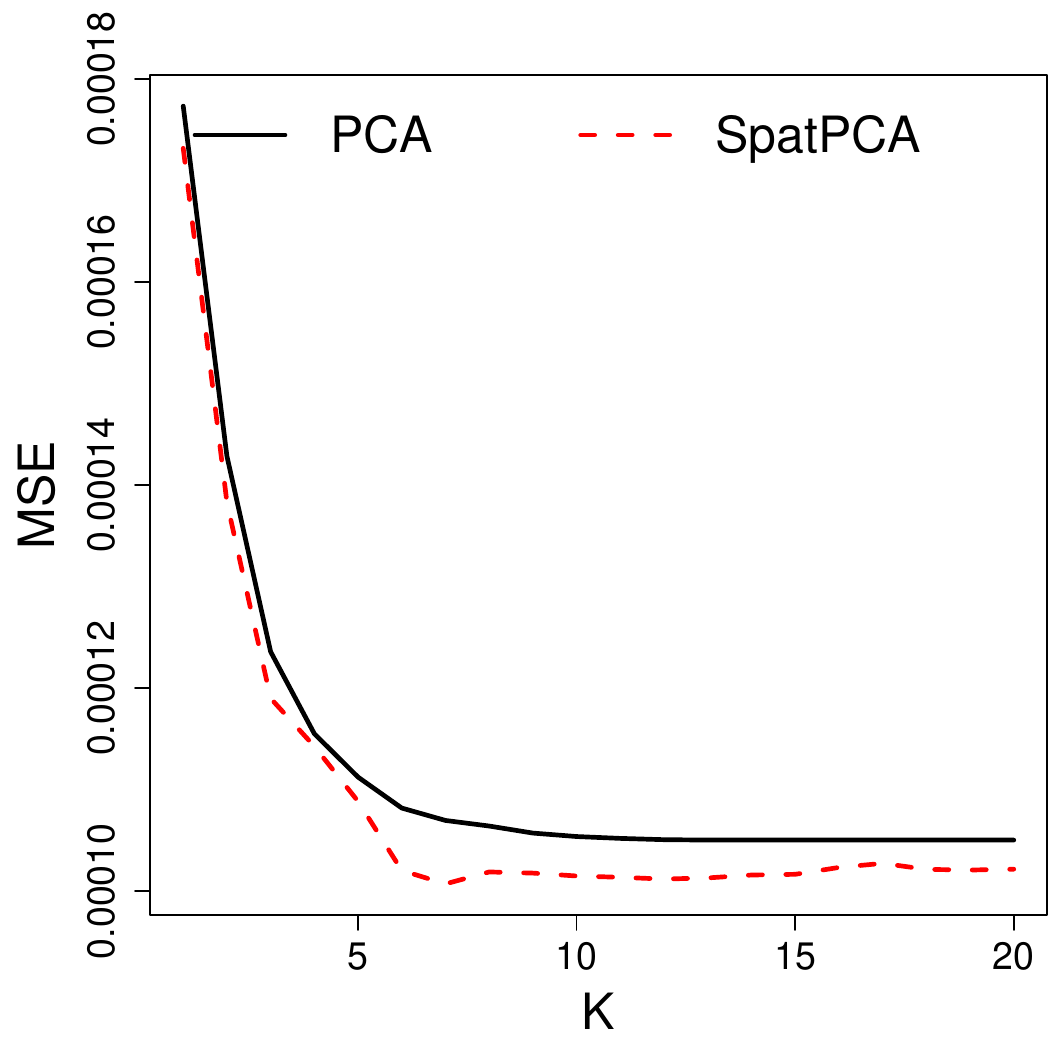}}
\caption{{Mean squared errors of covariance matrix estimation with respect to $K$ for PCA and SpatPCA.}}
\label{fig:sse}
\end{figure}

{We used the validation data to evaluate the performance between PCA and SpatPCA in terms of the mean squared error (MSE),
$\|\hat{\bm{\Sigma}}-\bm{S}_v\|_F^2/p^2$, where $\hat{\bm{\Sigma}}$ is a generic estimate of $\mathrm{var}(\bm{Y})$ based on the training data, and
$\bm{S}_v$ is the sample covariance matrix based on the validation data.
The resulting MSE for PCA is $1.05\times 10^{-4}$, which is slightly larger than $1.02\times 10^{-4}$ for SpatPCA.
Figure \ref{fig:sse} shows the MSEs with respect to various $K$ values for both PCA and SpatPCA. The results indicate that SpatPCA is not sensitive to the choice of $K$
as long as $K$ is sufficiently large. Our choice of $\hat{K}=6$ for SpatPCA based on $(\ref{eq:khat})$ appears to be effective, and is smaller than $\hat{K}=15$ for PCA.}

\subsection{Two-Dimensional Experiment II}
\label{sec:2d-2}

{To reflect a real-world situation, we generated data by mimicking the SST dataset analyzed in the previous subsection,
except we applied a larger noise variance.
Specifically, we generated data according to (\ref{eq:measurement})
with $K=2$, $\bm{\xi}_i\sim N(\bm{0}, \mathrm{diag}(\lambda_1,\lambda_2))$, $\bm{\epsilon}_i\sim N(\bm{0},\bm{I})$, $n=60$,
and at the same $2,780$ locations from the SST dataset.
Here $\phi_1(\cdot)$ and $\phi_2(\cdot)$ are given by $\hat{\phi}_1(\cdot)$ and $\hat{\phi}_2(\cdot)$ (see Figure~\ref{fig:sst_real})
and $(\lambda_1,\lambda_2)=(91.3,16.1)$ estimated by SpatPCA in the previous subsection.}

\begin{figure}\centering
{$\hat{\phi}_1(\cdot)$ from PCA\hspace{132pt}$\hat{\phi}_1(\cdot)$ from SpatPCA}
\includegraphics[scale=0.2]{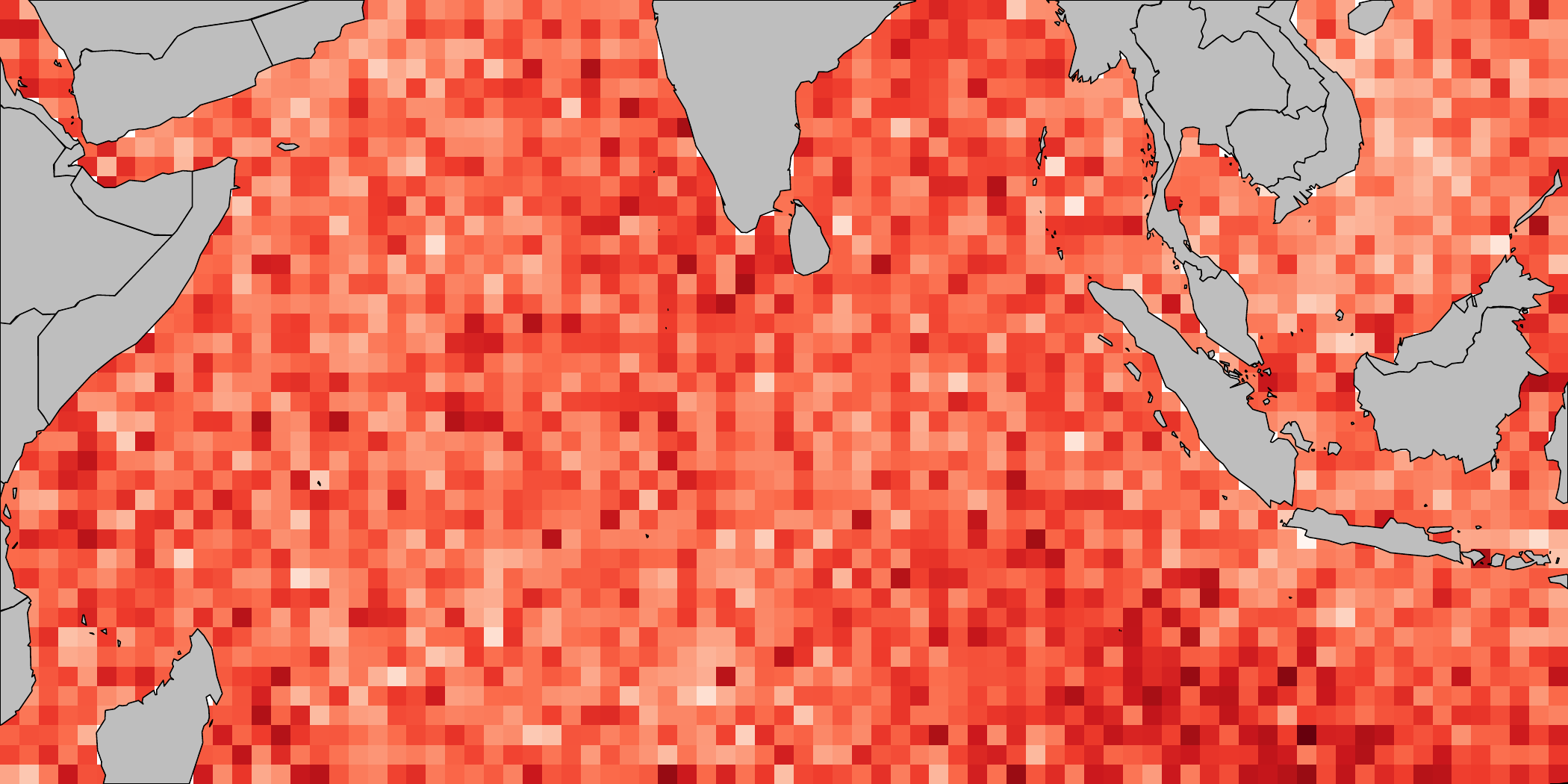}
\includegraphics[scale=0.2]{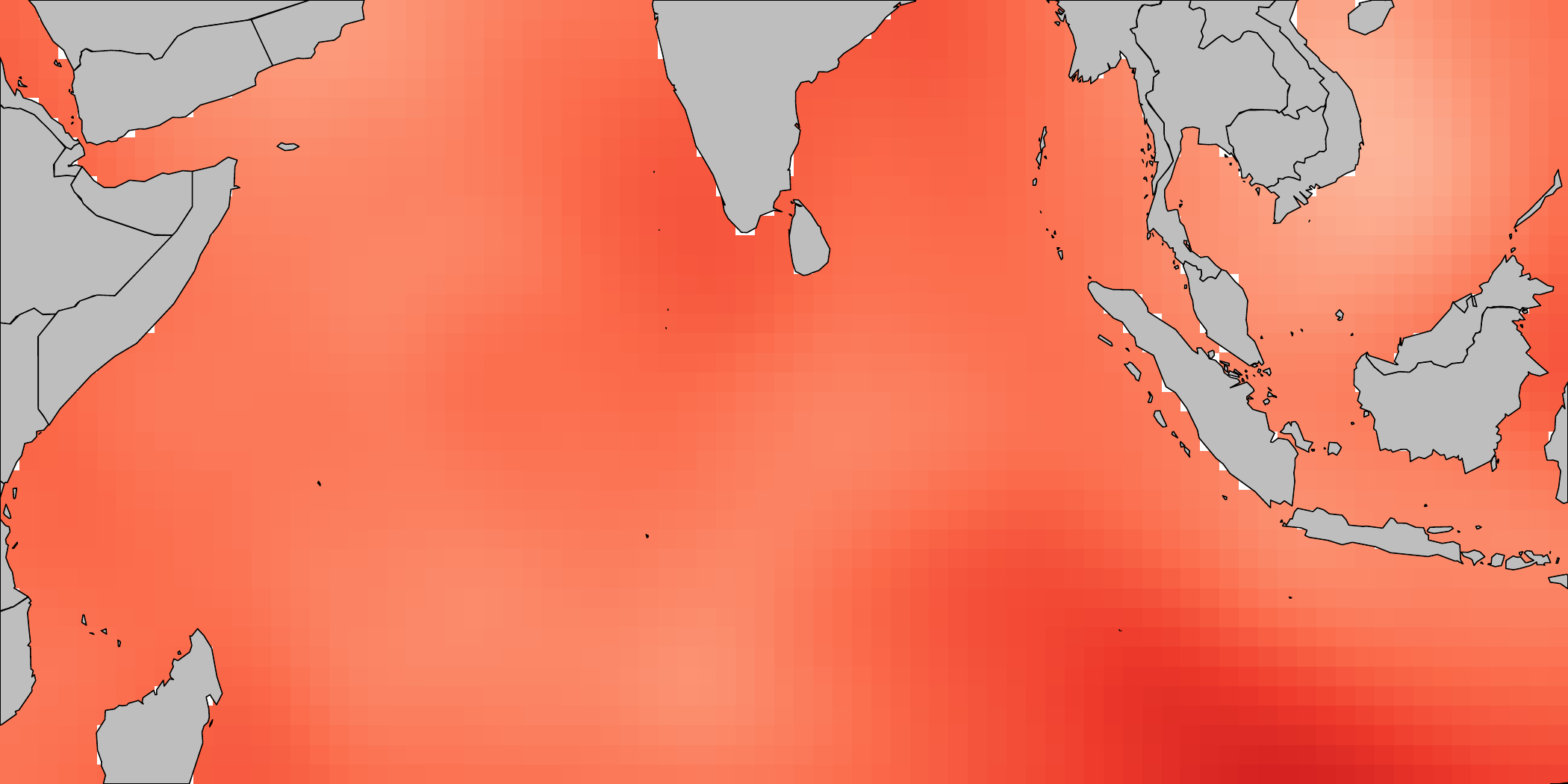}\\
\includegraphics[scale=0.2]{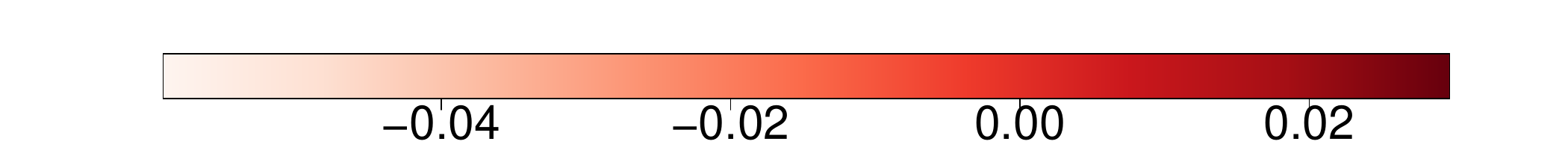}\\
{$\hat{\phi}_2(\cdot)$ from PCA\hspace{132pt}$\hat{\phi}_2(\cdot)$ from SpatPCA}
\includegraphics[scale=0.2]{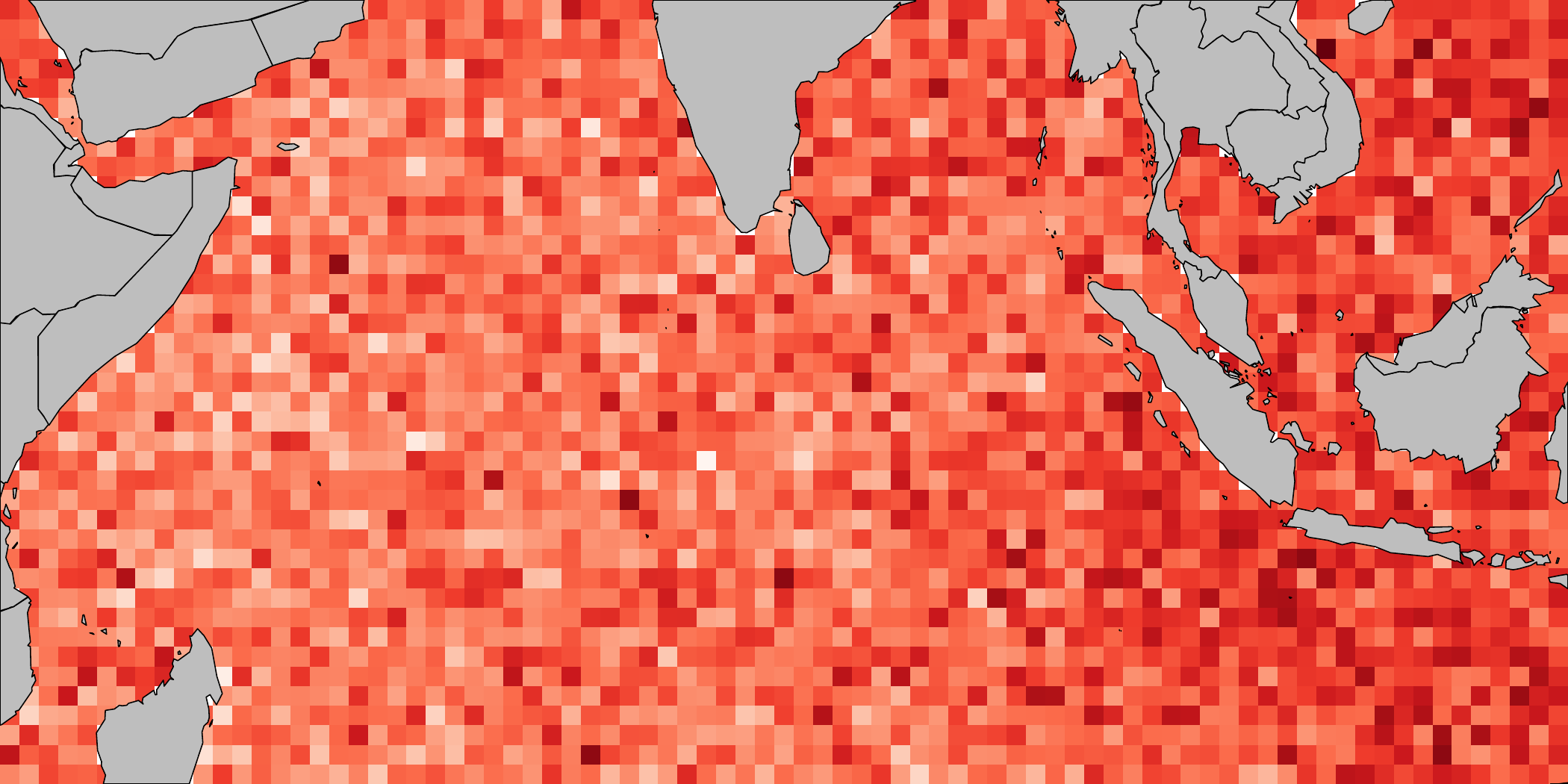}
\includegraphics[scale=0.2]{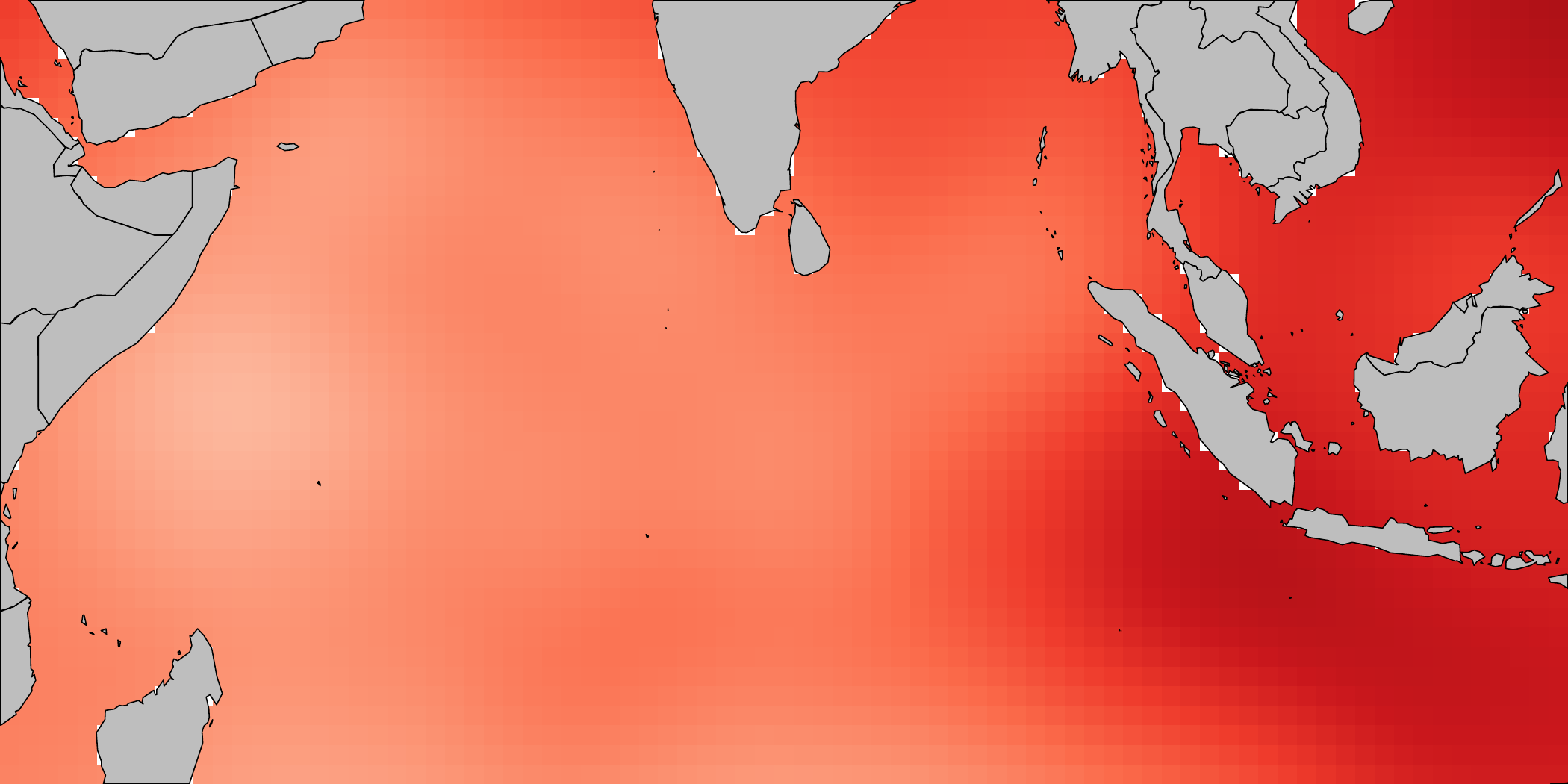}\\
\includegraphics[scale=0.2]{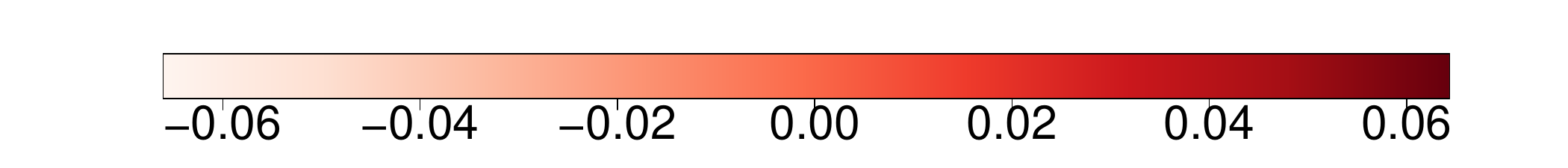}
\caption{Estimates of $\phi_1(\cdot)$ and $\phi_2(\cdot)$ {obtained from} PCA and SpatPCA in the two-dimensional experiment of Section \ref{sec:2d-2}
based on a randomly simulated dataset, where {the areas in gray are the land}.}
\label{fig:sst_realsim}
\end{figure}

\begin{figure}\centering
{
	\begin{tabular}{cc}
\includegraphics[scale=0.15]{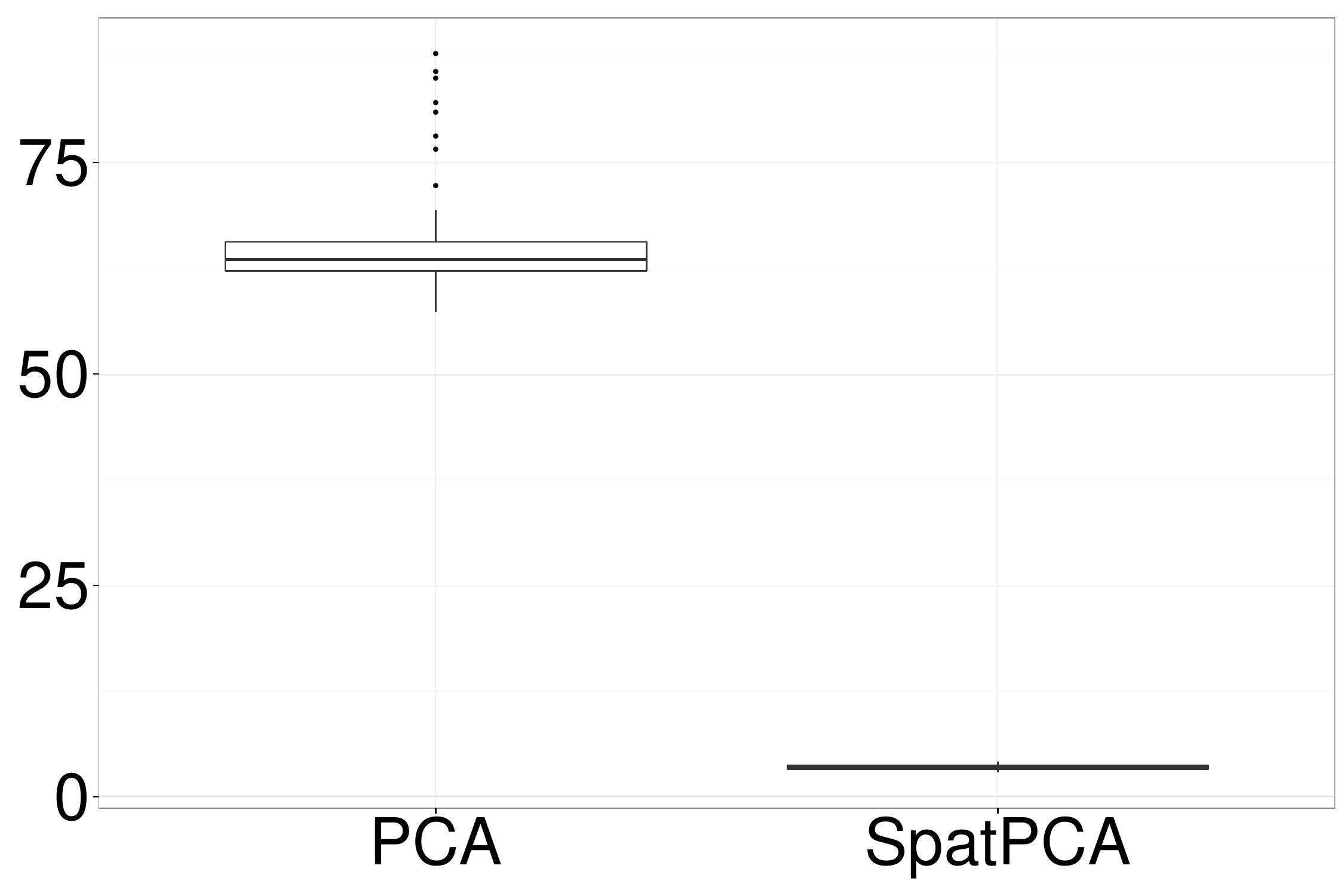}&
\includegraphics[scale=0.15]{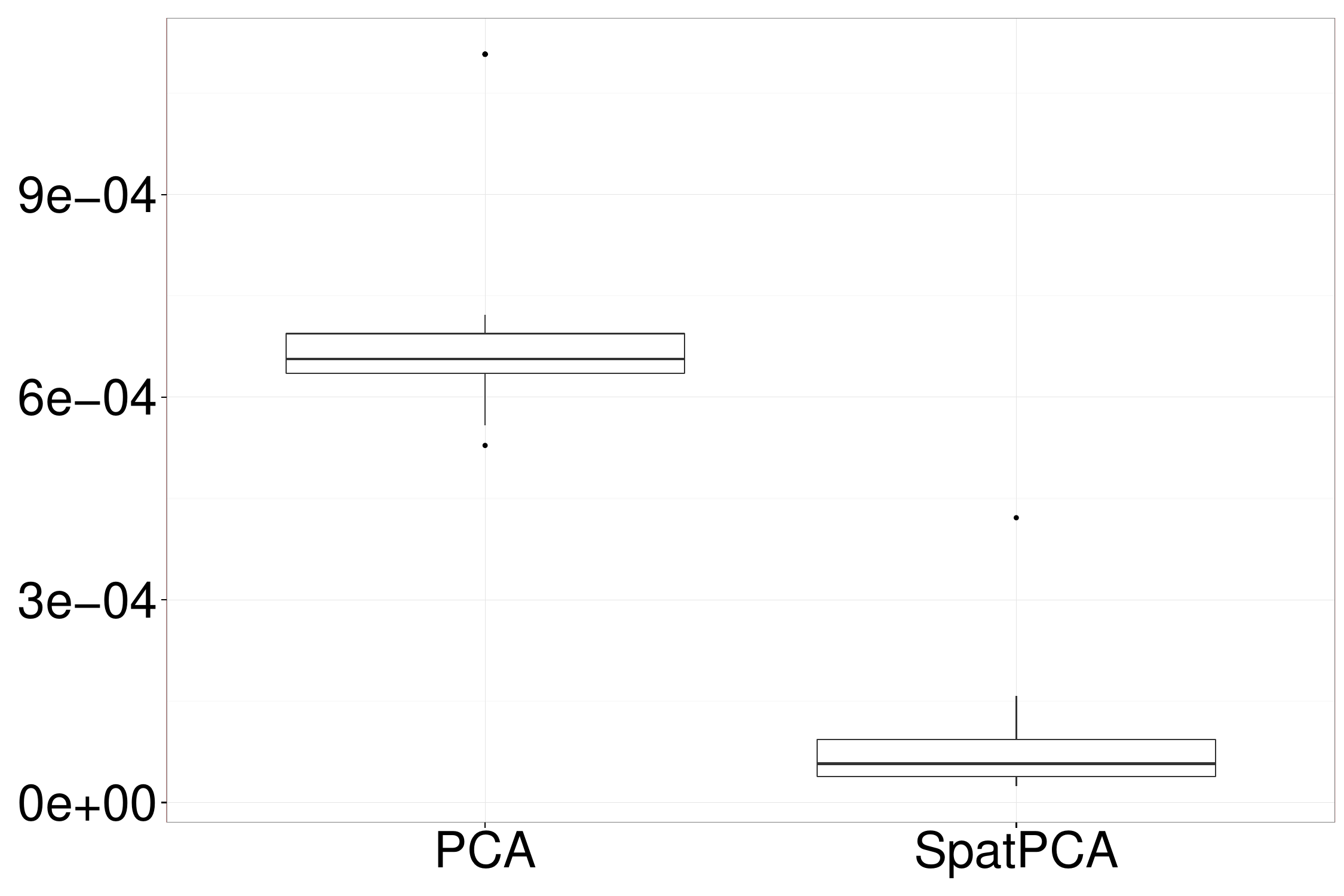}\\
{(a)}&{(b)}
\end{tabular}
}
\caption{{(a) Boxplots of loss function values for PCA and SpatPCA in the two-dimensional simulation experiment of
Section \ref{sec:2d-2} based on 50 simulation replicates: (a) Average squared prediction errors of (\ref{eq:loss_sim});
(b) Average squared estimation errors  of (\ref{eq:loss2_sim}). }}
\label{fig:loss2_realsim}
\end{figure}

{We applied the 5-fold CV of \eqref{eq:cv.gamma} and (\ref{eq:khat}) to select the tuning parameters $(\tau_1, \tau_2)$ and $K$
in the same way as in the previous subsection.}
Figure~\ref{fig:sst_realsim} shows the estimates of $\phi_1(\cdot)$ and $\phi_2(\cdot)$ for PCA and SpatPCA
based on a randomly generated dataset.
Because we consider a larger noise variance than those in the previous subsection,
the first two patterns estimated from PCA turn out to be very noisy.
In contrast, SpatPCA can still reconstruct the first two patterns very well with little noise. 
The results in terms of the loss functions of \eqref{eq:loss_sim} and \eqref{eq:loss2_sim} are summarized in Figure~\ref{fig:loss2_realsim}.
{Once again, SpatPCA outperforms PCA by a large margin.}

\section*{{Acknowledgements}}
{The authors are grateful to the associate editor and the two referees
for their insightful and constructive comments, which greatly improve the presentation of this paper.
This research was supported in part by ROC Ministry of Science and Technology grant MOST 103-2118-M-001-007-MY3.}
\ignore{
\section*{{Supplementary Material}}
{The supplementary contains Proof of Proposition~\ref{prop:sigma_Lambda_hat}.}
}
{
\section*{Appendix}
\label{sec: appendix}

\begin{proof}[Proof of Proposition~\ref{prop:sigma_Lambda_hat}]
First, we prove \eqref{eq:Lambda.hat}. From Corollary 1 of \citet{regularized_covariance}, the
minimizer of $h(\bm{\Lambda},\sigma^2)$ given $\sigma^2$ is
\begin{equation}
\hat{\bm{\Lambda}}(\sigma^2)=\hat{\bm{V}} \mathrm{diag}\big((\hat{d}_1-\sigma^2-\gamma)_+,\dots,(\hat{d}_{K}-\sigma^2-\gamma)_+\big)\hat{\bm{V}}'.
\label{eq:Lambda.hat2}
\end{equation}

\noindent Hence \eqref{eq:Lambda.hat} is obtained.

Next, we prove \eqref{eq:sigma.hat}. Rewrite the objective function of (\ref{eq:covariance.estimate}) as:
\begin{align}
	h(\bm{\Lambda}, \sigma^2)
=&~ \frac{1}{2}\|\hat{\bm{\Phi}}\hat{\bm{\Phi}}'\bm{S}\hat{\bm{\Phi}}\hat{\bm{\Phi}}' - \hat{\bm{\Phi}}\bm{\Lambda}\hat{\bm{\Phi}}' 
	-\sigma^2\bm{I}_p\|^2_F+\frac{1}{2}\|\bm{S} -\hat{\bm{\Phi}}\hat{\bm{\Phi}}'\bm{S}\hat{\bm{\Phi}}\hat{\bm{\Phi}}'\|^2_F \nonumber\\
&~ +\sigma^2\mathrm{tr}( \hat{\bm{\Phi}}\hat{\bm{\Phi}}'\bm{S}\hat{\bm{\Phi}}\hat{\bm{\Phi}}'-\bm{S})+\gamma\|\hat{\bm{\Phi}}
	\bm{\Lambda}\hat{\bm{\Phi}}'\|_*.
\label{eq:appendix_obj}
\end{align}

\noindent From \eqref{eq:Lambda.hat2} and \eqref{eq:appendix_obj}, we have
\begin{align*}
	h(\hat{\bm{\Lambda}}(\sigma^2), \sigma^2)
=&~ \frac{1}{2}\|\hat{\bm{\Phi}}\hat{\bm{\Phi}}'\bm{S}\hat{\bm{\Phi}}\hat{\bm{\Phi}}' - \hat{\bm{\Phi}}\hat{\bm{\Lambda}}(\sigma^2)\hat{\bm{\Phi}}' 
	-\sigma^2\bm{I}_p\|^2_F+\gamma\|\hat{\bm{\Phi}}\hat{\bm{\Lambda}}(\sigma^2)\hat{\bm{\Phi}}'\|_*\\
	&~+\frac{1}{2}\|\bm{S} -\hat{\bm{\Phi}}\hat{\bm{\Phi}}'\bm{S}\hat{\bm{\Phi}}\hat{\bm{\Phi}}'\|^2_F+
	\sigma^2\mathrm{tr}( \hat{\bm{\Phi}}\hat{\bm{\Phi}}'\bm{S}\hat{\bm{\Phi}}\hat{\bm{\Phi}}'-\bm{S})\\
=&~ \frac{1}{2}\sum_{k=1}^{K}\big\{\hat{d}_k^2-(\hat{d}_k-\sigma^2-\gamma)_+^2\big\} +\frac{p}{2}\sigma^4-\sigma^2\mathrm{tr}(\bm{S})
	+\frac{1}{2}\|\bm{S} -\hat{\bm{\Phi}}\hat{\bm{\Phi}}'\bm{S}\hat{\bm{\Phi}}\hat{\bm{\Phi}}'\|^2_F.
\end{align*}
\noindent Minimizing $h(\hat{\bm{\Lambda}}(\sigma^2), \sigma^2)$, we obtain
\begin{equation} 
\hat{\sigma}^2
=\mathop{\arg\min}_{\sigma^2 \geq 0}\bigg\{p\sigma^4-2\sigma^2\mathrm{tr}(\bm{S})-\sum_{k=1}^{K}
(\hat{d}_k-\sigma^2-\gamma)_+^2\bigg\}.
\label{eq:sigma.hat2}
\end{equation}

\noindent Clearly, if $\hat{d}_1\leq \gamma$, then $\hat{\sigma}^2=\displaystyle\frac{1}{p}\mathrm{tr}(\bm{S})$.
We remain to consider $\hat{d}_1>\gamma$. Let
\[
\hat{L}^*=\max\big\{L:\hat{d}_L-\gamma>\hat{\sigma}^2,\,L=1,\dots,K\big\}.
\]
From \eqref{eq:sigma.hat2}, $\hat{\sigma}^2=\displaystyle\frac{1}{p-\hat{L}^*}
\bigg(\mathrm{tr}(\bm{S})-\sum_{k=1}^{\hat{L}^*} (\hat{d}_k-\gamma)\bigg)$. It suffices to show that $\hat{L}^*=\hat{L}$.
Since $\hat{d}_{\hat{L}^*}-\gamma >\displaystyle\frac{1}{p-\hat{L}^*} \bigg(\mathrm{tr}(\bm{S})-\sum_{k=1}^{\hat{L}^*} (\hat{d}_k-\gamma)\bigg)$,
by the definition of $\hat{L}$, we have $\hat{L}\geq \hat{L}^*$, implying $\hat{d}_{\hat{L}}\geq\hat{d}_{\hat{L}^*}$.
Suppose that $\hat{L}>\hat{L}^*$. It immediately follows from the definition of
$\hat{L}^*$ that $\hat{d}_{\hat{L}}-\gamma\leq\hat{\sigma}^2<\hat{d}_{\hat{L}^*}-\gamma$, which contradicts to $\hat{d}_{\hat{L}}\geq
\hat{d}_{\hat{L}^*}$. Therefore, $\hat{L}=\hat{L}^*$. This completes the proof.
\end{proof}
}

\bibliographystyle{imsart-nameyear}
\bibliography{tps}	

\begin{thebibliography}{27}

\bibitem[\protect\citeauthoryear{Boyd et~al.}{2011}]{admm}
\begin{barticle}[author]
\bauthor{\bsnm{Boyd},~\bfnm{Stephen}\binits{S.}},
  \bauthor{\bsnm{Parikh},~\bfnm{Neal}\binits{N.}},
  \bauthor{\bsnm{Chu},~\bfnm{Eric}\binits{E.}},
  \bauthor{\bsnm{Peleato},~\bfnm{Borja}\binits{B.}} \AND
  \bauthor{\bsnm{Eckstein},~\bfnm{Jonathan}\binits{J.}}
(\byear{2011}).
\btitle{Distributed optimization and statistical learning via the alternating
  direction method of multipliers}.
\bjournal{Foundations and Trends in Machine Learning}
\bvolume{3}
\bpages{1-124}.
\end{barticle}
\endbibitem

\bibitem[\protect\citeauthoryear{Cressie and Johannesson}{2008}]{fixedrank}
\begin{barticle}[author]
\bauthor{\bsnm{Cressie},~\bfnm{Noel}\binits{N.}} \AND
  \bauthor{\bsnm{Johannesson},~\bfnm{Gardar}\binits{G.}}
(\byear{2008}).
\btitle{Fixed Rank Kriging for Very Large Spatial Data Sets}.
\bjournal{Journal of the Royal Statistical Society. Series B}
\bvolume{70}
\bpages{209-226}.
\end{barticle}
\endbibitem

\bibitem[\protect\citeauthoryear{d'Aspremont, Bach and Ghaoui}{2008}]{spca2}
\begin{barticle}[author]
\bauthor{\bsnm{d'Aspremont},~\bfnm{Alexandre}\binits{A.}},
  \bauthor{\bsnm{Bach},~\bfnm{Francis}\binits{F.}} \AND
  \bauthor{\bsnm{Ghaoui},~\bfnm{Laurent~El}\binits{L.~E.}}
(\byear{2008}).
\btitle{Optimal solutions for sparse principal component analysis}.
\bjournal{Journal of Machine Learning Research}
\bvolume{9}
\bpages{1269-1294}.
\end{barticle}
\endbibitem

\bibitem[\protect\citeauthoryear{Demsar et~al.}{2013}]{spatial_pca}
\begin{barticle}[author]
\bauthor{\bsnm{Demsar},~\bfnm{Urska}\binits{U.}},
  \bauthor{\bsnm{Harris},~\bfnm{Paul}\binits{P.}},
  \bauthor{\bsnm{Brunsdon},~\bfnm{Chris}\binits{C.}},
  \bauthor{\bsnm{Fotheringham},~\bfnm{A.~Stewart}\binits{A.~S.}} \AND
  \bauthor{\bsnm{McLoone},~\bfnm{Sean}\binits{S.}}
(\byear{2013}).
\btitle{Principal component analysis on spatial data: an overview}.
\bjournal{Annals of the Association of American Geographers}
\bvolume{103}
\bpages{106-128}.
\end{barticle}
\endbibitem

\bibitem[\protect\citeauthoryear{Deser et~al.}{2009}]{sst}
\begin{barticle}[author]
\bauthor{\bsnm{Deser},~\bfnm{Clara}\binits{C.}},
  \bauthor{\bsnm{Alexander},~\bfnm{Michael~A.}\binits{M.~A.}},
  \bauthor{\bsnm{Xie},~\bfnm{Shang-Ping}\binits{S.-P.}} \AND
  \bauthor{\bsnm{Phillips},~\bfnm{Adam~S.}\binits{A.~S.}}
(\byear{2009}).
\btitle{Sea surface temperature variability: patterns and mechanisms}.
\bjournal{Annual Review of Marine Science}
\bvolume{2}
\bpages{115-143}.
\end{barticle}
\endbibitem

\bibitem[\protect\citeauthoryear{Friedman, Hastie and Tibshirani}{2010}]{coord}
\begin{barticle}[author]
\bauthor{\bsnm{Friedman},~\bfnm{Jerome}\binits{J.}},
  \bauthor{\bsnm{Hastie},~\bfnm{Trevor}\binits{T.}} \AND
  \bauthor{\bsnm{Tibshirani},~\bfnm{Robert}\binits{R.}}
(\byear{2010}).
\btitle{Regularization paths for generalized linear models via coordinate
  descent}.
\bjournal{Journal of Statistical Software}
\bvolume{33}
\bpages{1-22}.
\end{barticle}
\endbibitem

\bibitem[\protect\citeauthoryear{Gabay and Mercier}{1976}]{amdd_2}
\begin{barticle}[author]
\bauthor{\bsnm{Gabay},~\bfnm{Daniel}\binits{D.}} \AND
  \bauthor{\bsnm{Mercier},~\bfnm{Bertrand}\binits{B.}}
(\byear{1976}).
\btitle{A dual algorithm for the solution of nonlinear variational problems via
  finite element approximation}.
\bjournal{Computer and Mathematics with Applications}
\bvolume{2}
\bpages{17-40}.
\end{barticle}
\endbibitem

\bibitem[\protect\citeauthoryear{Green and Silverman}{1994}]{nonparametric}
\begin{bbook}[author]
\bauthor{\bsnm{Green},~\bfnm{P.~J.}\binits{P.~J.}} \AND
  \bauthor{\bsnm{Silverman},~\bfnm{B.~W.}\binits{B.~W.}}
(\byear{1994}).
\btitle{Nonparametric regression and generalized linear model: a roughness
  penalty approach}.
\bpublisher{Chapman and Hall}.
\end{bbook}
\endbibitem

\bibitem[\protect\citeauthoryear{Guo et~al.}{2010}]{fused}
\begin{barticle}[author]
\bauthor{\bsnm{Guo},~\bfnm{Jian}\binits{J.}},
  \bauthor{\bsnm{James.},~\bfnm{Gareth}\binits{G.}},
  \bauthor{\bsnm{Levina},~\bfnm{Elizaveta}\binits{E.}},
  \bauthor{\bsnm{Michailidis},~\bfnm{George}\binits{G.}} \AND
  \bauthor{\bsnm{Zhu},~\bfnm{Ji}\binits{J.}}
(\byear{2010}).
\btitle{Principal component analysis with sparse fused loadings}.
\bjournal{Journal of Computational and Graphical Statistics}
\bvolume{19}
\bpages{930-946}.
\end{barticle}
\endbibitem

\bibitem[\protect\citeauthoryear{Hannachi, Jolliffe and
  Stephenson}{2007}]{eofreview}
\begin{barticle}[author]
\bauthor{\bsnm{Hannachi},~\bfnm{A.}\binits{A.}},
  \bauthor{\bsnm{Jolliffe},~\bfnm{Ian~T.}\binits{I.~T.}} \AND
  \bauthor{\bsnm{Stephenson},~\bfnm{D.~B.}\binits{D.~B.}}
(\byear{2007}).
\btitle{Empirical orthogonal functions and related techniques in atmospheric
  science: A review}.
\bjournal{International Journal of Climatology}
\bvolume{27}
\bpages{1119-1152}.
\end{barticle}
\endbibitem

\bibitem[\protect\citeauthoryear{Hong and Lian}{2013}]{hong}
\begin{barticle}[author]
\bauthor{\bsnm{Hong},~\bfnm{Zhaoping}\binits{Z.}} \AND
  \bauthor{\bsnm{Lian},~\bfnm{Heng}\binits{H.}}
(\byear{2013}).
\btitle{Sparse-smooth regularized singular value decomposition}.
\bjournal{Journal of Multivariate Analysis}
\bvolume{117}
\bpages{163-174}.
\end{barticle}
\endbibitem

\bibitem[\protect\citeauthoryear{Huang, Shen and Buja}{2008}]{fda}
\begin{barticle}[author]
\bauthor{\bsnm{Huang},~\bfnm{Jianhua~Z.}\binits{J.~Z.}},
  \bauthor{\bsnm{Shen},~\bfnm{Haipeng}\binits{H.}} \AND
  \bauthor{\bsnm{Buja},~\bfnm{Andreas}\binits{A.}}
(\byear{2008}).
\btitle{Functional principal components analysis via penalized rank one
  approximation}.
\bjournal{Electronic Journal of Statistics}
\bvolume{2}
\bpages{678-695}.
\end{barticle}
\endbibitem

\bibitem[\protect\citeauthoryear{Jolliffe}{1987}]{jolliffe1987rotation}
\begin{barticle}[author]
\bauthor{\bsnm{Jolliffe},~\bfnm{Ian~T.}\binits{I.~T.}}
(\byear{1987}).
\btitle{Rotation of principal components: Some comments}.
\bjournal{Journal of Climatology}
\bvolume{7}
\bpages{507--510}.
\end{barticle}
\endbibitem

\bibitem[\protect\citeauthoryear{Jolliffe}{2002}]{jolliffe2002principal}
\begin{bbook}[author]
\bauthor{\bsnm{Jolliffe},~\bfnm{Ian~T.}\binits{I.~T.}}
(\byear{2002}).
\btitle{Principal component analysis}.
\bpublisher{Wiley Online Library}.
\end{bbook}
\endbibitem

\bibitem[\protect\citeauthoryear{Jolliffe, Uddin and Vines}{2002}]{lassoeof}
\begin{barticle}[author]
\bauthor{\bsnm{Jolliffe},~\bfnm{Ian~T.}\binits{I.~T.}},
  \bauthor{\bsnm{Uddin},~\bfnm{Mudassir}\binits{M.}} \AND
  \bauthor{\bsnm{Vines},~\bfnm{S.~K.}\binits{S.~K.}}
(\byear{2002}).
\btitle{Simplified EOFs--three alternatives to rotation}.
\bjournal{Climate Research}
\bvolume{20}
\bpages{271-279}.
\end{barticle}
\endbibitem

\bibitem[\protect\citeauthoryear{Kang and Cressie}{2011}]{bayes}
\begin{barticle}[author]
\bauthor{\bsnm{Kang},~\bfnm{Emily~L.}\binits{E.~L.}} \AND
  \bauthor{\bsnm{Cressie},~\bfnm{Noel}\binits{N.}}
(\byear{2011}).
\btitle{Bayesian Inference for the Spatial Random Effects Model}.
\bjournal{Journal of the American Statistical Association}
\bvolume{106}
\bpages{972-983}.
\end{barticle}
\endbibitem

\bibitem[\protect\citeauthoryear{Karhunen}{1947}]{karhunen}
\begin{barticle}[author]
\bauthor{\bsnm{Karhunen},~\bfnm{Kari}\binits{K.}}
(\byear{1947}).
\btitle{{\"U}ber lineare methoden in der Wahrscheinlichkeitsrechnung}.
\bjournal{Annales Academi{\ae} Scientiarum Fennic{\ae} Series A}
\bvolume{37}
\bpages{1-79}.
\end{barticle}
\endbibitem

\bibitem[\protect\citeauthoryear{Lo{\`e}ve}{1978}]{loeve}
\begin{bbook}[author]
\bauthor{\bsnm{Lo{\`e}ve},~\bfnm{Michel}\binits{M.}}
(\byear{1978}).
\btitle{Probability theory}.
\bpublisher{Springer-Verlag, New York}.
\end{bbook}
\endbibitem

\bibitem[\protect\citeauthoryear{Lu and Zhang}{2012}]{spca3}
\begin{barticle}[author]
\bauthor{\bsnm{Lu},~\bfnm{Zhaosong}\binits{Z.}} \AND
  \bauthor{\bsnm{Zhang},~\bfnm{Yong}\binits{Y.}}
(\byear{2012}).
\btitle{An augmented Lagrangian approach for sparse principal component
  analysis}.
\bjournal{Mathematical Programming}
\bvolume{135}
\bpages{149--193}.
\end{barticle}
\endbibitem

\bibitem[\protect\citeauthoryear{Ramsay and Silverman}{2005}]{bfda}
\begin{bbook}[author]
\bauthor{\bsnm{Ramsay},~\bfnm{J.~O.}\binits{J.~O.}} \AND
  \bauthor{\bsnm{Silverman},~\bfnm{B.~W.}\binits{B.~W.}}
(\byear{2005}).
\btitle{Functional data analysis},
\bedition{2nd} ed.
\bpublisher{New York: Springer}.
\end{bbook}
\endbibitem

\bibitem[\protect\citeauthoryear{Richman}{1986}]{rotate}
\begin{barticle}[author]
\bauthor{\bsnm{Richman},~\bfnm{Michael~B.}\binits{M.~B.}}
(\byear{1986}).
\btitle{Rotation of principal components}.
\bjournal{Journal of Climatology}
\bvolume{6}
\bpages{293-335}.
\end{barticle}
\endbibitem

\bibitem[\protect\citeauthoryear{Richman}{1987}]{richman1987rotation}
\begin{barticle}[author]
\bauthor{\bsnm{Richman},~\bfnm{Michael~B.}\binits{M.~B.}}
(\byear{1987}).
\btitle{Rotation of principal components: A reply}.
\bjournal{Journal of Climatology}
\bvolume{7}
\bpages{511--520}.
\end{barticle}
\endbibitem

\bibitem[\protect\citeauthoryear{Shen and Huang}{2008}]{Shen}
\begin{barticle}[author]
\bauthor{\bsnm{Shen},~\bfnm{Haipeng}\binits{H.}} \AND
  \bauthor{\bsnm{Huang},~\bfnm{Jianhua~Z.}\binits{J.~Z.}}
(\byear{2008}).
\btitle{Sparse principal component analysis via regularized low rank matrix
  approximation}.
\bjournal{Journal of Multivariate Analysis}
\bvolume{99}
\bpages{1015-1034}.
\end{barticle}
\endbibitem

\bibitem[\protect\citeauthoryear{Tibshirani}{1996}]{lasso}
\begin{barticle}[author]
\bauthor{\bsnm{Tibshirani},~\bfnm{Robert}\binits{R.}}
(\byear{1996}).
\btitle{Regression shrinkage and selection via the Lasso}.
\bjournal{Journal of the Royal Statistical Society. Series B}
\bvolume{58}
\bpages{267-288}.
\end{barticle}
\endbibitem

\bibitem[\protect\citeauthoryear{Tzeng and
  Huang}{2015}]{regularized_covariance}
\begin{barticle}[author]
\bauthor{\bsnm{Tzeng},~\bfnm{ShengLi}\binits{S.}} \AND
  \bauthor{\bsnm{Huang},~\bfnm{Hsin-Cheng}\binits{H.-C.}}
(\byear{2015}).
\btitle{Non-stationary multivariate spatial covariance estimation via low-rank
  regularization}.
\bjournal{Statistical Sinica}
\bvolume{26}
\bpages{151-172}.
\end{barticle}
\endbibitem

\bibitem[\protect\citeauthoryear{Yao, Muller and Wang}{2005}]{fpca_long}
\begin{barticle}[author]
\bauthor{\bsnm{Yao},~\bfnm{Fang}\binits{F.}},
  \bauthor{\bsnm{Muller},~\bfnm{Hans-Georg}\binits{H.-G.}} \AND
  \bauthor{\bsnm{Wang},~\bfnm{Jane-Ling}\binits{J.-L.}}
(\byear{2005}).
\btitle{Functional data analysis for sparse longitudinal data}.
\bjournal{Journal of the American Statistical Association}
\bvolume{100}
\bpages{577-590}.
\end{barticle}
\endbibitem

\bibitem[\protect\citeauthoryear{Zou, Hastie and Tibshirani}{2006}]{spca}
\begin{barticle}[author]
\bauthor{\bsnm{Zou},~\bfnm{Hui}\binits{H.}},
  \bauthor{\bsnm{Hastie},~\bfnm{Trevor}\binits{T.}} \AND
  \bauthor{\bsnm{Tibshirani},~\bfnm{Robert}\binits{R.}}
(\byear{2006}).
\btitle{Sparse principal component analysis}.
\bjournal{Journal of Computational and Graphical Statistics}
\bvolume{15}
\bpages{265-286}.
\end{barticle}
\endbibitem

\end{thebibliography}
\end{document}